\documentclass[floatfix,nofootinbib,twocolumn,preprintnumbers,notitlepage]{revtex4-1}

\usepackage{times}
\usepackage{xcolor}
\usepackage{amsfonts}
\usepackage{amsmath}
\usepackage{amssymb}
\usepackage{bm}
\usepackage{dcolumn}
\usepackage{enumerate}
\usepackage{epsfig}
\usepackage{graphicx}
\usepackage{graphics}
\usepackage[latin1]{inputenc}
\usepackage{latexsym}
\usepackage{rotating}
\usepackage{hyperref}
\usepackage[caption=false]{subfig}
\usepackage[normalem]{ulem}

\newcommand{\<}{\begin{equation}}
\newcommand{\?}{\end{equation}}

\begin{document}

\title{Effect of Ignoring Eccentricity in Testing General Relativity with Gravitational Waves}
\author{Purnima Narayan}
\affiliation{Department of Physics and Astronomy, The University of Mississippi, University, Mississippi 38677, USA}
\author{Nathan~K.~Johnson-McDaniel}
\affiliation{Department of Physics and Astronomy, The University of Mississippi, University, Mississippi 38677, USA}
\author{Anuradha Gupta}
\affiliation{Department of Physics and Astronomy, The University of Mississippi, University, Mississippi 38677, USA}

\date{\today}

\begin{abstract}
Detections of gravitational waves emitted from binary black hole coalescences allow us to probe the strong-field dynamics of general relativity (GR). One can compare the observed gravitational-wave signals with theoretical waveform models to constrain possible deviations from GR. Any physics that is not included in these waveform models might show up as apparent GR deviations. The waveform models used in current tests of GR describe binaries on quasicircular orbits, since most of the binaries detected by ground-based gravitational-wave detectors are expected to have negligible eccentricities. Thus, a signal from an eccentric binary in GR is likely to show up as a deviation from GR in the current implementation of these tests. We study the response of four standard tests of GR to eccentric binary black hole signals with the forecast O4 sensitivity of the LIGO-Virgo network. Specifically, we consider two parameterized tests (TIGER and FTI), the modified dispersion relation test, and the inspiral-merger-ringdown consistency test. To model eccentric signals, we use non-spinning numerical relativity simulations from the SXS catalog with three mass ratios $(1,2,3)$, which we scale to a redshifted total mass of $80M_\odot$ and luminosity distance of $400$~Mpc. For each of these mass ratios, we consider signals with eccentricities of $\sim0.05$ and $\sim 0.1$ at $17$ Hz. We find that signals with larger eccentricity lead to very significant false GR deviations in most tests while signals having smaller eccentricity lead to significant deviations in some tests. For the larger eccentricity cases, one would even get a deviation from GR with TIGER at $\sim 90\%$ credibility at a distance of $\gtrsim 1.5$~Gpc. Thus, it will be necessary to exclude the possibility of an eccentric binary in order to make any claim about detecting a deviation from GR.
\end{abstract}

\maketitle

\section{Introduction}
\label{sec:intro}

At present, general relativity (GR) is the most successful theory of gravity as it explains current astronomical observations and laboratory experiments \cite{Will:2014kxa,Wex:2014nva}. GR has been rigorously tested over the years but no statistically significant deviation has been found yet when tested using solar system observations~\cite{Will:2014kxa}, binary pulsar observations \cite{Wex:2014nva,Weisberg:2004hi,Voisin:2020lqi,Kramer:2021jcw}, and gravitational-wave (GW) observations \cite{GW150914_TGR,GW170817_TGR,O2_TGR,O3a_TGR,O3b_TGR}. Testing GR with GWs from mergers of binary systems has a special significance since it allows us to probe gravity in the highly nonlinear and dynamical regime that is not probed by other tests. In these tests, one compares theoretical waveform models with the data collected by GW detectors such as LIGO~\cite{LIGOScientific:2014pky} and Virgo~\cite{VIRGO:2014yos}. Any disagreement between the models and the data may hint toward a possible deviation from GR (modulo any effects due to non-stationary and non-Gaussian noise in the data). Hence it is crucial to have waveform models that are as accurate as possible, i.e., they should include all known physics in GR and have systematic errors that are well below the statistical errors in the observations. 

The current tests of GR carried out by the LIGO-Virgo-KAGRA collaboration (LVK) that use waveform models that describe the entire signal are based on waveform models designed for coalescing binary black holes (BBHs) on quasicircular orbits and lack information on the eccentricity of the orbit. This is a reasonable choice since binaries formed through the isolated formation channel \cite{Mapelli:2021taw} get efficiently circularized by gravitational radiation \cite{PhysRev.136.B1224,2021arXiv210812210T} and hence are expected to have negligible eccentricities shortly before the merger when their GWs enter the frequency band of ground-based detectors. However, there are other pathways that can lead to a significant eccentricity at small binary separations. For instance, binary formation from primordial black holes (e.g., \cite{Cholis:2016kqi,Wang:2021qsu}), dynamical interactions in dense stellar environments such as galactic cores or globular clusters (e.g., \cite{Wen:2002km,OLeary:2008myb,Antonini:2015zsa,Samsing:2017rat,Samsing:2017xmd,Gondan:2017wzd,Rodriguez:2017pec,Samsing:2017oij,Zevin:2018kzq,Rodriguez:2018pss,Gondan:2020svr,DallAmico:2023neb}), active galactic nuclei (e.g., \cite{Samsing:2020tda,Tagawa:2020jnc}), and the evolution of isolated triple systems (e.g., \cite{Antonini:2013tea,Antognini:2013lpa,Antonini:2017ash}). In these scenarios, the eccentricity could be as high as $\sim 1$ at $10$~Hz.

Various $N$-body simulations on the evolution of binaries in globular clusters (e.g., \cite{Samsing:2017xmd, Rodriguez:2018pss}) suggest that $\gtrsim 5\%$ of binaries can have eccentricities $>0.1$ when their GWs enter the advanced LIGO frequency window. This suggests that at least a fraction of binaries detected by LIGO-Virgo detectors will have non-negligible eccentricity. Recent analyses of data from GWTC-3 \cite{GWTC-3_paper} events found evidence of eccentricity~\citep{Romero-Shaw:2020thy, Gayathri:2020coq,Romero-Shaw:2021ual,Romero-Shaw:2022xko}, but are not able to distinguish between the effects of spin precession and eccentricity at present (see, e.g.,~\cite{Romero-Shaw:2022fbf}) since there are no waveform models including the merger-ringdown portion of the waveform that contain both effects. Moreover, it has been shown that inferred binary parameters will be biased if the detected binaries are on eccentric orbits \cite{Romero-Shaw:2020thy,OShea:2021ugg,Favata:2021vhw}, and nonnegligible residual eccentricity in the ground-based detector band can also mimic a significant deviation from GR \cite{Saini:2022igm,Bhat:2022amc}.

In this paper, we study the effect of ignoring eccentricity while testing GR using some of the standard tests employed by the LVK. Specifically, we consider the Test Infrastructure for GEneral Relativity (TIGER) \cite{TIGER2014,Meidam:2017dgf}, Flexible--Theory-Independent (FTI) \cite{Mehta:2022pcn}, Modified Dispersion Relation (MDR) \cite{Mirshekari:2011yq}, and inspiral-merger-ringdown (IMR) consistency \cite{Ghosh:2016qgn,Ghosh:2017gfp} tests and check their response to simulated eccentric BBH GW signals in the LIGO-Virgo network at its forecast O4 sensitivity~\cite{Aasi:2013wya}. 
To simulate eccentric GW signals, we use numerical relativity (NR) waveforms from the Simulating eXtreme Spacetimes (SXS) catalog~\cite{Boyle:2019kee}. Our simulated observations have the following properties: all binaries are non-spinning, have three mass ratios ($q=1, 2, 3$) and a redshifted total mass of $80M_\odot$, and are observed face-on at a luminosity distance of $400$ Mpc. Moreover, for each mass ratio, we choose NR simulations (from~\cite{Hinder:2017sxy}) where the binary's eccentricity is $\sim0.05$ and $\sim 0.1$ at $17$ Hz for our total mass of $80M_\odot$. For comparison, we also consider a quasicircular NR waveform for each mass ratio.

We found that, as expected, all quasicircular signals are consistent with GR at $90\%$ credibility in all tests except for $q=2, 3$ in the IMR consistency test. We find that the biases obtained in the IMR consistency can be attributed to the inclusion of higher modes in the analysis for the face-on signals we consider, though the same set of higher modes is used in both the simulated signal and the recovery waveform, and this bias is even present when the same waveforms are used for both the simulated signal and recovery. Ongoing studies~\cite{Mukesh_ICTS} have found that these biases are only significant for binaries very close to face-on (or face-off). The signals with lower eccentricity show significant GR deviations in the TIGER and FTI tests for higher order post-Newtonian (PN) testing parameters while the higher-eccentricity signals show very significant GR deviations for almost all testing parameters. Both lower- and higher-eccentricity signals are found to be consistent with GR at $90\%$ credibility for almost all testing parameters in MDR test. On the contrary, the higher-eccentricity signals show strong GR deviations in the IMR consistency test even in an analysis without higher modes. We also study the scaling of the posterior probability distributions of testing parameters with luminosity distance for a few cases. This is because increasing the distance leads to fainter signals which in turn lead to broader posteriors, so any GR deviation that is present might be lost in the statistical error. We found that we can still observe GR deviations at $\sim 90\%$ credibility from eccentric signals placed at distances $\gtrsim 1.5$~Gpc, $\gtrsim 1.2$~Gpc, and $\gtrsim 0.5$~Gpc in the TIGER, FTI, and MDR tests, respectively.

This paper is organized as follows: In Sec.~\ref{sec:tgr}, we give the details of the four tests of GR we consider and in Sec.~\ref{sec:inj} we give the specifics of our simulated observations. In Sec.~\ref{sec:results}, we discuss the results from our analysis and we conclude in Sec.~\ref{sec:concl}. We use geometrized ($G = c = 1$) units throughout.
 
\section{Tests of GR}
\label{sec:tgr}

The tests we consider are all based on waveform models for quasicircular BBHs in GR, viz., IMRPhenomPv2~\cite{Hannam:2013oca,Khan:2015jqa,Bohe:PPv2} (TIGER), SEOBNRv4HM\_ROM~\cite{Cotesta:2018fcv,Cotesta:2020qhw} (FTI), and IMRPhenomXPHM~\cite{Pratten:2020ceb} (IMR consistency and MDR). IMRPhenomPv2 is a frequency-domain phenomenological model which only has the dominant $(l,m)=(2,\pm 2)$ modes in the coprecessing frame and a simple, single-spin model for precession. SEOBNRv4HM\_ROM is a frequency-domain reduced-order model of a (time-domain) aligned-spin effective-one-body model that includes the $(2,\pm 1)$, $(3, \pm 3)$, $(4, \pm 4)$, and $(5, \pm 5)$ modes in addition to the dominant $(2,\pm 2)$ modes. IMRPhenomXPHM is a frequency-domain phenomenological model that improves the accuracy of IMRPhenomPv2, including two-spin precession and the $(2,\pm 1)$, $(3, \pm 3)$, $(3, \pm 2)$, and $(4, \pm 4)$ subdominant modes in the coprecessing frame. The latest LVK testing GR catalog paper~\cite{O3b_TGR} also uses IMRPhenomXPHM for the IMR consistency test and the version without higher modes (IMRPhenomXP) for the MDR test. The latest LVK testing GR catalog paper does not include the TIGER test, but the previous testing GR catalog paper~\cite{O3a_TGR} also uses IMRPhenomPv2. The LVK FTI analyses use SEOBNRv4\_ROM (the version without higher modes) for most events, but the previous testing GR catalog paper~\cite{O3a_TGR} uses SEOBNRv4HM\_ROM when applying FTI to signals with significant evidence for higher modes.

\subsection{TIGER and FTI}
\label{ssec:par}

The TIGER test~\cite{TIGER2014,Meidam:2017dgf} introduces parameterized deviations in the frequency-domain phase of the BBH signal. The version used in, e.g.,~\cite{O2_TGR,O3a_TGR} modifies the phase of the aligned-spin dominant mode IMRPhenomD waveform model~\cite{Khan:2015jqa}, and then this modified phase is twisted up using GR spin precession (as in the unmodified IMRPhenomPv2) to obtain a modified version of the precessing IMRPhenomPv2 waveform. There is a new IMRPhenomXP-based version of TIGER that was not ready for inclusion in~\cite{O3b_TGR} and is not yet publicly available.
The parameterized deviations are introduced in the PN coefficients ($\varphi_k$, $\varphi_{kl}$) in the Fourier-domain inspiral phase of the waveform (leaving off additive constants and phase and time shifts)
\<
\Phi(f)=\frac{3}{128\eta v^5}\sum_{k=0}^7 (\varphi_k v^k + 3\varphi_{kl}v^k \ln v)
\?
(the factor of $3$ in the log term is due to the definition of PN coefficients used in TIGER)
as well as in the phenomenological coefficients in the late-inspiral and merger phases of the waveform ($\beta_k$ and $\alpha_k$). See Table~I in~\cite{GW150914_TGR} for a summary of the frequency dependence of these terms. In the frequency-domain phase expression, $\eta := m_1 m_2/(m_1+m_2)^2$ is the symmetric mass ratio, where $m_{1,2}$ are the binary's individual masses, and $v := (\pi M f)^{1/3}$, where $M := m_1 + m_2$ is the binary's (redshifted) total mass and $f$ is the GW frequency. Additionally, the logarithmic coefficients are only nonzero for $k \in \{5, 6\}$. 

The IMRPhenomD phase is constructed to be $C^1$, so the changes to one of the lower-frequency portions of the phase affects the remainder of the phase due to the $C^1$ matching. For example, the deviations in the PN coefficients also affect the late inspiral and merger-ringdown portions of the signal. Denoting any of these coefficients by $p_k$, the deviation parameter $\delta\hat{p}_k$ is introduced by the replacement $p_k \to (1 + \delta\hat{p}_k)p_k$, except for $\delta\hat{\varphi}_1$ 
which is zero in GR, so we just normalize by $0$PN coefficient. Additionally, for the PN coefficients, the deviation parameter is normalized by the nonspinning portion of the coefficient, to prevent degeneracies in cases where the spins can cause a coefficient to vanish. The deviation parameters are all zero in GR.

While one expects all PN coefficients after a given order to be modified in an alternative gravity theory, we only vary one parameter at a time in our application of TIGER, as in the LVK catalog analyses~\cite{O2_TGR,O3a_TGR}. We do this since one obtains uninformative results when allowing multiple parameters to vary simultaneously, as illustrated for GW150914 in~\cite{GW150914_TGR}, but one can still detect GR deviations that modify multiple PN coefficients (or other testing parameters) 
when varying a single one (even if this testing parameter is itself not modified), as illustrated in~\cite{Meidam:2017dgf,TGR_relation}. However, in the future, it will possible to constrain all PN coefficients at the same time with good accuracy using multiband observations of BBHs \cite{Gupta:2020lxa,Datta:2020vcj}. This is because degeneracies between parameters are removed when combining data from a ground-based detector such as Cosmic Explorer \cite{Evans:2021gyd} and a space-based detector such as LISA \cite{LISA}, which improves the measurement of parameters. Principal component analysis is another method to perform multiparameter tests which is shown to be effective with observations from current~\cite{Shoom:2021mdj,Saleem:2021nsb} and future~\cite{Datta:2022izc,Datta:2023muk} GW detectors.

The FTI test~\cite{Mehta:2022pcn} is similar to TIGER, except it only considers deviations in the PN coefficients and is applicable to any aligned-spin waveform model, though the current implementation of the higher-mode version is restricted to SEOBNRv4HM\_ROM. Additionally, it tapers the deviations to zero above a given frequency instead of letting them affect the rest of the signal. FTI also normalizes the deviation parameter using the full PN coefficient, including the spin contributions, but we reweight the results to the TIGER convention as in the LVK analyses~\cite{O2_TGR, O3a_TGR, O3b_TGR}, for easy comparison. 

\subsection{Modified dispersion relation}
\label{ssec:mdr}

The MDR test introduces a phenomenological dispersion relation, following~\cite{Mirshekari:2011yq}, which gives a frequency-dependent propagation of GWs. Specifically, it considers
\<
E^2 = p^2 + A_\alpha p^\alpha,
\?
where $E$ and $p$ are the energy and momentum of the GWs, while $A_\alpha$ and $\alpha$ are phenomenological parameters that determine the strength of the GR deviation and the frequency dependence of the dispersion, respectively. For $\alpha = 0$ and $A_0 > 0$, this corresponds to the dispersion relation of a massive graviton. As discussed in~\cite{O2_TGR}, it is a good assumption to take the waveform close to the source to be that given by GR to a very good approximation, and the only modification to the waveform is due to the dispersive propagation. In general, this modification is an addition $\propto A_\alpha f^{\alpha - 1}$ to the waveform's frequency-domain phase,\footnote{While the MDR dephasing for $\alpha = 0$ has the same frequency dependance as the $\delta\hat{\varphi}_2$ TIGER/FTI and $\delta\hat{\alpha}_2$ TIGER testing parameters, the MDR dephasing affects the entire signal, while the TIGER and FTI dephasing is only restricted to certain frequencies.} and the magnitude of the dephasing increases with distance (see, e.g.,~\cite{O2_TGR} for the specific expressions).\footnote{The exponent in Eq.~(4) of~\cite{O2_TGR} should be $1/(2 - \alpha)$, as pointed out in~\cite{O3a_TGR}. Additionally, as in, e.g.,~\cite{O3b_TGR}, we use the TT+lowP+lensing+ext cosmological parameters from~\cite{Ade:2015xua} in calculating the dephasing.} As in the LVK analyses (e.g.,~\cite{O3b_TGR}), we consider $\alpha \in\{0, 0.5, 1.5, 2.5, 3, 3.5, 4\}$, where $\alpha = 2$ is omitted since there is no dispersion in this case. We also omit $\alpha = 1$ since the current implementation gives the logarithmic dephasing one gets using the particle velocity
considered in~\cite{Mirshekari:2011yq}, while the expression using the group velocity~\cite{Ezquiaga:2022nak} (which it makes more sense to consider) gives a constant dephasing. Such a constant dephasing is detectable with waveforms including higher modes, but is not implemented in the current implementation of the test in LALSuite~\cite{LALSuite}. Also as in the LVK analyses, we sample in an effective wavelength parameter (given in~\cite{O2_TGR}) and consider the positive and negative $A_\alpha$ cases separately. We then combine together the results for the two different signs of $A_\alpha$ and reweight to a flat prior in $A_\alpha$, as described in~\cite{O2_TGR}.

\subsection{IMR consistency test}
\label{ssec:imrct}

The IMR consistency test~\cite{Ghosh:2016qgn,Ghosh:2017gfp} checks the consistency of the low- and high-frequency portions of a BBH signal. The division between these portions of the signal is made at the median of the ($|m| = 2$) GW frequency of the innermost stable circular orbit (ISCO) of the final Kerr black hole~\cite{Bardeen:1972fi} obtained from the GR analysis of the full signal.
The LVK analysis uses a more involved procedure to obtain the cutoff frequency using the medians of the individual masses and spins. However, we found that this gives negligible differences (at most $1$~Hz) compared to the more straightforward calculation we use. Thus, considering the dominant $(2, \pm 2)$ modes of the waveform, the low- and high-frequency portions of the signal correspond to the inspiral and postinspiral stages of the binary's coalescence.

The test assesses the consistency of the two portions of the signal by inferring the (redshifted) final mass $M_f$ and spin $\chi_f$ from each portion, giving deviation parameters
\<
\frac{\Delta M_f}{\bar{M}_f} := 2\frac{M_f^\text{insp} - M_f^\text{postinsp}}{M_f^\text{insp} + M_f^\text{postinsp}}, \quad \frac{\Delta \chi_f}{\bar{\chi}_f} := 2\frac{\chi_f^\text{insp} - \chi_f^\text{postinsp}}{\chi_f^\text{insp} + \chi_f^\text{postinsp}},
\?
where the ``insp'' and ``postinsp'' superscripts correspond to the low- and high-frequency portions of the signal. These deviation parameters should both be zero if the signal is consistent with the waveform model used in the analysis (which is a quasicircular BBH merger in GR in all current applications).
The final mass and spin are computed as follows. One first performs parameter estimation analysis for each portion of the signal using a standard BBH waveform (in our case, IMRPhenomXPHM) parameterized by the binary's initial masses and spins. One then computes the final mass and spin using an average of fits to NR simulations~\cite{Hofmann:2016yih,Healy:2016lce,Jimenez-Forteza:2016oae}.\footnote{We augment the aligned-spin final spin fits with the contribution from in-plane spins~\cite{spinfit-T1600168}, but as in~\cite{O2_TGR, O3a_TGR, O3b_TGR}, we do not evolve the initial spins before applying the fits.} As in~\cite{O3a_TGR, O3b_TGR}, we reweight to a flat prior in the deviation parameters to obtain the final results.

\section{Simulated observations and parameter estimation setup}
\label{sec:inj}

We consider simulated BBH observations in the LIGO-Virgo network with the forecast O4 sensitivity~\cite{Aasi:2013wya}---we use the more sensitive LIGO noise curve and do not include KAGRA since it is expected to be much less sensitive than LIGO and Virgo in O4~\cite{timeline_graphic}. We also do not include noise in our simulated observations (i.e., taking the zero realization of Gaussian noise) in order to avoid biases due to specific noise realizations. We model the BBH waveforms using a selection of nonspinning NR simulations from the SXS catalog~\cite{Boyle:2019kee}. In particular, all the eccentric waveforms are ones used in~\cite{Hinder:2017sxy}, since that paper gives the eccentricities at a fixed dimensionless frequency obtained by comparison with a PN waveform. We consider cases with eccentricities around $0.05$ and $0.1$, for comparison. We do not consider the higher eccentricity simulations from that paper, since they are not long enough to include all the power starting at $20$~Hz for our chosen total mass of $80M_\odot$.
We also consider quasicircular waveforms with the same mass ratios, for comparison. We give the properties of all the simulations we consider in Table~\ref{tab:sims}. We use the $N = 2$ extrapolated waveforms and the highest resolution simulation available in the SXS catalog.

\begin{table}
\caption{\label{tab:sims} The SXS simulations we consider and their properties. All of these simulations have negligible spins. The eccentricities given are those from~\cite{Hinder:2017sxy}, which are quoted at a PN velocity squared of $0.075$ (so an $|m| = 2$ GW frequency of $\sim 17$~Hz for the $80M_\odot$ binaries we consider), except for the ones that give an upper bound of $10^{-4}$. For these, we quote an upper bound that is greater than the eccentricities quoted in the SXS metadata. Those eccentricities come from the eccentricity reduction procedure, which is not designed to measure nonzero values of eccentricity.}
\begin{tabular}{*{4}{c}}
\hline\hline
ID & & mass ratio & eccentricity\\
\hline
SXS:BBH:1155 & & $1$ & $< 10^{-4}$\\
SXS:BBH:1355 & & $1$ & $0.053$\\
SXS:BBH:1357 & & $1$ & $0.097$\\[0.5em]
SXS:BBH:1222 & & $2$ & $< 10^{-4}$\\
SXS:BBH:1364 & & $2$ & $0.044$\\
SXS:BBH:1368 & & $2$ & $0.097$\\[0.5em]
SXS:BBH:2265 & & $3$ & $< 10^{-4}$\\
SXS:BBH:1371 & & $3$ & $0.055$\\
SXS:BBH:1373 & & $3$ & $0.093$\\
\hline\hline
\end{tabular}
\end{table}

We choose a (redshifted) total mass of $80M_\odot$ so that the simulation is long enough that the binary's entire signal is in the detectors' sensitive band starting from a Fourier frequency of $20$~Hz. We consider a face-on signal (inclination angle $0$) so that only the $m = 2$ [spin-$(-2)$-weighted] spherical harmonic modes of the signal contribute, and thus we do not have to worry about the $m > 2$ modes, which would require significantly longer NR simulations in order for the signal to include all the power down to $20$~Hz without a much higher total mass ($\sim 120M_\odot$ for the current simulations even if only including the $|m| = 3$ modes). For the other extrinsic parameters, we choose a luminosity distance of $400$~Mpc, similar to GW150914~\cite{GW150914}, and a randomly chosen sky location (right ascension, declination of $3.19$, $-0.14$~rad), polarization angle ($1.53$~rad), and coalescence GPS time ($1129708949$). We also consider selected cases at a larger distance, for comparison, as discussed in Sec.~\ref{ssec:low_snr}. In GW data analysis terminology we often refer to these simulated GW observations as {\it injections}, which we will henceforth use in the paper.

We choose the same (spin-weighted spherical harmonic) mode content in the injections as the (coprecessing frame) modes present in the waveform models used in the tests, to avoid any biases due to missing modes. Since only the $m = 2$ mode contributes in our face-on case, that means that we just include the $(2, 2)$ mode in the injections to which we apply TIGER and FTI tests (using IMRPhenomPv2 and SEOBNRv4HM\_ROM, respectively) and also include the $(3, 2)$ mode in the injections to which we apply the MDR and IMR consistency tests (using IMRPhenomXPHM). The $(3,2)$ mode makes a $\sim 10\%$ correction to the $(2,2)$ mode's amplitude; in general the maximum amplitudes of the $(\ell, 2)$ modes scale roughly as $10^{2-\ell}$ with respect to the $(2,2)$ mode, so the $(4,2)$ and higher modes make a $\sim1\%$ correction. Thus, we do not expect significant differences between our results and the results one would obtain when applying these tests to a real eccentric signal with the same parameters as our signals. We obtain network signal-to-noise ratios (SNRs) for our $(2,2)$ + $(3,2)$ [$(2,2)$ only] injections of about $120$ ($116$), $106$ ($103$), and $90$ ($87$) for mass ratios $q=m_1/m_2=1$, $2$, and $3$, respectively. The SNR values given are those for the lower-eccentricity injections, rounded to the nearest integer. The SNRs of the quasicircular and higher-eccentricity injections differ by $<1$.

We perform the parameter estimation using the implementation of nested sampling~\cite{Skilling2004a} in the LALInference code~\cite{Veitch:2014wba} in the LALSuite software library~\cite{LALSuite}. We use a lower frequency of $20$~Hz and upper frequency of $512$~Hz in analyzing the injections, except for the different upper and lower frequencies used in the inspiral and postinspiral analyses, respectively, for the IMR consistency test.
We use the same priors on the GR parameters as in the LVK applications of these tests, which are the same as the GR parameter estimation analyses (see, e.g., the discussion in Appendix~E of~\cite{GWTC-3_paper}), except with larger ranges in some cases to account for correlations with non-GR parameters and with a prior on the luminosity distance that is uniform in Euclidean volume, instead of the more complicated prior uniform in comoving frame merger rate used in the LVK GR parameter estimation analyses. Specifically, we use uniform priors in redshifted masses and spin magnitudes as well as isotropic priors in spin directions, binary orientation, and sky location. The priors on the non-GR parameters are all flat.

\section{Results}
\label{sec:results}

We now discuss the results obtained when performing the four tests of GR described in Sec.~\ref{sec:tgr} on the simulated eccentric signals discussed in Sec.~\ref{sec:inj}. 

\subsection{TIGER}
\label{ssec:tiger}

\begin{figure*}
\includegraphics[width=0.95\linewidth]{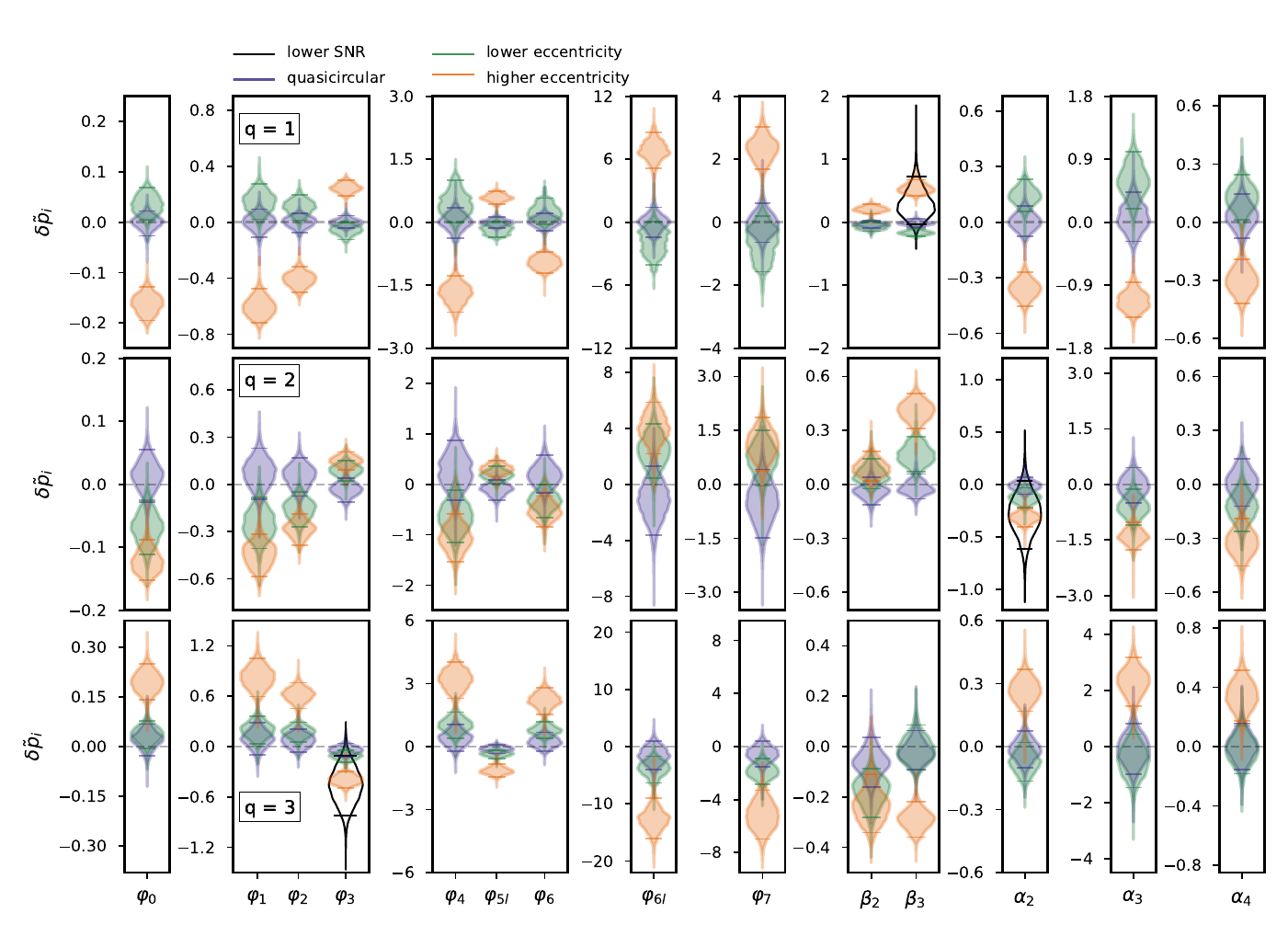}
\caption{\label{fig:TIGER_plot}The results of the TIGER test on the quasicircular, lower-eccentricity, and higher-eccentricity simulated injections of mass ratios $1$, $2$, and $3$ in the top, middle and bottom panels, respectively. The posteriors of the testing parameters are presented as violin plots and the associated $90\%$ credible intervals are labelled as horizontal bars. We mark the GR value of zero with dashed lines. We also show the results for the lower SNR injections, simulated by scaling the distance of selected higher-eccentricity runs for each mass ratio, as black unfilled violin plots. Details about the lower SNR cases are given in Sec.~\ref{ssec:low_snr}.}

\end{figure*}																					

We give the posterior probability distributions (henceforth posterior distributions or posteriors) of the TIGER testing parameters for all our injections in Fig.~\ref{fig:TIGER_plot}.
As expected, results from quasicircular injections for all three mass ratios are consistent with GR at $90\%$ credibility. For the lower-eccentricity injections, we find that GR is excluded at $\gtrsim 90\%$ credibility in almost all cases. The higher-eccentricity injections all show strong deviations from GR, with GR excluded at $>90\%$ credibility (often well above this) for all three mass ratios and all testing parameters.

\begin{figure*}
\includegraphics[width=0.95\textwidth]{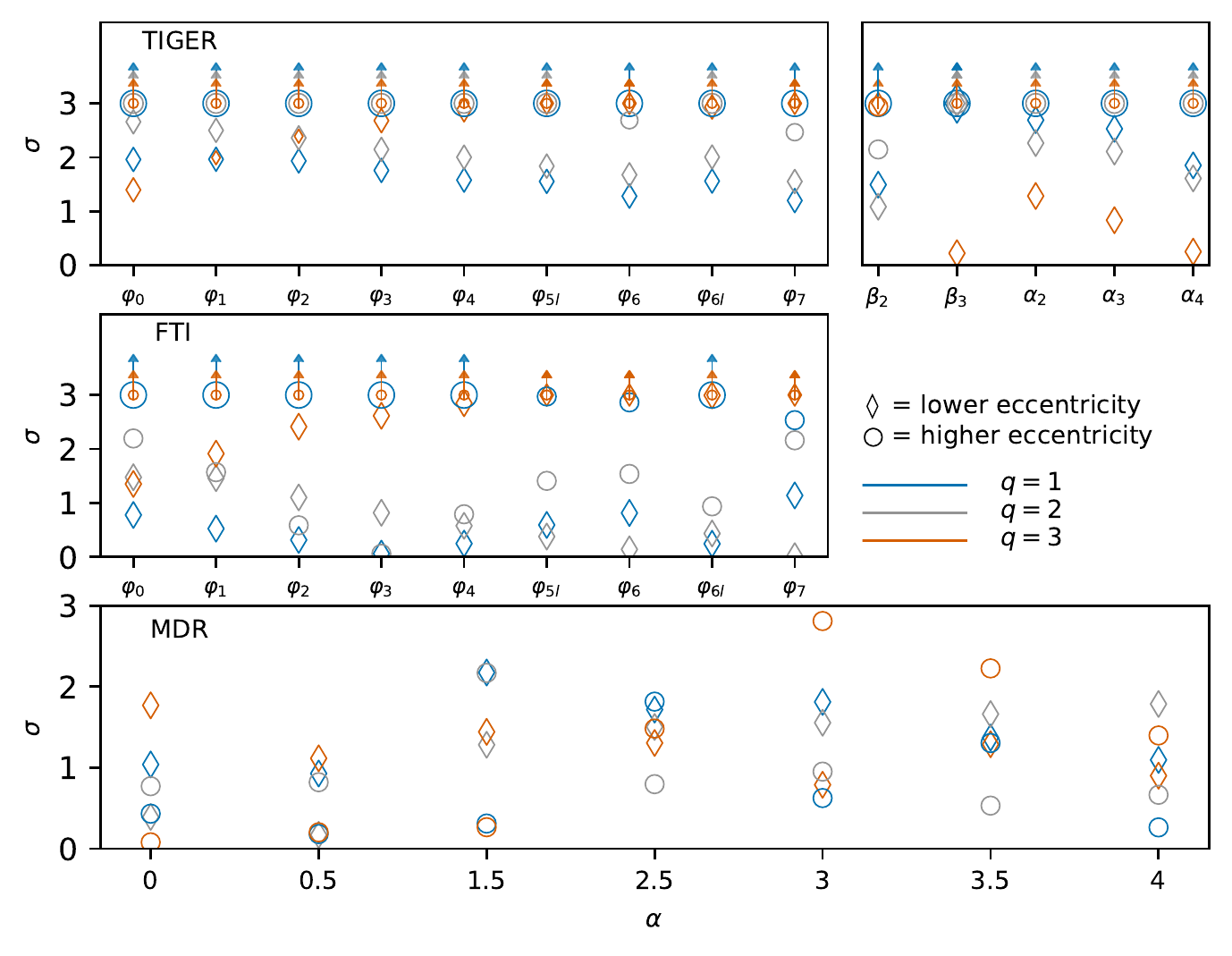}
\caption{\label{fig:quantile_plot} The Gaussian sigma value at which GR is excluded for the eccentric injections with the TIGER, FTI and MDR tests. A lower bound of $3\sigma$ is denoted using an upward arrow, above which the values cannot be stated with certainty from the order of $10^4$ posterior samples in our analyses. The sizes of different markers and lengths of arrows have no significance and are varied to avoid overlaps with other data points as much as possible.   
}
\end{figure*}

We summarize the statistical level at which GR is excluded in Fig.~\ref{fig:quantile_plot}, giving the equivalent Gaussian sigmas.
However, we find that GR is excluded at such high credible levels in some cases that we cannot trust that the GR quantile is estimated accurately with the $\sim 10^4$ posterior samples we obtain. In order to estimate an appropriate lower bound in such cases, we drew $1.8\times 10^4$ samples from a Gaussian and compared the analytically computed Gaussian sigma values with the ones obtained using the same kernel density estimator (KDE) calculation applied to the results of the tests of GR. We chose this number of samples to be similar to (and on the lower side of) the number of samples we obtain for many of our analyses. We also varied the mean and standard deviation of the Gaussian to produce different GR quantiles and to reproduce the rough properties of the posteriors we obtain for the testing parameters. We found that Gaussian sigma values above around $3\sigma$ had absolute errors (comparing the KDE and analytic results) of more than $0.1$, so we quote a lower bound of $3\sigma$ on significances. 

We find that GR is excluded at $>3\sigma$ for all testing parameters for the $q=1$ higher-eccentricity injections. GR is also excluded with $>3\sigma$ for the $q = 2,3$ higher-eccentricity injections with the exception of $\delta\hat{\varphi}_6$, $\delta\hat{\varphi}_7$, and $\delta\hat{\beta}_2$ for $q=2$ and $\delta\hat{\beta}_2$ for $q=3$, though in all of these cases GR is excluded at $>2\sigma$ and close to $3\sigma$ in some cases. The lower-eccentricity $q=1$ and $q=2$ injections exclude GR at $<3\sigma$ for all testing parameters with the exception of $\delta\hat{\beta}_{3}$, where it is excluded at $>3\sigma$. For the $q=3$ lower-eccentricity injection, GR is excluded at $>3\sigma$ only for the $\delta\hat{\varphi}_{5l}$, $\delta\hat{\varphi}_6$, $\delta\hat{\varphi}_7$, and $\delta\hat{\beta}_2$ testing parameters.

In Fig.~\ref{fig:TIGER_plot}, we notice that the sign of the deviation from GR for a given mass ratio and eccentricity is different for different testing parameters. This is due to the PN coefficients and phenomenological parameters used in the normalization themselves having different signs. We also see that all the testing parameters are on the opposite sides of zero for the lower- and higher-eccentricity injections for $q = 1$. This can be attributed to these cases being well outside of the linear regime of the test's response to eccentricity, and these cases generally have significantly different sets of GR and non-GR parameters giving the best agreement with the observed signal for the two eccentricities. For instance, for $\delta\hat\varphi_0$ and $q = 1$, the chirp mass is biased to larger values for the smaller eccentricity and smaller values for the larger eccentricity, while for the same testing parameter and $q = 2$, the bias in the GR parameters generally increases monotonically from the smaller to the larger eccentricity, as does the value of the testing parameter.  

Additionally, we find that the sign of a given testing parameter is different for different mass ratios. For the higher-eccentricity cases, there is generally a monotonic dependence of the value of the testing parameter on mass ratio, but for the lower-eccentricity cases, the signs of the PN coefficient testing parameters are the same for $q = 1$ and $q = 3$ and opposite those for $q = 2$. We investigate this difference in signs as follows: We first take the $3.5$PN accurate quasicircular TaylorF2 nonspinning inspiral phase \cite{Blanchet:2004ek,Buonanno:2009zt} and add the TIGER testing parameters at each PN order. We compare this phase with the eccentric PN inspiral phase from Moore \emph{et al.}~\cite{Moore:2016qxz} which incorporates the effect of eccentricity to $3$PN order (but is $3.5$PN accurate in the quasicircular terms) and the leading-order (quadratic) terms in eccentricity. For each testing parameter, we obtain the value that minimizes the least-squares difference between the two phases with all the GR parameters fixed to the same values. However, we do not find any indication of a sign flip, and also find that this analysis returns values of the testing parameter of the order $10^{-3}$, significantly smaller than what we find in the full analysis, suggesting that the merger-ringdown portion of the signal is quite important here. Nevertheless, we do find that the frequency derivative of the eccentric PN phase depends nonmonotonically on mass ratio (the ordering of the value by mass ratio is $1$, $3$, $2$), showing that there is some nonmonotonicity present in the PN results.

We now compare our results with those from Saini \emph{et al.}~\cite{Saini:2022igm} which also studied the effect of ignoring eccentricity when performing the parameterized test of PN coefficients (though they do not consider the $\delta\hat{\varphi}_{1}$ testing parameter). The authors compare the expected value of the deviation parameters due to eccentricity using the formalism from~\cite{Cutler:2007mi}, which is based on the Fisher matrix approach \cite{Cutler:1994ys,Poisson:1995ef} that is used to obtain a prediction for the statistical errors in the deviation parameters. They only consider the inspiral portion of the signal and model the eccentric waveforms by adding the nonspinning eccentric contribution to the $3$PN frequency domain phase from Moore \emph{et al.}~\cite{Moore:2016qxz} to the aligned-spin $3.5$PN GR phase~\cite{Blanchet:2004ek,Arun:2008kb,Buonanno:2009zt,Mishra:2016whh}, which is also what they use to model the GR signals. Out of all the cases that Saini \emph{et al.}\ considered, we make comparisons only for binaries with total redshifted masses of $72M_\odot$ and $99M_\odot$ (source-frame masses of $65M_\odot$ and $90M_\odot$) since they are the closest to the $80M_\odot$ used for our injections. Saini \emph{et al.}\ only consider a mass ratio of $2$, include aligned spins of $0.5$ and $0.4$, place the binaries at a distance of $500$ Mpc, and specify the binaries' eccentricity at $10$~Hz. Since we are using the eccentricity values from~\cite{Hinder:2017sxy}, which are given at the PN velocity squared of $0.075$, corresponding to $\sim 17$~Hz for our $80M_\odot$ binary, we use Eqs.~(4.17) in~\cite{Moore:2016qxz} to obtain an estimate of our lower and higher eccentricities at $10$~Hz, giving $0.08$ and $0.18$, respectively, for the $q = 2$ case. We also have to convert the deviation parameters from the FTI convention (i.e., including the spinning terms in the PN coefficient used for the scaling) used by Saini \emph{et al.}\ to the TIGER convention (i.e., only scaling by the nonspinning PN terms) that we use. We do this roughly using their injected parameters, which give scaling factors of $0.70$, $0.75$, $0.23$, $0.029$, and $-1.2$ (ratios of full to nonspinning PN coefficients) for $\delta\hat{\varphi}_{3}$, $\delta\hat{\varphi}_{4}$, $\delta\hat{\varphi}_{5l}$, $\delta\hat{\varphi}_{6}$, and $\delta\hat{\varphi}_{7}$. For the other cases, the PN coefficients have no dependence on the spin, so no rescaling is necessary.

Comparing the standard deviations of our posterior distributions of inspiral testing parameters for $q = 2$ to the statistical biases shown in Fig.~1 of Saini \emph{et al.}, scaling their results to our SNRs, we found that the Saini \emph{et al.}\ statistical errors are larger by a factor of $\sim 2$--$9$ (smallest for $\delta\hat{\varphi}_{3}$ and largest for $\delta\hat{\varphi}_{4}$) with the exception of $\delta\hat{\varphi}_{6}$, $\delta\hat{\varphi}_{6l}$, and $\delta\hat{\varphi}_{7}$, where the Saini \emph{et al.}\ statistical errors are smaller than our standard deviations by a factor of $\sim 3$--$5$. (When making these comparisons here and below we always quote the smaller of the two differences between our results and the Saini \emph{et al.}\ results for total masses of $72M_\odot$ and $99M_\odot$.) For these high-PN order coefficients, the smaller errors found by Saini \emph{et al.}\ may be because they take the inspiral to extend up to the ISCO frequency of the final black hole ($Mf \simeq 0.06$ in their case), while the TIGER testing parameters are only applied up to the end of the IMRPhenomD inspiral phase ($Mf = 0.018$). Additionally, while we scale the Saini \emph{et al.}\ statistical errors to our SNR to make the results more comparable, there are many differences between our two analyses, so the significant differences we find are likely not unexpected. In particular, in addition to the differences in statistical methods and waveforms, Saini \emph{et al.}\ also only consider a single LIGO detector with a slightly older noise curve, and use a lower-frequency cutoff of $10$~Hz, while we use $20$~Hz.

We also compare the median of our $q = 2$ posteriors with the systematic biases shown in Fig.~1 of Saini \emph{et al.} In general, we find that the agreement is better for the lower PN coefficients ($k \leq 4$). Specifically, for both eccentricities, we found that our median is contained within the Saini \emph{et al.}\ range of systematic biases for the two total masses for $\delta\hat{\varphi}_{0}$ and $\delta\hat{\varphi}_{4}$. Additionally, for the larger eccentricity our median is only $\sim 5\%$ smaller than the result for the larger total mass for $\delta\hat{\varphi}_{2}$ and $\sim 20\%$ larger than the result for the smaller total mass for $\delta\hat{\varphi}_{3}$. For $\delta\hat{\varphi}_{5l}$, the Saini \emph{et al.}\ systematic bias is $\sim 5$ times ($\sim 20\%$) larger than our median for the larger (smaller) eccentricity. For all the $\delta\hat{\varphi}_{k}$ not yet mentioned for a given eccentricity, the Saini \emph{et al.}\ systematic bias is smaller than our median.
For the larger eccentricity, the Saini \emph{et al.}\ systematic bias is smaller by a factor of $\sim 10$ for $\delta\hat{\varphi}_{6}$ and $\delta\hat{\varphi}_{7}$, while it is smaller by a factor of $\sim 20$ for $\delta\hat{\varphi}_{6l}$. For the smaller eccentricity, the Saini \emph{et al.}\ systematic bias is smaller by factors of $\sim 2$ (for $\delta\hat{\varphi}_{3}$) to $\sim 60$ (for $\delta\hat{\varphi}_{6l}$).

It is also useful to compare the statistical level at which GR is excluded in the two analyses. Here we can compare the ratio of the systematic bias to statistical error (scaled to our SNR) from Saini \emph{et al.}\ with the Gaussian sigma equivalents to our GR quantiles (as given in Fig.~\ref{fig:quantile_plot}). We find that both analyses agree that there is a significant GR deviation for the larger eccentricity and $\delta\hat{\varphi}_{0}$, $\delta\hat{\varphi}_{2}$, and $\delta\hat{\varphi}_{4}$. However, the Saini \emph{et al.}\ systematic biases are smaller or comparable to their (SNR scaled) statistical errors for the other PN coefficients and the higher eccentricity, as well as for the smaller eccentricity, even though we find significant ($>2\sigma$) GR deviations for most testing parameters in those cases. Nevertheless, the Saini \emph{et al.}\ study finds that one can obtain a significant GR deviation for an eccentricity of $\sim 0.1$ for the smaller total masses (particularly $15M_\odot$) for which their inspiral-only analysis is more reliable, so their overall conclusions are in agreement with those of our study.

Finally, we consider the biases in the GR parameters: As mentioned above for $\delta\hat{\varphi}_0$, we find that for $q = 1$, the chirp mass is biased in different directions for the smaller and larger eccentricity cases; we also find that the bias is in the opposite direction with the $\delta\hat{\alpha}_k$ deviation parameters than with the other deviation parameters. This bias primarily comes from a bias in the total mass, and we also find that there are biases in the total mass and chirp mass for the other mass ratios, though the biases are in the same direction for both eccentricity values for almost all testing parameters, and for $q = 3$, the total mass is consistent with the injected value for the PN deviation parameters. We also find biases in the effective spin~\cite{Racine:2008qv,Santamaria:2010yb}, and find significant support for nonzero values of the effective precession spin parameter $\chi_p$~\cite{Hannam:2013oca, Schmidt:2014iyl} for many cases, particularly for the higher eccentricities and $\delta\hat{\varphi}_k$ deviation parameters. There are even no samples near $\chi_p = 0$ for a few testing parameters in the higher-eccentricity cases. However, we do not find that the cases with larger support for precession have smaller GR deviations, as one might think would be the case (i.e., that the precession was absorbing some of the GR deviation). In fact, we usually finds both more support for precession and a larger GR deviation when increasing the eccentricity, though we generally do not find correlations between the testing parameter and $\chi_p$. However, we find significant correlations between the testing parameter and $\chi_\text{eff}$. We also find biases in the distance (both to larger and smaller values) and inclination angle, particularly for the larger eccentricity cases.

\subsection{FTI}

\begin{figure*}
\includegraphics[width=0.95\linewidth]{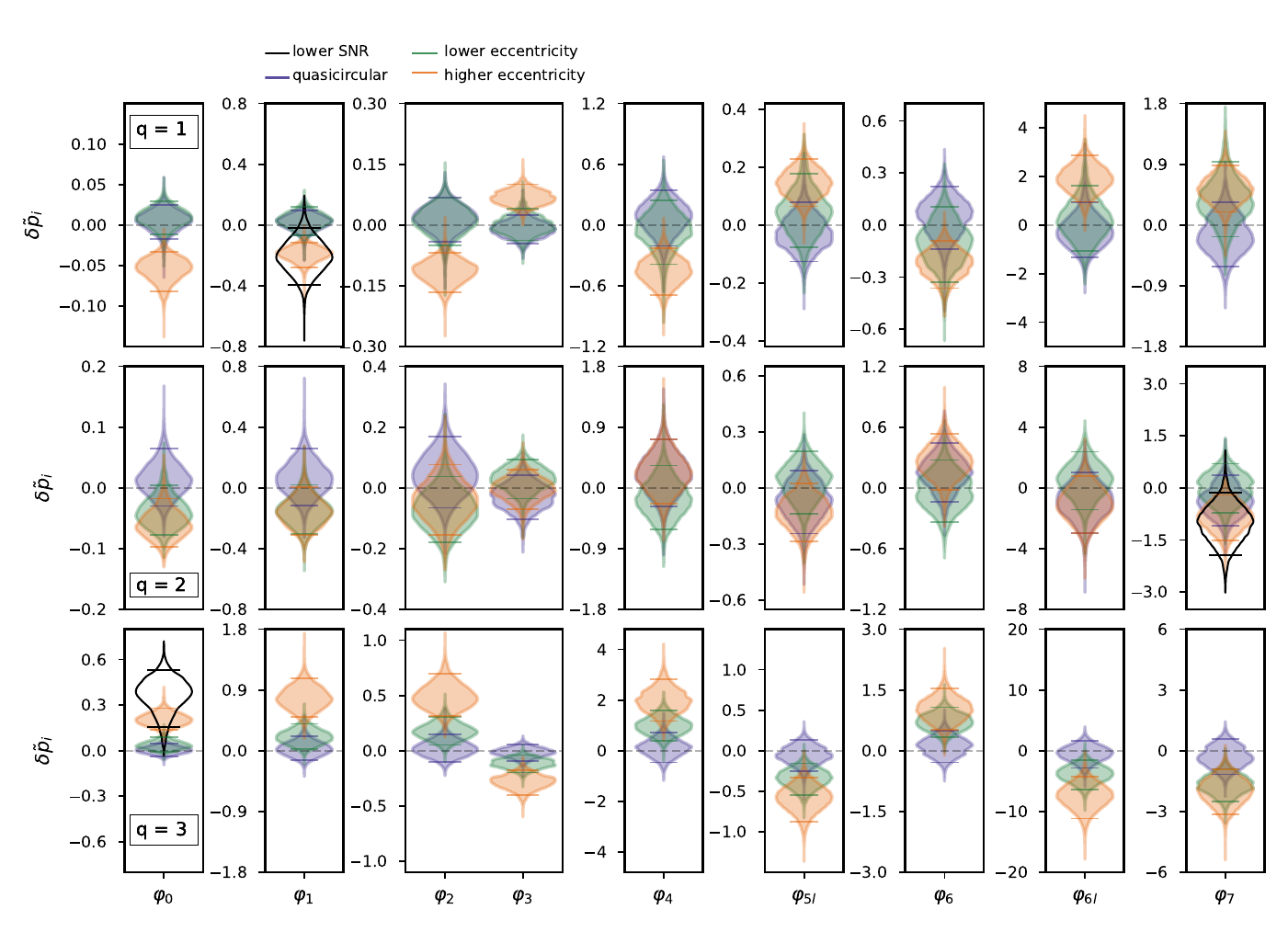}
\caption{\label{fig:FTI_plot} Violin plots for FTI testing parameters. The color scheme and the layout of the subplots are similar to Fig.~\ref{fig:TIGER_plot}.}
\end{figure*}

Similar to Fig.~\ref{fig:TIGER_plot}, Fig.~\ref{fig:FTI_plot} displays the posterior distributions of FTI testing parameters as violin plots.
As we anticipate, all quasicircular injections are consistent with GR at $90\%$ credibility. The $q = 1$ and $2$ lower-eccentricity injections are both consistent with GR at $90\%$ credibility. However, for the $q = 3$ lower-eccentricity injection, GR is excluded at $> 90\%$ credibility for all but $\delta\hat{\varphi}_{0}$, with the credible level at which GR is excluded increasing with increasing PN order. 
For the higher-eccentricity injections, the testing parameters show significant deviations from GR for $q=1$ and $q=3$ while moderate to no deviation for $q=2$, which is consistent with GR at $90\%$ credibility for most testing parameters. We refer to Fig.~\ref{fig:quantile_plot} again for the corresponding GR quantiles for all of our injections. We find that for the $q=1$ higher-eccentricity injection, GR is excluded at $>3\sigma$ for all testing parameters, except for $\delta\hat{\varphi}_{5l}$, $\delta\hat{\varphi}_{6}$, and $\delta\hat{\varphi}_{7}$. For the $q=2$ higher-eccentricity injection, there is consistency with GR at $<2\sigma$, except for the $\delta\hat{\varphi}_{0}$ and $\delta\hat{\varphi}_{7}$ testing parameters, where GR is excluded at slightly above $2\sigma$.\footnote{We checked that the significant difference between the TIGER and FTI results for $q = 2$ is not due to the inclusion of higher modes in the FTI analysis by applying FTI for $\delta\hat{\varphi}_3$ with just the dominant $l = |m| = 2$ modes to an injection containing just these modes and found that the posterior still peaks at $0$.} For $q=3$, we again find that GR is excluded at $>3\sigma$ for all testing parameters for the higher-eccentricity injection and for $\delta\hat{\varphi}_{5l}$, $\delta\hat{\varphi}_{6}$ for the lower-eccentricity injection. We see the same general monotonic dependence of the value of the testing parameters with mass ratio seen for TIGER, but do not see the difference in signs for the lower- and higher-eccentricity cases seen for $q = 1$ with TIGER.

Comparing the standard deviations and medians of our FTI posterior distributions with the results from Saini \emph{et al.}\ as we did for TIGER, we find the same relation between the statistical errors that we did with TIGER. The medians we obtain with FTI are all smaller in magnitude than those obtained with TIGER, so we find that for the smaller eccentricity they are contained between the values for the two (redshifted) total masses considered by Saini \emph{et al.}\ ($72M_\odot$ and $99M_\odot$) for $\delta\hat{\varphi}_{0}$, $\delta\hat{\varphi}_{2}$, and $\delta\hat{\varphi}_{4}$. Our median is $\sim 2$ times larger for $\delta\hat{\varphi}_{3}$ and $\delta\hat{\varphi}_{6}$, $\sim 30$ times larger for $\delta\hat{\varphi}_{6l}$, and $\sim 6$ ($\sim 2$) times smaller for $\delta\hat{\varphi}_{5l}$ ($\delta\hat{\varphi}_{7}$). For the larger eccentricity, only the median for $\delta\hat{\varphi}_{0}$ is between the two boundaries. For $\delta\hat{\varphi}_{6}$, $\delta\hat{\varphi}_{6l}$, and $\delta\hat{\varphi}_{7}$, our median is $\sim 5$--$8$ times larger than the systematic bias obtained by Saini \emph{et al.}, but in all the other cases our median is smaller than the Saini \emph{et al.}\ systematic bias, only by a factor of $\sim 3$ for $\delta\hat{\varphi}_{4}$, but up to a factor of $\sim 20$ for $\delta\hat{\varphi}_{3}$. Since the FTI GR deviations are not as large as the TIGER ones for $q = 2$, there are not many cases as for TIGER where we find a significant GR deviation with FTI but Saini \emph{et al.}\ find a systematic bias that is comparable to or smaller than the statistical error scaled to our SNR, though $\delta\hat{\varphi}_{7}$ larger eccentricity is a notable such case.

We find in general the same sorts of biases in the GR parameters as for TIGER, except mostly smaller (as is likely expected for the smaller parameter space of nonprecessing systems considered here), and without any significant bias in the inclination angle. However, we do find that there are notable biases to larger total masses, more equal mass ratios, and larger distances for the $q = 3$ higher-eccentricity case.

\subsection{MDR}
\begin{figure*}
\includegraphics[width=0.95\linewidth]{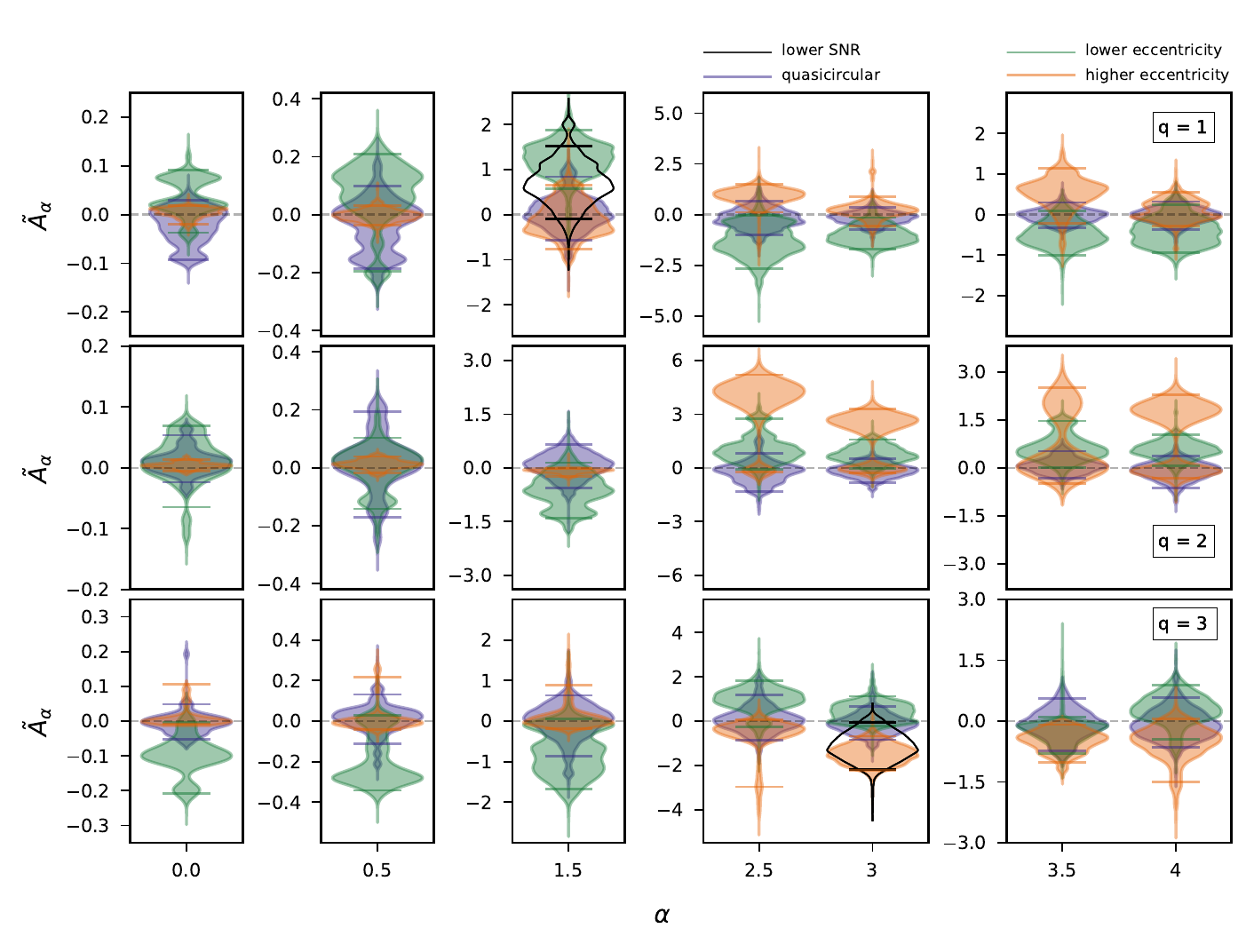}
\caption{\label{fig:MDR_plot} Violin plots for the modified dispersion relation dimensionless parameter $\tilde{A}_\alpha:=10^{43-12\alpha}A_\alpha/{\rm{eV}^{2-\alpha}}$ for different values of the modified dispersion relation exponent $\alpha$, where the scaling is chosen to keep the plotted results of order unity. The color scheme and layout of subplots are similar to Fig.~\ref{fig:TIGER_plot}. The $q=2$ lower-SNR result is omitted due to difficulty in obtaining reliable results.
}
\end{figure*}

Figure~\ref{fig:MDR_plot} illustrates the posteriors of the modified dispersion parameter $\tilde{A}_\alpha$ for different values of the modified dispersion relation exponent $\alpha$. All the cases considered are consistent with GR at $90\%$ credibility except for the $q=2$, $\alpha = 1.5$; $q=3$, $\alpha = 3$; and $q=3$, $\alpha = 3.5$ higher-eccentricity and $q=1$, $\alpha = 1.5$ lower-eccentricity cases where GR is excluded at $>2\sigma$. However, GR is excluded at $<3\sigma$ for all testing parameters as illustrated in Fig.~\ref{fig:quantile_plot}. The lack of significant GR deviation in MDR test means that the modification to the waveform from the modified dispersion is mostly orthogonal to the modification introduced due to eccentricity (compared to the manifold of quasicircular waveforms). We find that in most cases, there is significant support for precession (usually well-constrained $\chi_p$ posteriors with no posterior samples near zero), except in a few lower-eccentricity $q = 2$ cases. These $\chi_p$ posteriors are generally considerably narrower than those for TIGER. There are also small biases in the effective spin (in both directions) in some cases, as well as biases in the total mass to slightly larger values.

\subsection{IMR Consistency Test}
\begin{figure*}
\includegraphics[width=0.45\textwidth]{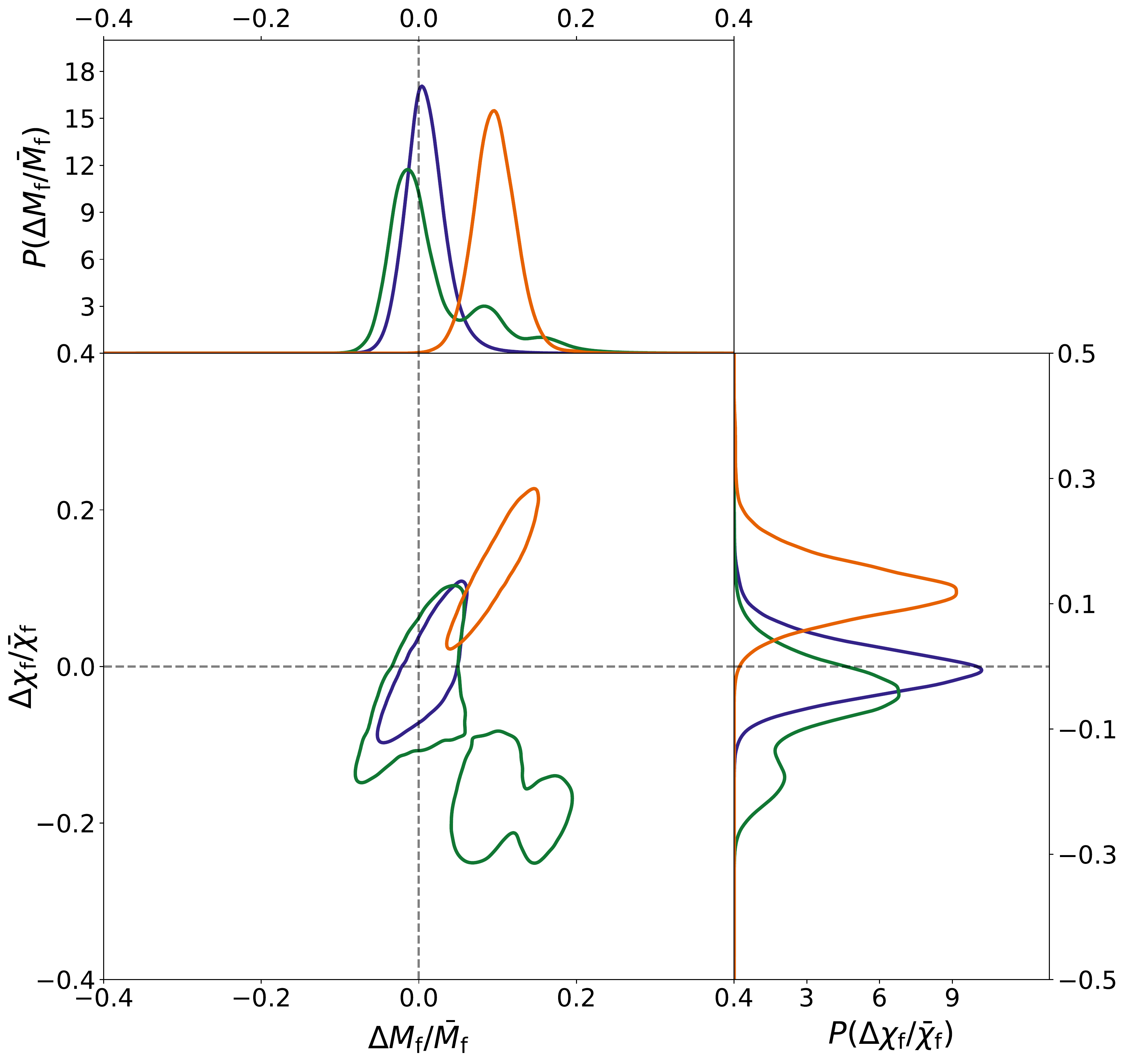}\quad\quad
\includegraphics[width=0.45\textwidth]{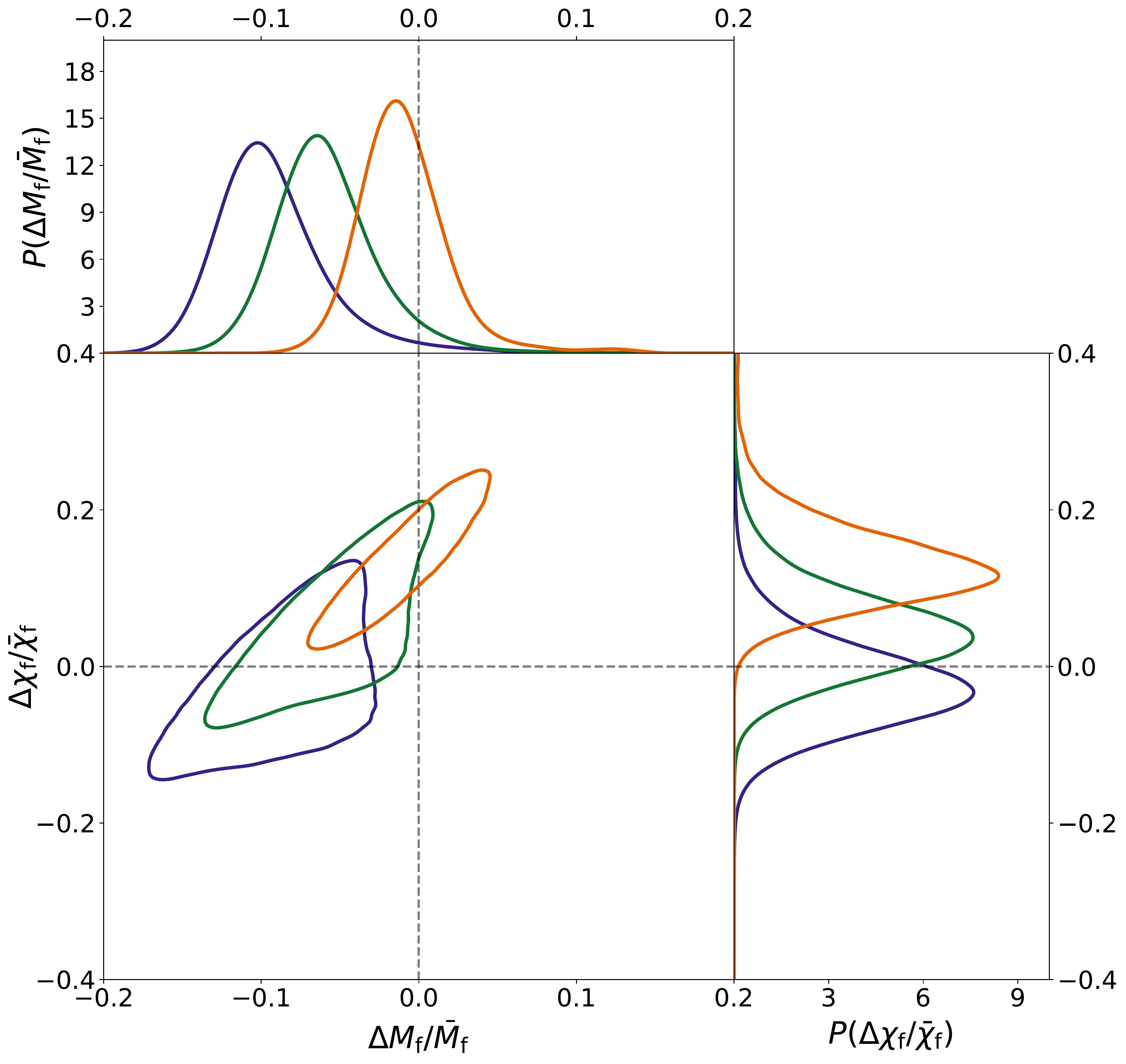}\\[2em]
\includegraphics[width=0.45\textwidth]{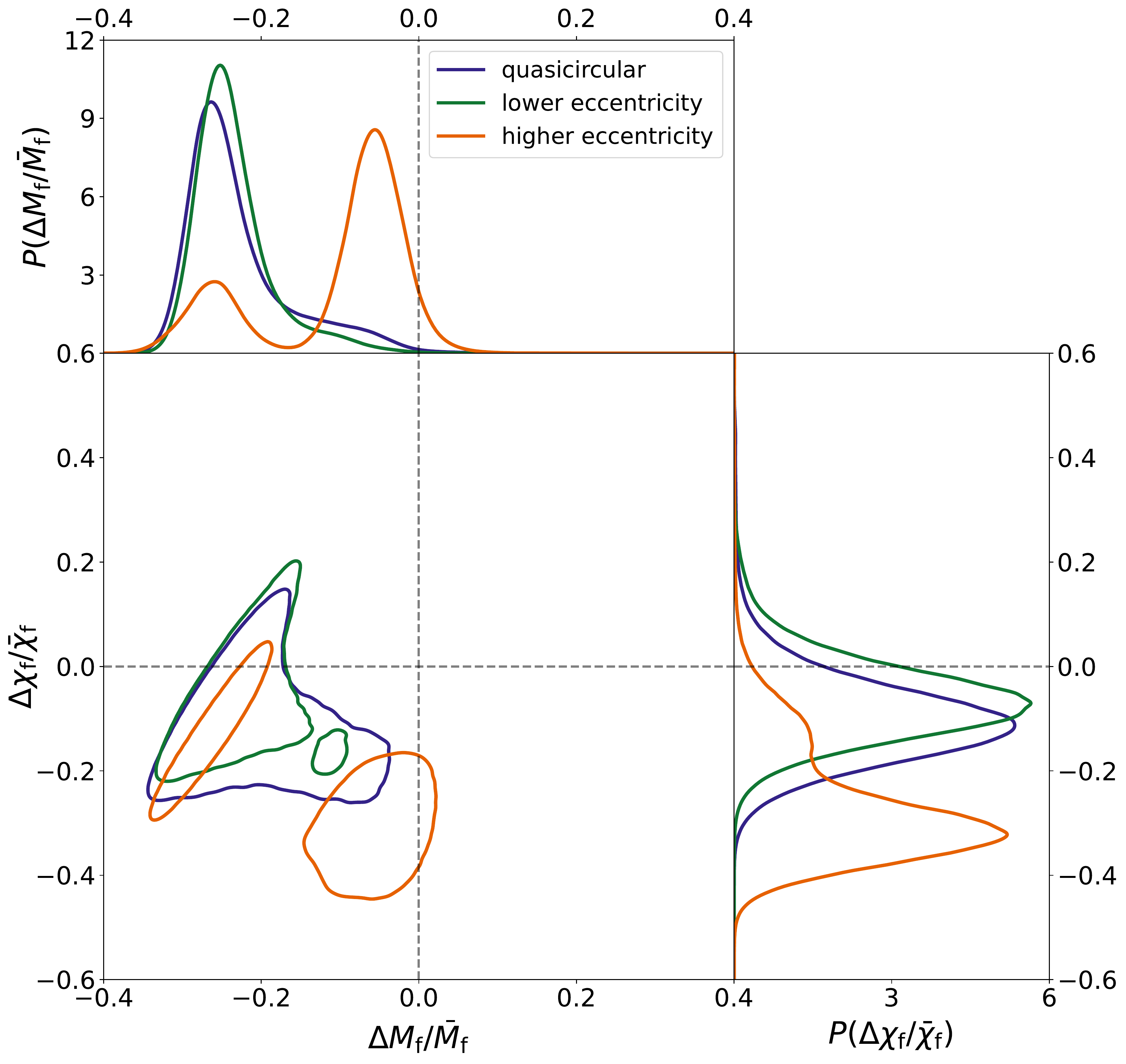}
\caption{\label{fig:IMRCT_plot}The results from the IMR consistency test as the $90\%$ credible regions of the joint posterior distributions of the recovered final mass and spin deviation parameters for quasicircular, lower-eccentricity, and higher-eccentricity injections for mass ratios $1$ (top left), $2$ (top right), and $3$ (bottom). We also show the one-dimensional distributions for the marginalized deviation parameters. The color scheme is the same as in Fig.~\ref{fig:TIGER_plot}. Note that the range of the horizontal axes is smaller for the $q = 2$ plot than for the other two cases and the vertical axis range is larger for the $q = 3$ plot than for the other two cases.}
\end{figure*}
\begin{figure*}
\includegraphics[width=0.45\textwidth]{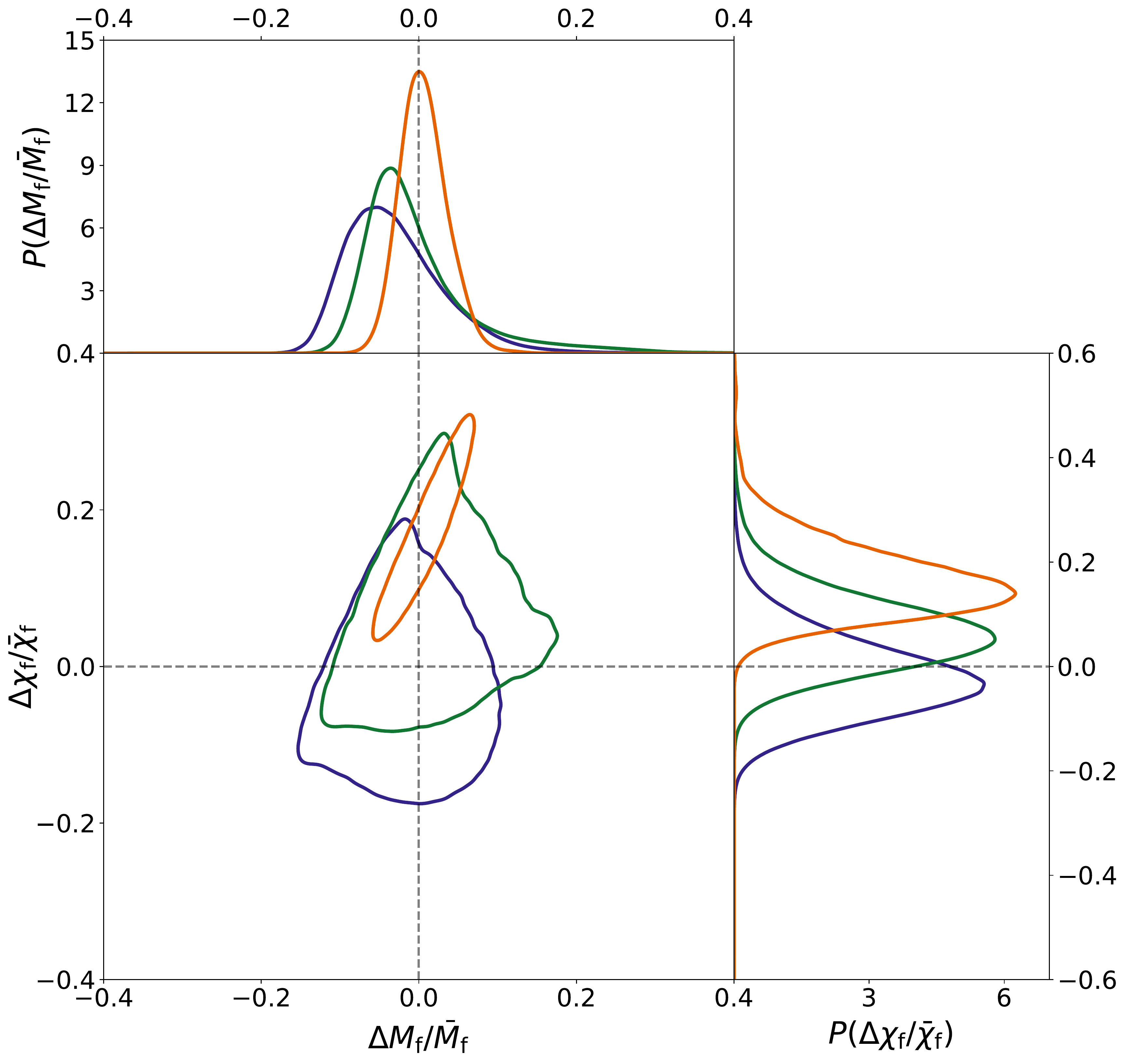}\quad\quad
\includegraphics[width=0.45\textwidth]{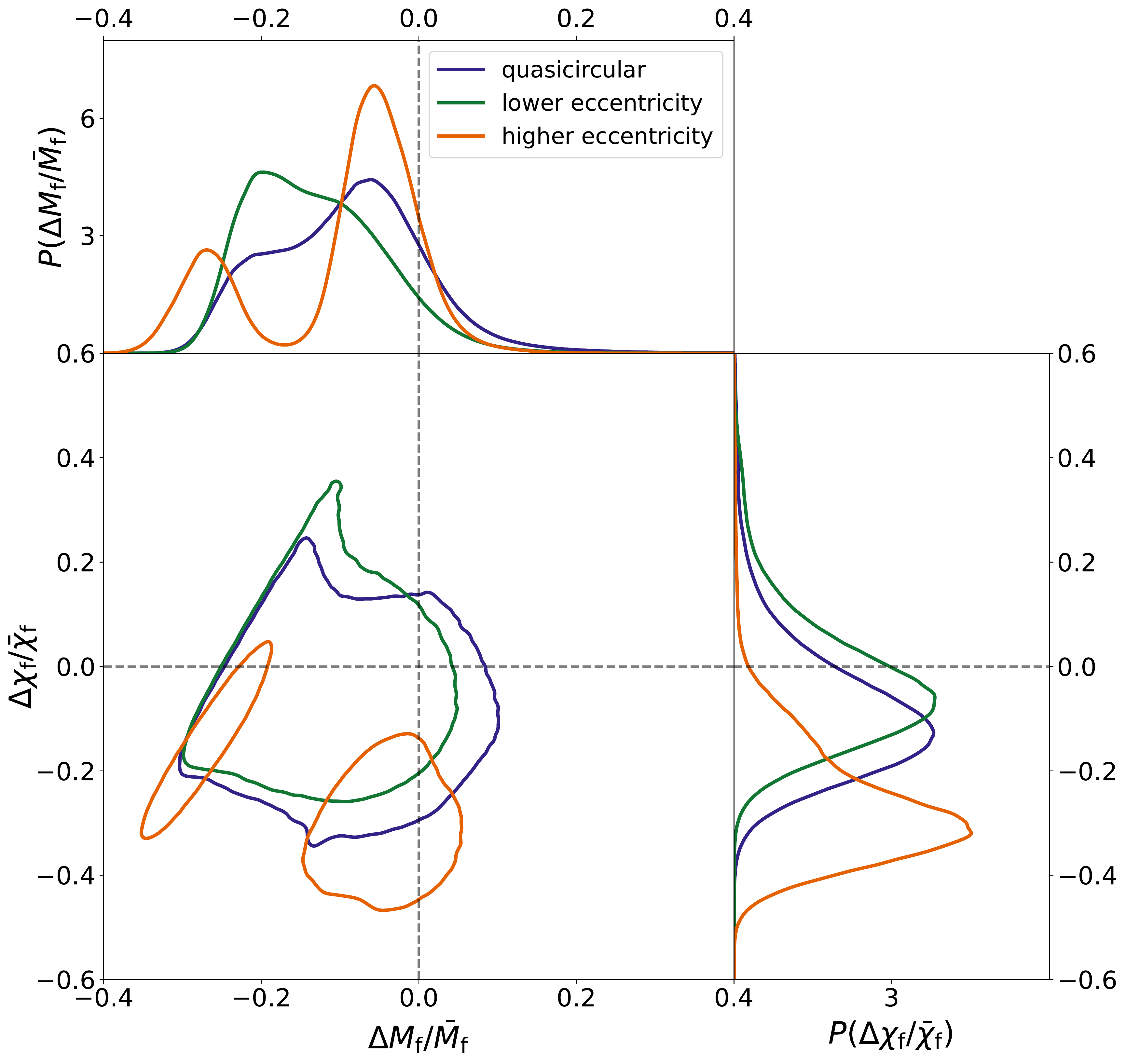}
\caption{\label{fig:IMRCT_22mode_plot} The same as Fig.~\ref{fig:IMRCT_plot} except for just the $q = 2,3$ cases and only including the $(2,\pm2)$ (coprecessing frame) modes in the injections and recovery waveform model. Here the $q = 2$ plot now has the same horizontal axis range as the other plots (here and in Fig.~\ref{fig:IMRCT_plot}) and the $q = 3$ plot has a larger range for the vertical axis.}
\end{figure*}

In Fig.~\ref{fig:IMRCT_plot} we show our results from the IMR consistency test.\footnote{The cutoff frequencies are $\{120, 119, 126\}$~Hz, $\{107, 107, 108\}$~Hz, and $\{95, 96, 94\}$~Hz for $q = 1$, $2$, and $3$, giving the values as $\{\text{quasicircular}, \text{lower-eccentricity}, \text{higher-eccentricity}\}$. These are obtained from a GR analysis of the signals using IMRPhenomXPHM.} We find a significant deviation from GR for all three mass ratios for the higher-eccentricity injections (GR quantiles of $99.9\%$, $\sim$$100\%$, and $99.4\%$ for $q = 1$, $2$, and $3$, respectively)\footnote{We have not checked the extent to which these large GR quantiles are reliable given the order of $10^4$ posterior samples we have.} and also for the lower-eccentricity injections for $q = 2, 3$ (GR quantiles of $93\%$ and $99.8\%$, respectively).\footnote{Both inspiral and postinspiral analyses prefer equal masses, while the total mass is biased to lower values in the inspiral and to higher values in the postinspiral.} Surprisingly, we also find that $q=2, 3$ quasicircular injections give noticeable GR deviations (GR quantiles of $96\%$ and $99\%$, respectively).
We find the same biases when applying the test to injections with the same parameters created using IMRPhenomXPHM instead of NR waveforms, so this is not due to waveform systematics. Thus, we suspect that this is due to the presence of higher modes in the recovery waveform model (i.e., IMRPhenomXPHM), since there are not yet extensive tests of the IMR consistency test using waveform models including higher order modes. However, while there are applications of the test to detected signals using IMRPhenomXPHM~\cite{O3b_TGR}, the methods papers~\cite{Ghosh:2016qgn,Ghosh:2017gfp} do not consider waveform models with higher modes.

To verify this hypothesis, for these cases we applied the IMR consistency test to the $(2,2)$-mode-only
NR injections using IMRPhenomXP, which is the same as IMRPhenomXPHM, except it only has the $(2,\pm2)$ modes in the coprecessing frame, as the recovery waveform model. We use the same cutoff frequencies as in the IMRPhenomXPHM analysis. We show the results from these analyses in Fig.~\ref{fig:IMRCT_22mode_plot}. As expected, we now find that the quasicircular injections indeed agree with GR (as do the lower-eccentricity injections) while the higher-eccentricity injections still show a GR deviation (GR quantiles of $\sim 100\%$ and $99\%$ for $q = 2$ and $3$, respectively). Therefore, we conclude that it is necessary to have a better understanding of the IMR consistency test when using waveforms with higher order modes. The studies necessary to obtain such a better understanding are currently underway, and they have already found that the bias is only significant when the binary is quite close to face-on (or face-off), though the bias does increase monotonically as the inclination varies from edge-on to face-on/face-off~\cite{Mukesh_ICTS}.

Bhat \emph{et al.}~\cite{Bhat:2022amc} also studied the effect of missing eccentricity in the recovery waveform model when performing the IMR consistency test. They used the same Fisher matrix approach in this analysis as in the PN parameter analysis in Saini \emph{et al.}, but here they use a full waveform model (the non-precessing dominant mode IMRPhenomD model \cite{Khan:2015jqa}), instead of just restricting to the PN inspiral waveform. They model the eccentric inspiral signal by adding the PN eccentric frequency domain phase contribution to the IMRPhenomD phasing. They also assume that the eccentricity has a negligible effect on the merger-ringdown part of the signal (as one expects will be the case for small eccentricities, since eccentricity decreases during the inspiral), and ignored its effects on the mapping between the inspiral parameters and the final mass $M_{\rm f}$ and spin $\chi_{\rm f}$ of the merger remnant. Bhat \emph{et al.}\ considered only a mass ratio of $2$ (and aligned spins of $0.4$ and $0.3$) and specified the binary's eccentricity at $10$~Hz. Thus, as described in Sec.~\ref{ssec:tiger}, we approximately obtain the eccentricity of our $q = 2$ injections at $10$~Hz, giving $\sim 0.08$ and $\sim 0.18$ at $10$~Hz. 

We make a rough comparison between our $q = 2$ results for a (redshifted) total mass of $80M_\odot$ with the Bhat \emph{et al.}\ results for redshifted total masses of $72M_\odot$ and $111M_\odot$ (source-frame masses of $65M_\odot$ and $100M_\odot$). While the $111M_\odot$ total mass is considerably further from our $80M_\odot$ total mass than the $72M_\odot$ mass is, we still consider it to bracket our total mass, particularly since the spins in the Bhat \emph{et al.}\ signal make the dominant ($\ell = m = 2$, $n = 0$) quasinormal mode (QNM) frequency of the final black hole in the $111M_\odot$ case closer to the QNM frequency of the final black hole in our injections. Specifically, the QNM frequency in our $q = 2$ injections is $211$~Hz (the same to this accuracy for all three simulations), while the QNM frequency for the $111M_\odot$ and $72M_\odot$ total mass Bhat \emph{et al.}\ cases is $173$~Hz and $267$~Hz, respectively, so $38$~Hz less and $56$~Hz greater, respectively. We compute the QNM frequencies from the final masses and spins using the fit from~\cite{London:2018nxs}. For the Bhat \emph{et al.}\ case, the final mass and spin are $0.950$ times the total mass and $0.764$, respectively (computed using the average of fits to NR results used in the IMR consistency test).

Since Bhat \emph{et al.}\ just give $\Delta M_f$ and $\Delta \chi_f$, we divide these by the injected values of the final mass and spin (quoted above) for the purposes of this comparison. We compare our IMRPhenomXP results (see Fig.~\ref{fig:IMRCT_22mode_plot}) and the results in Fig.~3 of Bhat \emph{et al.}, which gives results for eccentricities of $0.08$ and $0.15$.\footnote{The point for an eccentricity of $0.15$ and a source-frame total mass of $65M_\odot$ is not included in the plotted region in Fig.~3 of Bhat \emph{et al.}, but it has values of $\Delta M_f = 1.42M_\odot$, $\Delta\chi_f = 0.25$~\cite{Saini_PC}.} As in the comparison with Saini \emph{et al.}, we scale their statistical errors to our SNRs (using the SNR of the entire signal, noting that the SNRs of the inspiral and postinspiral portions that Bhat \emph{et al.}\ quote add in quadrature). We find that the scaled Bhat \emph{et al.}\ statistical errors bracket the widths of the $68\%$ credible intervals we obtain for the deviation parameters except for the larger eccentricity final mass, where our error is $\sim 30\%$ smaller. Comparing our medians with the Bhat \emph{et al.}\ systematic biases, we find that for the lower eccentricity, the median of our final mass posterior is about $2$ times larger than the systematic bias found by Bhat \emph{et al.}, but our final spin median is contained within the range for the two masses. For the larger eccentricity, our final mass median is $\sim 5$ times smaller than the one from Bhat \emph{et al.}, but only $\sim 30\%$ smaller for the final spin, though this comparison underestimates the difference between the two results, since the Bhat \emph{et al.}\ results with which we are comparing are for a somewhat smaller eccentricity ($0.15$ vs.\ $\sim 0.18$), and there is a relatively steep dependence of the systematic error on the eccentricity. As in the Saini \emph{et al.}\ comparison, here we compare with the Bhat \emph{et al.}\ total mass that gives the smaller differences (of the two total masses that bracket our total mass).

In general, we find that the Bhat \emph{et al.}\ results qualitatively agree with ours. Specifically, we find that while the Bhat \emph{et al.}\ statistical errors scaled to our SNR would give a significant GR deviation for the lower eccentricity and the lower total mass, the larger total mass would give consistency with GR in this case, in agreement with our results. Bhat \emph{et al.}\ also find a significant GR deviation for an eccentricity of $0.15$ for both total masses, in agreement with our result for an eccentricity of $\sim 0.18$.

\subsection{Checks of the scaling with SNR}
\label{ssec:low_snr}

As noted earlier, our injections with total mass $80M_\odot$ and luminosity distance $400$~Mpc have network SNRs $\sim90$--$120$ at the forecast O4 sensitivity we are considering, which are on the higher side and way above the minimum SNRs of $\sim 10$ one obtains for the high-significance signals considered in the LVK testing GR analyses~\cite{O2_TGR, O3a_TGR, O3b_TGR}.
There will be larger errors on testing parameters (broader posterior distributions) for lower-SNR signals, so apparent GR deviations due to eccentricity can be lost in the statistical error at smaller SNRs. In the high-SNR limit, the width of the posteriors scales like $1/\text{SNR}$; we want to check if this scaling works well for our injections. We do not consider a low-SNR version of each of our injections (given in Table~\ref{tab:sims}) but make at least one for each of the higher-eccentricity cases, as well as for one lower-eccentricity case. We prepare our low-SNR injections as follows. For each injection and TIGER, FTI, and MDR testing parameter, we compute the luminosity distance at which we would expect to exclude GR at $90\%$ credibility. We compute this by finding the scaling of the width of the $90\%$ credible interval for which the edge of the scaled credible interval would just touch the GR value of zero, keeping the median of the posterior the same. We then apply this scaling to the injected luminosity distance, and for each test and mass ratio pick the injection and testing parameter that gives the largest luminosity distance for our low-SNR injection. The low-SNR injections then use this scaled luminosity distance while keeping the other binary parameters the same as for the high-SNR ones. We then apply the test to that injection just for the specific testing parameter used to find the scaling. For this study, we do not perform the IMR consistency test due to the significant systematic bias we find for that test due to the presence of higher order modes. The injected SNR in the low-SNR cases can be found by scaling the SNRs quoted in Sec.~\ref{sec:inj} by the appropriate distance scaling.

\paragraph*{TIGER:} For the TIGER test, it is the higher-eccentricity case that gives the largest GR quantiles for all three mass ratios. For $q=1$, $2$, and $3$, we obtained the largest scaling factors for $\delta\hat{\beta}_{3}$, $\delta\hat{\alpha}_{2}$, and $\delta\hat{\varphi}_{3}$, giving $4.9$, $3.7$, and $3.8$, respectively. The results are shown as black unfilled violins in Fig.~\ref{fig:TIGER_plot}. As expected, these posterior distributions are all broader than their high-SNR counterparts, and the $90\%$ bound is very close to the GR value in the $q = 1,2$ cases, though the median is also notably shifted to lower values in the $q=1$ case, with a smaller shift in the $q=2$ case. For $q=3$, GR is still excluded at $2.2\sigma$ due to a shift in the median away from zero. These shifts in the median are due to degeneracies between the testing parameter and the chirp mass and between the chirp mass and distance. The distance prior we have chosen favors larger distances and thus larger chirp masses, while larger chirp masses are correlated with smaller values for these three testing parameters. Since a lower SNR allows for a larger mismatch of the model waveform with the injection, the degeneracy makes the testing parameter posterior peak at smaller values. As discussed for FTI in~\cite{Mehta:2022pcn}, the degeneracy between the testing parameter and chirp mass is most prominent for $\delta\hat{\varphi}_0$ and other low-PN-order parameters, but it is also present for other testing parameters.

\paragraph*{FTI:} For FTI test, it is again the higher-eccentricity case that gives the largest GR quantiles for all three mass ratios. For $q=1$, $2$, and $3$, we obtained the largest scaling factors for $\delta\hat{\varphi}_{3}$, $\delta\hat{\varphi}_{7}$, and $\delta\hat{\varphi}_{0}$, giving $2.5$, $1.3$, and $3.0$, respectively. The results are shown as black unfilled violins in Fig.~\ref{fig:FTI_plot}. We observe that the scaling works very well for the $q=1$ case, with the $90\%$ bound almost exactly at the GR value, and fairly well for the $q=2$ case, though GR is still excluded at $1.9\sigma$. However, while the posterior for $q=3$ is broadened, as expected, there is also a significant shift away from zero in the median, and GR is still excluded beyond $3\sigma$, as it is in the high-SNR case. This is due to the significant degeneracy between $\delta\hat{\varphi}_{0}$ and chirp mass mentioned above, where larger $\delta\hat{\varphi}_{0}$ values are correlated with larger chirp masses, which are correlated with the larger distances favored by our distance prior.

\paragraph*{MDR:} For the MDR test, we find that the $q=1$ lower-eccentricity $\alpha=1.5$, $q=2$ higher-eccentricity $\alpha=1.5$, and $q=3$ higher-eccentricity $\alpha=3$ cases give the largest scaling factors, specifically $1.9$, $1.3$, and $1.9$, respectively. The results for $q = 1$ and $q = 3$ are shown as black unfilled violins in Fig.~\ref{fig:MDR_plot}. We do not plot the $q = 2$ results, since we were unable to obtain reliable results in this case, finding that the posterior peaks at significantly larger values of $A_\alpha$ than the $A_\alpha$ values that give the largest likelihood (the maximum likelihood is $\sim 4$ orders of magnitude larger than the likelihood values near the peak of the posterior), while there is only a factor of $\lesssim 2$ difference in the prior values. We get the expected broadening of posteriors for $q=1$ and $q=3$, and also see shifts of the median to smaller values of $A_\alpha$, which we expect, since a given $A_\alpha$ causes a larger dephasing on the waveform at a larger distance. To check that the shifts in the median are indeed the expected ones, we compare the posteriors on the dephasing in the low- and high-SNR cases. However, we find that the dephasing posterior peaks at zero for both eccentricities, it has a secondary peak at nonzero values in the high-SNR cases, making it difficult to draw any conclusions.

\section{Summary and Conclusions}
\label{sec:concl}

The waveform models employed in the LVK's current tests of GR do not account for the effects of eccentricity and are only applicable to BBHs on quasicircular orbits. This has not been a serious issue since the binaries formed through isolated formation channels~\cite{Mapelli:2021taw} are expected to have negligible eccentricity by the time their signals enter the sensitive band of ground-based GW detectors. However, there are many other formation pathways (such as dynamical formation, e.g.,~\cite{Rodriguez:2017pec,Samsing:2017xmd,Rodriguez:2018pss}) that can lead to non-negligible eccentricities in the frequency band of ground-based GW detectors for a small fraction of detected signals. 
Thus, as we detect more and more GW signals as current GW detectors improve in sensitivity~\cite{Aasi:2013wya}, it is anticipated that a fraction of these binaries may be eccentric in nature. The mismatch between an eccentric GW signal in the data and the quasicircular waveform model used in the tests of GR can lead to a false GR deviation (see \cite{Saini:2022igm,Bhat:2022amc}). In this paper, we investigate the effect of ignoring eccentricity when performing some of the LVK's standard tests of GR on realistic eccentric BBH signals.

Specifically, we consider the TIGER \cite{TIGER2014,Meidam:2017dgf}, FTI \cite{Mehta:2022pcn}, MDR \cite{Mirshekari:2011yq}, and IMR consistency \cite{Ghosh:2016qgn,Ghosh:2017gfp} tests and check their response to simulated eccentric BBH GW signals in the LIGO-Virgo network at the forecast O4 sensitivity \cite{Aasi:2013wya}. Our eccentric GW signals are modeled using NR waveforms from the SXS catalog~\cite{Boyle:2019kee}; all waveforms are non-spinning, and we consider three mass ratios ($q=1,2,3$). We inject the signals with a total mass of $80M_\odot$ and at a luminosity distance of $400$~Mpc, giving SNRs of about $120$, $105$, and $90$ for the three mass ratios. We choose the SXS simulations such that the binary's eccentricity is $\sim0.05$ and $\sim0.1$ at $17$~Hz. For each mass ratio, we also consider a quasicircular SXS waveform, to compare our results with eccentric cases. We inject these NR waveforms into zero noise.

As expected, all quasicircular injections are consistent with GR at $90\%$ credibility when subjected to the TIGER, FTI, and MDR tests. However, for the IMR consistency test, the $q=2$ and $3$ quasicircular injections show a significant GR deviation (GR excluded at $>2\sigma$). We find that this is attributable to the use of higher modes in the recovery waveform model (i.e., IMRPhenomXPHM). In particular, when we keep only the $(2,2)$ mode in the quasicircular $q=2,3$ injections and use IMRPhenomXP (which does not contain higher modes) to perform the IMR consistency test, no GR deviation is found.

For the TIGER test, we found that the lower-eccentricity injections are consistent with GR at $<3\sigma$ except for the higher-PN-order parameters for $q = 3$ and $\delta\hat{\beta}_3$ for $q = 1,2$ where they exclude GR at $>3\sigma$. We found very significant GR deviations ($>3\sigma$ in almost all cases) with TIGER for the higher-eccentricity injections. For the FTI test, we found the lower-eccentricity injections to be consistent with GR at $2\sigma$, except for the higher-PN-order parameters in the $q = 3$ case, where these are $>3\sigma$ deviations. Higher-eccentricity injections show large GR deviations in many cases, though not as large as in the TIGER analysis. In the MDR test, both lower- and higher-eccentricity injections are found to be consistent with GR at $3\sigma$, with only three cases where GR is excluded at $>2\sigma$. Further, the IMR consistency test with higher mode analysis reports strong GR deviations ($>2.7\sigma$) for both lower- and higher-eccentricity injections, except for the lower-eccentricity case for $q = 1$, which is consistent with GR at $90\%$ credibility. However, the analysis without higher modes for the $q = 2,3$ cases finds that GR is excluded at $90\%$ credibility (indeed $>2.5\sigma$) only for the higher-eccentricity cases.

We also checked the scaling of our results with distance for the TIGER, FTI, and MDR results in a few cases that we expected to still give a significant GR deviation at much larger distances. Here we found that one will still exclude GR at the $\sim 90\%$ credible level for at least one testing parameter at a distance of $\sim 2$~Gpc ($\sim 1.5$~Gpc) for TIGER and the $q = 1$ ($q = 2,3$) higher-eccentricity injections; at distances of $\sim 1$, $0.5$, and $1.2$~Gpc for FTI and the $q = 1$, $2$, and $3$ higher-eccentricity injections; and at distances of $\sim 0.7$~Gpc ($\sim 0.5$~Gpc) for MDR and the $q = 1$ lower-eccentricity and $q = 3$ higher-eccentricity ($q = 2$ higher-eccentricity) injections. We found that the $q = 1$ lower-eccentricity injection gives a larger GR deviation with MDR than the higher-eccentricity injection does.

The results we obtained in this paper suggest that one will obtain strong GR deviations when applying standard current LVK tests of GR to GW signals from binaries with non-negligible eccentricity ($\sim0.05-0.1$ at $17$~Hz). While we have only considered a small portion of binary parameter space, in particular just one total mass, the Fisher matrix results in~\cite{Saini:2022igm,Bhat:2022amc} suggest that this is a fairly general conclusion. Therefore, the possibility of the signal being from an eccentric binary needs to be ruled out before one can make any claims of GR violation.

Ruling out an eccentric binary as a possible cause of an apparent GR violation will require analysis of the signal using waveforms for eccentric BBHs in GR, and likely the implementation of tests of GR (those used in this paper or others) using eccentric GR waveform models as a baseline. It will be necessary to use waveforms including both eccentricity and precession in such analyses. While there are not yet any full IMR models for BBHs with both eccentricity and precession, there is a PN inspiral model with both these effects~\cite{Klein:2021jtd}, as well as full IMR effective-one-body models for eccentric BBHs with aligned spins~\cite{Nagar:2021gss,Ramos-Buades:2021adz} and an NR surrogate model for nonspinning eccentric BBHs~\cite{Islam:2021mha}. Thus, the prospects for having waveform models for precessing eccentric BBHs in the near future seem good, though even when these are available, it will be necessary to perform careful studies to determine the extent to which one can distinguish various possible GR deviations from the effects of eccentricity.

Additionally, there are several other physical effects that are missing in the current waveform models employed by the LVK's tests of GR, e.g., gravitational lensing and environmental effects, and could be important.
For instance, \cite{Ezquiaga:2020gdt} showed that strong lensing can modify the lensed GW signal in such a way that it can become inconsistent with unlensed GR GW signal and \cite{Vijaykumar:2022dlp} showed that this can indeed lead to biases in estimating parameters of lensed signal if the recovery waveform model does not account for lensing effects. Furthermore, even if the magnitude of the environmental effects are expected to be small \cite{Cardoso:2019rou} for ground-based detectors \cite{Barausse:2014tra,Bonvin:2016qxr}, there could be a possibility of detecting a GW signal with the effect of a third body at the forecast O4 sensitivity (see, e.g.,~\cite{Vijaykumar:2023tjg}).  
Therefore, it will be interesting to see if and how such effects can potentially mimic a GR violation at the sensitivities that can be expected in the near future~\cite{Aasi:2013wya}.


\acknowledgments

We thank K.~G.~Arun and Pankaj Saini for useful comments, Mukesh Kumar Singh for sharing the results of his investigations into the IMR consistency test bias, and Archisman Ghosh for the code used to create the injections.
We also thank all the LIGO-Virgo-KAGRA testing GR group members who implemented these tests in publicly available code.
NKJ-M is supported by NSF grant AST-2205920. AG is supported in part by NSF grants PHY-2308887 and AST-2205920. The authors are grateful for computational resources provided by the LIGO Laboratory and the Leonard E Parker Center for Gravitation, Cosmology and Astrophysics at the University of Wisconsin-Milwaukee and supported by National Science Foundation Grants PHY-0757058, PHY-0823459, PHY-1700765, and PHY-1626190. 

This study used the software packages LALSuite~\cite{LALSuite}, Matplotlib~\cite{Hunter:2007ouj}, NumPy~\cite{Harris:2020xlr}, PESummary~\cite{Hoy:2020vys}, Positive~\cite{London:2018nxs}, SciPy~\cite{Virtanen:2019joe}, and Seaborn~\cite{Waskom:2021psk}.

This is LIGO document number P2300161.

\bibliography{ecc_tgr_refs}

\begin{thebibliography}{109}%
\makeatletter
\providecommand \@ifxundefined [1]{%
 \@ifx{#1\undefined}
}%
\providecommand \@ifnum [1]{%
 \ifnum #1\expandafter \@firstoftwo
 \else \expandafter \@secondoftwo
 \fi
}%
\providecommand \@ifx [1]{%
 \ifx #1\expandafter \@firstoftwo
 \else \expandafter \@secondoftwo
 \fi
}%
\providecommand \natexlab [1]{#1}%
\providecommand \enquote  [1]{``#1''}%
\providecommand \bibnamefont  [1]{#1}%
\providecommand \bibfnamefont [1]{#1}%
\providecommand \citenamefont [1]{#1}%
\providecommand \href@noop [0]{\@secondoftwo}%
\providecommand \href [0]{\begingroup \@sanitize@url \@href}%
\providecommand \@href[1]{\@@startlink{#1}\@@href}%
\providecommand \@@href[1]{\endgroup#1\@@endlink}%
\providecommand \@sanitize@url [0]{\catcode `\\12\catcode `\$12\catcode
  `\&12\catcode `\#12\catcode `\^12\catcode `\_12\catcode `\%12\relax}%
\providecommand \@@startlink[1]{}%
\providecommand \@@endlink[0]{}%
\providecommand \url  [0]{\begingroup\@sanitize@url \@url }%
\providecommand \@url [1]{\endgroup\@href {#1}{\urlprefix }}%
\providecommand \urlprefix  [0]{URL }%
\providecommand \Eprint [0]{\href }%
\@ifxundefined \urlstyle {%
  \providecommand \doi  [0]{\begingroup \@sanitize@url \@doi}%
  \providecommand \@doi [1]{\endgroup \@@startlink {\doibase
  #1}doi:\discretionary {}{}{}#1\@@endlink }%
}{%
  \providecommand \doi  [0]{doi:\discretionary{}{}{}\begingroup
  \urlstyle{rm}\Url }%
}%
\providecommand \doibase [0]{http://dx.doi.org/}%
\providecommand \Doi [0]{\begingroup \@sanitize@url \@Doi }%
\providecommand \@Doi  [1]{\endgroup\@@startlink{\doibase#1}\@@Doi}%
\providecommand \@@Doi [1]{#1\@@endlink}%
\providecommand \selectlanguage [0]{\@gobble}%
\providecommand \bibinfo  [0]{\@secondoftwo}%
\providecommand \bibfield  [0]{\@secondoftwo}%
\providecommand \translation [1]{[#1]}%
\providecommand \BibitemOpen [0]{}%
\providecommand \bibitemStop [0]{}%
\providecommand \bibitemNoStop [0]{.\EOS\space}%
\providecommand \EOS [0]{\spacefactor3000\relax}%
\providecommand \BibitemShut  [1]{\csname bibitem#1\endcsname}%
\bibitem [{\citenamefont {Will}(2014)}]{Will:2014kxa}%
  \BibitemOpen
  \bibfield  {author} {\bibinfo {author} {\bibfnamefont {C.~M.}\ \bibnamefont
  {Will}},\ }\Doi {10.12942/lrr-2014-4} {\bibfield  {journal} {\bibinfo
  {journal} {Living Rev. Relativity},\ }\textbf {\bibinfo {volume} {17}},\
  \bibinfo {pages} {4} (\bibinfo {year} {2014})},\ \Eprint
  {http://arxiv.org/abs/1403.7377} {arXiv:1403.7377 [gr-qc]} \BibitemShut
  {NoStop}%
\bibitem [{\citenamefont {Wex}(2014)}]{Wex:2014nva}%
  \BibitemOpen
  \bibfield  {author} {\bibinfo {author} {\bibfnamefont {N.}~\bibnamefont
  {Wex}},\ }\enquote {\bibinfo {title} {{Testing Relativistic Gravity with
  Radio Pulsars}},}\ in\ \Doi {https://doi.org/10.1515/9783110345667} {\emph
  {\bibinfo {booktitle} {{Frontiers in Relativistic Celestial Mechanics, Volume
  2: Applications and Experiments}}}},\ \bibinfo {editor} {edited by\ \bibinfo
  {editor} {\bibfnamefont {S.~M.}\ \bibnamefont {Kopeikin}}}\ (\bibinfo
  {publisher} {De Gruyter},\ \bibinfo {address} {Berlin},\ \bibinfo {year}
  {2014})\ \Eprint {http://arxiv.org/abs/1402.5594} {arXiv:1402.5594 [gr-qc]}
  \BibitemShut {NoStop}%
\bibitem [{\citenamefont {Weisberg}\ and\ \citenamefont
  {Taylor}(2005)}]{Weisberg:2004hi}%
  \BibitemOpen
  \bibfield  {author} {\bibinfo {author} {\bibfnamefont {J.~M.}\ \bibnamefont
  {Weisberg}}\ and\ \bibinfo {author} {\bibfnamefont {J.~H.}\ \bibnamefont
  {Taylor}},\ }\href@noop {} {\bibfield  {journal} {\bibinfo  {journal} {ASP
  Conf. Ser.},\ }\textbf {\bibinfo {volume} {328}},\ \bibinfo {pages} {25}
  (\bibinfo {year} {2005})},\ \Eprint {http://arxiv.org/abs/astro-ph/0407149}
  {arXiv:astro-ph/0407149} \BibitemShut {NoStop}%
\bibitem [{\citenamefont {Voisin}\ \emph {et~al.}(2020)\citenamefont {Voisin},
  \citenamefont {Cognard}, \citenamefont {Freire}, \citenamefont {Wex},
  \citenamefont {Guillemot}, \citenamefont {Desvignes}, \citenamefont
  {Kramer},\ and\ \citenamefont {Theureau}}]{Voisin:2020lqi}%
  \BibitemOpen
  \bibfield  {author} {\bibinfo {author} {\bibfnamefont {G.}~\bibnamefont
  {Voisin}}, \bibinfo {author} {\bibfnamefont {I.}~\bibnamefont {Cognard}},
  \bibinfo {author} {\bibfnamefont {P.~C.~C.}\ \bibnamefont {Freire}}, \bibinfo
  {author} {\bibfnamefont {N.}~\bibnamefont {Wex}}, \bibinfo {author}
  {\bibfnamefont {L.}~\bibnamefont {Guillemot}}, \bibinfo {author}
  {\bibfnamefont {G.}~\bibnamefont {Desvignes}}, \bibinfo {author}
  {\bibfnamefont {M.}~\bibnamefont {Kramer}}, \ and\ \bibinfo {author}
  {\bibfnamefont {G.}~\bibnamefont {Theureau}},\ }\Doi
  {10.1051/0004-6361/202038104} {\bibfield  {journal} {\bibinfo  {journal}
  {Astron. Astrophys.},\ }\textbf {\bibinfo {volume} {638}},\ \bibinfo {pages}
  {A24} (\bibinfo {year} {2020})},\ \Eprint {http://arxiv.org/abs/2005.01388}
  {arXiv:2005.01388 [gr-qc]} \BibitemShut {NoStop}%
\bibitem [{\citenamefont {Kramer}\ \emph {et~al.}(2021)\citenamefont {Kramer}
  \emph {et~al.}}]{Kramer:2021jcw}%
  \BibitemOpen
  \bibfield  {author} {\bibinfo {author} {\bibfnamefont {M.}~\bibnamefont
  {Kramer}} \emph {et~al.},\ }\Doi {10.1103/PhysRevX.11.041050} {\bibfield
  {journal} {\bibinfo  {journal} {Phys. Rev. X},\ }\textbf {\bibinfo {volume}
  {11}},\ \bibinfo {pages} {041050} (\bibinfo {year} {2021})},\ \Eprint
  {http://arxiv.org/abs/2112.06795} {arXiv:2112.06795 [astro-ph.HE]}
  \BibitemShut {NoStop}%
\bibitem [{\citenamefont {Abbott}\ \emph {et~al.}(2016)\citenamefont {Abbott}
  \emph {et~al.}}]{GW150914_TGR}%
  \BibitemOpen
  \bibfield  {author} {\bibinfo {author} {\bibfnamefont {B.~P.}\ \bibnamefont
  {Abbott}} \emph {et~al.} (\bibinfo {collaboration} {LIGO Scientific
  Collaboration and Virgo Collaboration}),\ }\Doi
  {10.1103/PhysRevLett.116.221101} {\bibfield  {journal} {\bibinfo  {journal}
  {Phys. Rev. Lett.},\ }\textbf {\bibinfo {volume} {116}},\ \bibinfo {pages}
  {221101} (\bibinfo {year} {2016})},\ \Eprint
  {http://arxiv.org/abs/1602.03841} {arXiv:1602.03841 [gr-qc]} \BibitemShut
  {NoStop}%
\bibitem [{\citenamefont {Abbott}\ \emph
  {et~al.}(2019){\natexlab{a}}\citenamefont {Abbott} \emph
  {et~al.}}]{GW170817_TGR}%
  \BibitemOpen
  \bibfield  {author} {\bibinfo {author} {\bibfnamefont {B.~P.}\ \bibnamefont
  {Abbott}} \emph {et~al.} (\bibinfo {collaboration} {LIGO Scientific
  Collaboration and Virgo Collaboration}),\ }\Doi
  {10.1103/PhysRevLett.123.011102} {\bibfield  {journal} {\bibinfo  {journal}
  {Phys. Rev. Lett.},\ }\textbf {\bibinfo {volume} {123}},\ \bibinfo {pages}
  {011102} (\bibinfo {year} {2019}{\natexlab{a}})},\ \Eprint
  {http://arxiv.org/abs/1811.00364} {arXiv:1811.00364 [gr-qc]} \BibitemShut
  {NoStop}%
\bibitem [{\citenamefont {Abbott}\ \emph
  {et~al.}(2019){\natexlab{b}}\citenamefont {Abbott} \emph {et~al.}}]{O2_TGR}%
  \BibitemOpen
  \bibfield  {author} {\bibinfo {author} {\bibfnamefont {B.~P.}\ \bibnamefont
  {Abbott}} \emph {et~al.} (\bibinfo {collaboration} {LIGO Scientific
  Collaboration and Virgo Collaboration}),\ }\Doi {10.1103/PhysRevD.100.104036}
  {\bibfield  {journal} {\bibinfo  {journal} {Phys. Rev. D},\ }\textbf
  {\bibinfo {volume} {100}},\ \bibinfo {pages} {104036} (\bibinfo {year}
  {2019}{\natexlab{b}})},\ \Eprint {http://arxiv.org/abs/1903.04467}
  {arXiv:1903.04467 [gr-qc]} \BibitemShut {NoStop}%
\bibitem [{\citenamefont {Abbott}\ \emph
  {et~al.}(2021){\natexlab{a}}\citenamefont {Abbott} \emph {et~al.}}]{O3a_TGR}%
  \BibitemOpen
  \bibfield  {author} {\bibinfo {author} {\bibfnamefont {R.}~\bibnamefont
  {Abbott}} \emph {et~al.} (\bibinfo {collaboration} {LIGO Scientific
  Collaboration and Virgo Collaboration}),\ }\Doi {10.1103/PhysRevD.103.122002}
  {\bibfield  {journal} {\bibinfo  {journal} {Phys. Rev. D},\ }\textbf
  {\bibinfo {volume} {103}},\ \bibinfo {pages} {122002} (\bibinfo {year}
  {2021}{\natexlab{a}})},\ \Eprint {http://arxiv.org/abs/2010.14529}
  {arXiv:2010.14529 [gr-qc]} \BibitemShut {NoStop}%
\bibitem [{\citenamefont {Abbott}\ \emph
  {et~al.}(2021){\natexlab{b}}\citenamefont {Abbott} \emph {et~al.}}]{O3b_TGR}%
  \BibitemOpen
  \bibfield  {author} {\bibinfo {author} {\bibfnamefont {R.}~\bibnamefont
  {Abbott}} \emph {et~al.} (\bibinfo {collaboration} {LIGO Scientific
  Collaboration, Virgo Collaboration, and KAGRA Collaboration}),\ }\href@noop
  {} {} (\bibinfo {year} {2021}{\natexlab{b}}),\ \Eprint
  {http://arxiv.org/abs/2112.06861} {arXiv:2112.06861 [gr-qc]} \BibitemShut
  {NoStop}%
\bibitem [{\citenamefont {Aasi}\ \emph {et~al.}(2015)\citenamefont {Aasi} \emph
  {et~al.}}]{LIGOScientific:2014pky}%
  \BibitemOpen
  \bibfield  {author} {\bibinfo {author} {\bibfnamefont {J.}~\bibnamefont
  {Aasi}} \emph {et~al.} (\bibinfo {collaboration} {LIGO Scientific
  Collaboration}),\ }\Doi {10.1088/0264-9381/32/7/074001} {\bibfield  {journal}
  {\bibinfo  {journal} {Classical Quantum Gravity},\ }\textbf {\bibinfo
  {volume} {32}},\ \bibinfo {pages} {074001} (\bibinfo {year} {2015})},\
  \Eprint {http://arxiv.org/abs/1411.4547} {arXiv:1411.4547 [gr-qc]}
  \BibitemShut {NoStop}%
\bibitem [{\citenamefont {Acernese}\ \emph {et~al.}(2015)\citenamefont
  {Acernese} \emph {et~al.}}]{VIRGO:2014yos}%
  \BibitemOpen
  \bibfield  {author} {\bibinfo {author} {\bibfnamefont {F.}~\bibnamefont
  {Acernese}} \emph {et~al.} (\bibinfo {collaboration} {Virgo Collaboration}),\
  }\Doi {10.1088/0264-9381/32/2/024001} {\bibfield  {journal} {\bibinfo
  {journal} {Classical Quantum Gravity},\ }\textbf {\bibinfo {volume} {32}},\
  \bibinfo {pages} {024001} (\bibinfo {year} {2015})},\ \Eprint
  {http://arxiv.org/abs/1408.3978} {arXiv:1408.3978 [gr-qc]} \BibitemShut
  {NoStop}%
\bibitem [{\citenamefont {Mapelli}(2021)}]{Mapelli:2021taw}%
  \BibitemOpen
  \bibfield  {author} {\bibinfo {author} {\bibfnamefont {M.}~\bibnamefont
  {Mapelli}},\ }\enquote {\bibinfo {title} {{Formation Channels of Single and
  Binary Stellar-Mass Black Holes}},}\ in\ \Doi
  {10.1007/978-981-15-4702-7_16-1} {\emph {\bibinfo {booktitle} {Handbook of
  Gravitational Wave Astronomy}}},\ \bibinfo {editor} {edited by\ \bibinfo
  {editor} {\bibfnamefont {C.}~\bibnamefont {Bambi}}, \bibinfo {editor}
  {\bibfnamefont {S.}~\bibnamefont {Katsanevas}}, \ and\ \bibinfo {editor}
  {\bibfnamefont {K.~D.}\ \bibnamefont {Kokkotas}}}\ (\bibinfo  {publisher}
  {Springer Singapore},\ \bibinfo {address} {Singapore},\ \bibinfo {year}
  {2021})\ \Eprint {http://arxiv.org/abs/2106.00699} {arXiv:2106.00699
  [astro-ph.HE]} \BibitemShut {NoStop}%
\bibitem [{\citenamefont {Peters}(1964)}]{PhysRev.136.B1224}%
  \BibitemOpen
  \bibfield  {author} {\bibinfo {author} {\bibfnamefont {P.~C.}\ \bibnamefont
  {Peters}},\ }\Doi {10.1103/PhysRev.136.B1224} {\bibfield  {journal} {\bibinfo
   {journal} {Phys. Rev.},\ }\textbf {\bibinfo {volume} {136}},\ \bibinfo
  {pages} {B1224} (\bibinfo {year} {1964})}\BibitemShut {NoStop}%
\bibitem [{\citenamefont {{Tucker}}\ and\ \citenamefont
  {{Will}}(2021)}]{2021arXiv210812210T}%
  \BibitemOpen
  \bibfield  {author} {\bibinfo {author} {\bibfnamefont {A.}~\bibnamefont
  {{Tucker}}}\ and\ \bibinfo {author} {\bibfnamefont {C.~M.}\ \bibnamefont
  {{Will}}},\ }\Doi {10.1103/PhysRevD.104.104023} {\bibfield  {journal}
  {\bibinfo  {journal} {\prd},\ }\textbf {\bibinfo {volume} {104}},\ \bibinfo
  {eid} {104023} (\bibinfo {year} {2021})},\ \Eprint
  {http://arxiv.org/abs/2108.12210} {arXiv:2108.12210 [gr-qc]} \BibitemShut
  {NoStop}%
\bibitem [{\citenamefont {Cholis}\ \emph {et~al.}(2016)\citenamefont {Cholis},
  \citenamefont {Kovetz}, \citenamefont {Ali-Ha\"\i{}moud}, \citenamefont
  {Bird}, \citenamefont {Kamionkowski}, \citenamefont {Mu\~noz},\ and\
  \citenamefont {Raccanelli}}]{Cholis:2016kqi}%
  \BibitemOpen
  \bibfield  {author} {\bibinfo {author} {\bibfnamefont {I.}~\bibnamefont
  {Cholis}}, \bibinfo {author} {\bibfnamefont {E.~D.}\ \bibnamefont {Kovetz}},
  \bibinfo {author} {\bibfnamefont {Y.}~\bibnamefont {Ali-Ha\"\i{}moud}},
  \bibinfo {author} {\bibfnamefont {S.}~\bibnamefont {Bird}}, \bibinfo {author}
  {\bibfnamefont {M.}~\bibnamefont {Kamionkowski}}, \bibinfo {author}
  {\bibfnamefont {J.~B.}\ \bibnamefont {Mu\~noz}}, \ and\ \bibinfo {author}
  {\bibfnamefont {A.}~\bibnamefont {Raccanelli}},\ }\Doi
  {10.1103/PhysRevD.94.084013} {\bibfield  {journal} {\bibinfo  {journal}
  {Phys. Rev. D},\ }\textbf {\bibinfo {volume} {94}},\ \bibinfo {pages}
  {084013} (\bibinfo {year} {2016})},\ \Eprint
  {http://arxiv.org/abs/1606.07437} {arXiv:1606.07437 [astro-ph.HE]}
  \BibitemShut {NoStop}%
\bibitem [{\citenamefont {Wang}\ and\ \citenamefont
  {Nitz}(2021)}]{Wang:2021qsu}%
  \BibitemOpen
  \bibfield  {author} {\bibinfo {author} {\bibfnamefont {Y.-F.}\ \bibnamefont
  {Wang}}\ and\ \bibinfo {author} {\bibfnamefont {A.~H.}\ \bibnamefont
  {Nitz}},\ }\Doi {10.3847/1538-4357/abe939} {\bibfield  {journal} {\bibinfo
  {journal} {Astrophys. J.},\ }\textbf {\bibinfo {volume} {912}},\ \bibinfo
  {pages} {53} (\bibinfo {year} {2021})},\ \Eprint
  {http://arxiv.org/abs/2101.12269} {arXiv:2101.12269 [astro-ph.HE]}
  \BibitemShut {NoStop}%
\bibitem [{\citenamefont {Wen}(2003)}]{Wen:2002km}%
  \BibitemOpen
  \bibfield  {author} {\bibinfo {author} {\bibfnamefont {L.}~\bibnamefont
  {Wen}},\ }\Doi {10.1086/378794} {\bibfield  {journal} {\bibinfo  {journal}
  {Astrophys. J.},\ }\textbf {\bibinfo {volume} {598}},\ \bibinfo {pages} {419}
  (\bibinfo {year} {2003})},\ \Eprint {http://arxiv.org/abs/astro-ph/0211492}
  {arXiv:astro-ph/0211492} \BibitemShut {NoStop}%
\bibitem [{\citenamefont {O'Leary}\ \emph {et~al.}(2009)\citenamefont
  {O'Leary}, \citenamefont {Kocsis},\ and\ \citenamefont
  {Loeb}}]{OLeary:2008myb}%
  \BibitemOpen
  \bibfield  {author} {\bibinfo {author} {\bibfnamefont {R.~M.}\ \bibnamefont
  {O'Leary}}, \bibinfo {author} {\bibfnamefont {B.}~\bibnamefont {Kocsis}}, \
  and\ \bibinfo {author} {\bibfnamefont {A.}~\bibnamefont {Loeb}},\ }\Doi
  {10.1111/j.1365-2966.2009.14653.x} {\bibfield  {journal} {\bibinfo  {journal}
  {Mon. Not. R. Astron. Soc.},\ }\textbf {\bibinfo {volume} {395}},\ \bibinfo
  {pages} {2127} (\bibinfo {year} {2009})},\ \Eprint
  {http://arxiv.org/abs/0807.2638} {arXiv:0807.2638 [astro-ph]} \BibitemShut
  {NoStop}%
\bibitem [{\citenamefont {Antonini}\ \emph {et~al.}(2016)\citenamefont
  {Antonini}, \citenamefont {Chatterjee}, \citenamefont {Rodriguez},
  \citenamefont {Morscher}, \citenamefont {Pattabiraman}, \citenamefont
  {Kalogera},\ and\ \citenamefont {Rasio}}]{Antonini:2015zsa}%
  \BibitemOpen
  \bibfield  {author} {\bibinfo {author} {\bibfnamefont {F.}~\bibnamefont
  {Antonini}}, \bibinfo {author} {\bibfnamefont {S.}~\bibnamefont
  {Chatterjee}}, \bibinfo {author} {\bibfnamefont {C.~L.}\ \bibnamefont
  {Rodriguez}}, \bibinfo {author} {\bibfnamefont {M.}~\bibnamefont {Morscher}},
  \bibinfo {author} {\bibfnamefont {B.}~\bibnamefont {Pattabiraman}}, \bibinfo
  {author} {\bibfnamefont {V.}~\bibnamefont {Kalogera}}, \ and\ \bibinfo
  {author} {\bibfnamefont {F.~A.}\ \bibnamefont {Rasio}},\ }\Doi
  {10.3847/0004-637X/816/2/65} {\bibfield  {journal} {\bibinfo  {journal}
  {Astrophys. J.},\ }\textbf {\bibinfo {volume} {816}},\ \bibinfo {pages} {65}
  (\bibinfo {year} {2016})},\ \Eprint {http://arxiv.org/abs/1509.05080}
  {arXiv:1509.05080 [astro-ph.GA]} \BibitemShut {NoStop}%
\bibitem [{\citenamefont {Samsing}\ and\ \citenamefont
  {Ramirez-Ruiz}(2017)}]{Samsing:2017rat}%
  \BibitemOpen
  \bibfield  {author} {\bibinfo {author} {\bibfnamefont {J.}~\bibnamefont
  {Samsing}}\ and\ \bibinfo {author} {\bibfnamefont {E.}~\bibnamefont
  {Ramirez-Ruiz}},\ }\Doi {10.3847/2041-8213/aa6f0b} {\bibfield  {journal}
  {\bibinfo  {journal} {Astrophys. J. Lett.},\ }\textbf {\bibinfo {volume}
  {840}},\ \bibinfo {pages} {L14} (\bibinfo {year} {2017})},\ \Eprint
  {http://arxiv.org/abs/1703.09703} {arXiv:1703.09703 [astro-ph.HE]}
  \BibitemShut {NoStop}%
\bibitem [{\citenamefont {Samsing}(2018)}]{Samsing:2017xmd}%
  \BibitemOpen
  \bibfield  {author} {\bibinfo {author} {\bibfnamefont {J.}~\bibnamefont
  {Samsing}},\ }\Doi {10.1103/PhysRevD.97.103014} {\bibfield  {journal}
  {\bibinfo  {journal} {Phys. Rev. D},\ }\textbf {\bibinfo {volume} {97}},\
  \bibinfo {pages} {103014} (\bibinfo {year} {2018})},\ \Eprint
  {http://arxiv.org/abs/1711.07452} {arXiv:1711.07452 [astro-ph.HE]}
  \BibitemShut {NoStop}%
\bibitem [{\citenamefont {Gond\'an}\ \emph {et~al.}(2018)\citenamefont
  {Gond\'an}, \citenamefont {Kocsis}, \citenamefont {Raffai},\ and\
  \citenamefont {Frei}}]{Gondan:2017wzd}%
  \BibitemOpen
  \bibfield  {author} {\bibinfo {author} {\bibfnamefont {L.}~\bibnamefont
  {Gond\'an}}, \bibinfo {author} {\bibfnamefont {B.}~\bibnamefont {Kocsis}},
  \bibinfo {author} {\bibfnamefont {P.}~\bibnamefont {Raffai}}, \ and\ \bibinfo
  {author} {\bibfnamefont {Z.}~\bibnamefont {Frei}},\ }\Doi
  {10.3847/1538-4357/aabfee} {\bibfield  {journal} {\bibinfo  {journal}
  {Astrophys. J.},\ }\textbf {\bibinfo {volume} {860}},\ \bibinfo {pages} {5}
  (\bibinfo {year} {2018})},\ \Eprint {http://arxiv.org/abs/1711.09989}
  {arXiv:1711.09989 [astro-ph.HE]} \BibitemShut {NoStop}%
\bibitem [{\citenamefont {Rodriguez}\ \emph
  {et~al.}(2018){\natexlab{a}}\citenamefont {Rodriguez}, \citenamefont
  {Amaro-Seoane}, \citenamefont {Chatterjee},\ and\ \citenamefont
  {Rasio}}]{Rodriguez:2017pec}%
  \BibitemOpen
  \bibfield  {author} {\bibinfo {author} {\bibfnamefont {C.~L.}\ \bibnamefont
  {Rodriguez}}, \bibinfo {author} {\bibfnamefont {P.}~\bibnamefont
  {Amaro-Seoane}}, \bibinfo {author} {\bibfnamefont {S.}~\bibnamefont
  {Chatterjee}}, \ and\ \bibinfo {author} {\bibfnamefont {F.~A.}\ \bibnamefont
  {Rasio}},\ }\Doi {10.1103/PhysRevLett.120.151101} {\bibfield  {journal}
  {\bibinfo  {journal} {Phys. Rev. Lett.},\ }\textbf {\bibinfo {volume}
  {120}},\ \bibinfo {pages} {151101} (\bibinfo {year} {2018}{\natexlab{a}})},\
  \Eprint {http://arxiv.org/abs/1712.04937} {arXiv:1712.04937 [astro-ph.HE]}
  \BibitemShut {NoStop}%
\bibitem [{\citenamefont {Samsing}\ \emph {et~al.}(2018)\citenamefont
  {Samsing}, \citenamefont {Askar},\ and\ \citenamefont
  {Giersz}}]{Samsing:2017oij}%
  \BibitemOpen
  \bibfield  {author} {\bibinfo {author} {\bibfnamefont {J.}~\bibnamefont
  {Samsing}}, \bibinfo {author} {\bibfnamefont {A.}~\bibnamefont {Askar}}, \
  and\ \bibinfo {author} {\bibfnamefont {M.}~\bibnamefont {Giersz}},\ }\Doi
  {10.3847/1538-4357/aaab52} {\bibfield  {journal} {\bibinfo  {journal}
  {Astrophys. J.},\ }\textbf {\bibinfo {volume} {855}},\ \bibinfo {pages} {124}
  (\bibinfo {year} {2018})},\ \Eprint {http://arxiv.org/abs/1712.06186}
  {arXiv:1712.06186 [astro-ph.HE]} \BibitemShut {NoStop}%
\bibitem [{\citenamefont {Zevin}\ \emph {et~al.}(2019)\citenamefont {Zevin},
  \citenamefont {Samsing}, \citenamefont {Rodriguez}, \citenamefont {Haster},\
  and\ \citenamefont {Ramirez-Ruiz}}]{Zevin:2018kzq}%
  \BibitemOpen
  \bibfield  {author} {\bibinfo {author} {\bibfnamefont {M.}~\bibnamefont
  {Zevin}}, \bibinfo {author} {\bibfnamefont {J.}~\bibnamefont {Samsing}},
  \bibinfo {author} {\bibfnamefont {C.}~\bibnamefont {Rodriguez}}, \bibinfo
  {author} {\bibfnamefont {C.-J.}\ \bibnamefont {Haster}}, \ and\ \bibinfo
  {author} {\bibfnamefont {E.}~\bibnamefont {Ramirez-Ruiz}},\ }\Doi
  {10.3847/1538-4357/aaf6ec} {\bibfield  {journal} {\bibinfo  {journal}
  {Astrophys. J.},\ }\textbf {\bibinfo {volume} {871}},\ \bibinfo {pages} {91}
  (\bibinfo {year} {2019})},\ \Eprint {http://arxiv.org/abs/1810.00901}
  {arXiv:1810.00901 [astro-ph.HE]} \BibitemShut {NoStop}%
\bibitem [{\citenamefont {Rodriguez}\ \emph
  {et~al.}(2018){\natexlab{b}}\citenamefont {Rodriguez}, \citenamefont
  {Amaro-Seoane}, \citenamefont {Chatterjee}, \citenamefont {Kremer},
  \citenamefont {Rasio}, \citenamefont {Samsing}, \citenamefont {Ye},\ and\
  \citenamefont {Zevin}}]{Rodriguez:2018pss}%
  \BibitemOpen
  \bibfield  {author} {\bibinfo {author} {\bibfnamefont {C.~L.}\ \bibnamefont
  {Rodriguez}}, \bibinfo {author} {\bibfnamefont {P.}~\bibnamefont
  {Amaro-Seoane}}, \bibinfo {author} {\bibfnamefont {S.}~\bibnamefont
  {Chatterjee}}, \bibinfo {author} {\bibfnamefont {K.}~\bibnamefont {Kremer}},
  \bibinfo {author} {\bibfnamefont {F.~A.}\ \bibnamefont {Rasio}}, \bibinfo
  {author} {\bibfnamefont {J.}~\bibnamefont {Samsing}}, \bibinfo {author}
  {\bibfnamefont {C.~S.}\ \bibnamefont {Ye}}, \ and\ \bibinfo {author}
  {\bibfnamefont {M.}~\bibnamefont {Zevin}},\ }\Doi
  {10.1103/PhysRevD.98.123005} {\bibfield  {journal} {\bibinfo  {journal}
  {Phys. Rev. D},\ }\textbf {\bibinfo {volume} {98}},\ \bibinfo {pages}
  {123005} (\bibinfo {year} {2018}{\natexlab{b}})},\ \Eprint
  {http://arxiv.org/abs/1811.04926} {arXiv:1811.04926 [astro-ph.HE]}
  \BibitemShut {NoStop}%
\bibitem [{\citenamefont {Gond\'an}\ and\ \citenamefont
  {Kocsis}(2021)}]{Gondan:2020svr}%
  \BibitemOpen
  \bibfield  {author} {\bibinfo {author} {\bibfnamefont {L.}~\bibnamefont
  {Gond\'an}}\ and\ \bibinfo {author} {\bibfnamefont {B.}~\bibnamefont
  {Kocsis}},\ }\Doi {10.1093/mnras/stab1722} {\bibfield  {journal} {\bibinfo
  {journal} {Mon. Not. R. Astron. Soc.},\ }\textbf {\bibinfo {volume} {506}},\
  \bibinfo {pages} {1665} (\bibinfo {year} {2021})},\ \Eprint
  {http://arxiv.org/abs/2011.02507} {arXiv:2011.02507 [astro-ph.HE]}
  \BibitemShut {NoStop}%
\bibitem [{\citenamefont {Dall'Amico}\ \emph {et~al.}(2023)\citenamefont
  {Dall'Amico}, \citenamefont {Mapelli}, \citenamefont {Torniamenti},\ and\
  \citenamefont {Arca~Sedda}}]{DallAmico:2023neb}%
  \BibitemOpen
  \bibfield  {author} {\bibinfo {author} {\bibfnamefont {M.}~\bibnamefont
  {Dall'Amico}}, \bibinfo {author} {\bibfnamefont {M.}~\bibnamefont {Mapelli}},
  \bibinfo {author} {\bibfnamefont {S.}~\bibnamefont {Torniamenti}}, \ and\
  \bibinfo {author} {\bibfnamefont {M.}~\bibnamefont {Arca~Sedda}},\
  }\href@noop {} {} (\bibinfo {year} {2023}),\ \Eprint
  {http://arxiv.org/abs/2303.07421} {arXiv:2303.07421 [astro-ph.HE]}
  \BibitemShut {NoStop}%
\bibitem [{\citenamefont {Samsing}\ \emph {et~al.}(2022)\citenamefont
  {Samsing}, \citenamefont {Bartos}, \citenamefont {D'Orazio}, \citenamefont
  {Haiman}, \citenamefont {Kocsis}, \citenamefont {Leigh}, \citenamefont {Liu},
  \citenamefont {Pessah},\ and\ \citenamefont {Tagawa}}]{Samsing:2020tda}%
  \BibitemOpen
  \bibfield  {author} {\bibinfo {author} {\bibfnamefont {J.}~\bibnamefont
  {Samsing}}, \bibinfo {author} {\bibfnamefont {I.}~\bibnamefont {Bartos}},
  \bibinfo {author} {\bibfnamefont {D.~J.}\ \bibnamefont {D'Orazio}}, \bibinfo
  {author} {\bibfnamefont {Z.}~\bibnamefont {Haiman}}, \bibinfo {author}
  {\bibfnamefont {B.}~\bibnamefont {Kocsis}}, \bibinfo {author} {\bibfnamefont
  {N.~W.~C.}\ \bibnamefont {Leigh}}, \bibinfo {author} {\bibfnamefont
  {B.}~\bibnamefont {Liu}}, \bibinfo {author} {\bibfnamefont {M.~E.}\
  \bibnamefont {Pessah}}, \ and\ \bibinfo {author} {\bibfnamefont
  {H.}~\bibnamefont {Tagawa}},\ }\Doi {10.1038/s41586-021-04333-1} {\bibfield
  {journal} {\bibinfo  {journal} {Nature (London)},\ }\textbf {\bibinfo
  {volume} {603}},\ \bibinfo {pages} {237} (\bibinfo {year} {2022})},\ \Eprint
  {http://arxiv.org/abs/2010.09765} {arXiv:2010.09765 [astro-ph.HE]}
  \BibitemShut {NoStop}%
\bibitem [{\citenamefont {Tagawa}\ \emph {et~al.}(2021)\citenamefont {Tagawa},
  \citenamefont {Kocsis}, \citenamefont {Haiman}, \citenamefont {Bartos},
  \citenamefont {Omukai},\ and\ \citenamefont {Samsing}}]{Tagawa:2020jnc}%
  \BibitemOpen
  \bibfield  {author} {\bibinfo {author} {\bibfnamefont {H.}~\bibnamefont
  {Tagawa}}, \bibinfo {author} {\bibfnamefont {B.}~\bibnamefont {Kocsis}},
  \bibinfo {author} {\bibfnamefont {Z.}~\bibnamefont {Haiman}}, \bibinfo
  {author} {\bibfnamefont {I.}~\bibnamefont {Bartos}}, \bibinfo {author}
  {\bibfnamefont {K.}~\bibnamefont {Omukai}}, \ and\ \bibinfo {author}
  {\bibfnamefont {J.}~\bibnamefont {Samsing}},\ }\Doi
  {10.3847/2041-8213/abd4d3} {\bibfield  {journal} {\bibinfo  {journal}
  {Astrophys. J. Lett.},\ }\textbf {\bibinfo {volume} {907}},\ \bibinfo {pages}
  {L20} (\bibinfo {year} {2021})},\ \Eprint {http://arxiv.org/abs/2010.10526}
  {arXiv:2010.10526 [astro-ph.HE]} \BibitemShut {NoStop}%
\bibitem [{\citenamefont {Antonini}\ \emph {et~al.}(2014)\citenamefont
  {Antonini}, \citenamefont {Murray},\ and\ \citenamefont
  {Mikkola}}]{Antonini:2013tea}%
  \BibitemOpen
  \bibfield  {author} {\bibinfo {author} {\bibfnamefont {F.}~\bibnamefont
  {Antonini}}, \bibinfo {author} {\bibfnamefont {N.}~\bibnamefont {Murray}}, \
  and\ \bibinfo {author} {\bibfnamefont {S.}~\bibnamefont {Mikkola}},\ }\Doi
  {10.1088/0004-637X/781/1/45} {\bibfield  {journal} {\bibinfo  {journal}
  {Astrophys. J.},\ }\textbf {\bibinfo {volume} {781}},\ \bibinfo {pages} {45}
  (\bibinfo {year} {2014})},\ \Eprint {http://arxiv.org/abs/1308.3674}
  {arXiv:1308.3674 [astro-ph.HE]} \BibitemShut {NoStop}%
\bibitem [{\citenamefont {Antognini}\ \emph {et~al.}(2014)\citenamefont
  {Antognini}, \citenamefont {Shappee}, \citenamefont {Thompson},\ and\
  \citenamefont {Amaro-Seoane}}]{Antognini:2013lpa}%
  \BibitemOpen
  \bibfield  {author} {\bibinfo {author} {\bibfnamefont {J.~M.}\ \bibnamefont
  {Antognini}}, \bibinfo {author} {\bibfnamefont {B.~J.}\ \bibnamefont
  {Shappee}}, \bibinfo {author} {\bibfnamefont {T.~A.}\ \bibnamefont
  {Thompson}}, \ and\ \bibinfo {author} {\bibfnamefont {P.}~\bibnamefont
  {Amaro-Seoane}},\ }\Doi {10.1093/mnras/stu039} {\bibfield  {journal}
  {\bibinfo  {journal} {Mon. Not. R. Astron. Soc.},\ }\textbf {\bibinfo
  {volume} {439}},\ \bibinfo {pages} {1079} (\bibinfo {year} {2014})},\ \Eprint
  {http://arxiv.org/abs/1308.5682} {arXiv:1308.5682 [astro-ph.HE]} \BibitemShut
  {NoStop}%
\bibitem [{\citenamefont {Antonini}\ \emph {et~al.}(2017)\citenamefont
  {Antonini}, \citenamefont {Toonen},\ and\ \citenamefont
  {Hamers}}]{Antonini:2017ash}%
  \BibitemOpen
  \bibfield  {author} {\bibinfo {author} {\bibfnamefont {F.}~\bibnamefont
  {Antonini}}, \bibinfo {author} {\bibfnamefont {S.}~\bibnamefont {Toonen}}, \
  and\ \bibinfo {author} {\bibfnamefont {A.~S.}\ \bibnamefont {Hamers}},\ }\Doi
  {10.3847/1538-4357/aa6f5e} {\bibfield  {journal} {\bibinfo  {journal}
  {Astrophys. J.},\ }\textbf {\bibinfo {volume} {841}},\ \bibinfo {pages} {77}
  (\bibinfo {year} {2017})},\ \Eprint {http://arxiv.org/abs/1703.06614}
  {arXiv:1703.06614 [astro-ph.GA]} \BibitemShut {NoStop}%
\bibitem [{\citenamefont {Abbott}\ \emph
  {et~al.}(2021){\natexlab{c}}\citenamefont {Abbott} \emph
  {et~al.}}]{GWTC-3_paper}%
  \BibitemOpen
  \bibfield  {author} {\bibinfo {author} {\bibfnamefont {R.}~\bibnamefont
  {Abbott}} \emph {et~al.} (\bibinfo {collaboration} {LIGO Scientific
  Collaboration, Virgo Collaboration, and KAGRA Collaboration}),\ }\href@noop
  {} {} (\bibinfo {year} {2021}{\natexlab{c}}),\ \Eprint
  {http://arxiv.org/abs/2111.03606} {arXiv:2111.03606 [gr-qc]} \BibitemShut
  {NoStop}%
\bibitem [{\citenamefont {Romero-Shaw}\ \emph {et~al.}(2020)\citenamefont
  {Romero-Shaw}, \citenamefont {Lasky}, \citenamefont {Thrane},\ and\
  \citenamefont {Bustillo}}]{Romero-Shaw:2020thy}%
  \BibitemOpen
  \bibfield  {author} {\bibinfo {author} {\bibfnamefont {I.~M.}\ \bibnamefont
  {Romero-Shaw}}, \bibinfo {author} {\bibfnamefont {P.~D.}\ \bibnamefont
  {Lasky}}, \bibinfo {author} {\bibfnamefont {E.}~\bibnamefont {Thrane}}, \
  and\ \bibinfo {author} {\bibfnamefont {J.~C.}\ \bibnamefont {Bustillo}},\
  }\Doi {10.3847/2041-8213/abbe26} {\bibfield  {journal} {\bibinfo  {journal}
  {Astrophys. J. Lett.},\ }\textbf {\bibinfo {volume} {903}},\ \bibinfo {pages}
  {L5} (\bibinfo {year} {2020})},\ \Eprint {http://arxiv.org/abs/2009.04771}
  {arXiv:2009.04771 [astro-ph.HE]} \BibitemShut {NoStop}%
\bibitem [{\citenamefont {Gayathri}\ \emph {et~al.}(2022)\citenamefont
  {Gayathri}, \citenamefont {Healy}, \citenamefont {Lange}, \citenamefont
  {O'Brien}, \citenamefont {Szczepanczyk}, \citenamefont {Bartos},
  \citenamefont {Campanelli}, \citenamefont {Klimenko}, \citenamefont
  {Lousto},\ and\ \citenamefont {O'Shaughnessy}}]{Gayathri:2020coq}%
  \BibitemOpen
  \bibfield  {author} {\bibinfo {author} {\bibfnamefont {V.}~\bibnamefont
  {Gayathri}}, \bibinfo {author} {\bibfnamefont {J.}~\bibnamefont {Healy}},
  \bibinfo {author} {\bibfnamefont {J.}~\bibnamefont {Lange}}, \bibinfo
  {author} {\bibfnamefont {B.}~\bibnamefont {O'Brien}}, \bibinfo {author}
  {\bibfnamefont {M.}~\bibnamefont {Szczepanczyk}}, \bibinfo {author}
  {\bibfnamefont {I.}~\bibnamefont {Bartos}}, \bibinfo {author} {\bibfnamefont
  {M.}~\bibnamefont {Campanelli}}, \bibinfo {author} {\bibfnamefont
  {S.}~\bibnamefont {Klimenko}}, \bibinfo {author} {\bibfnamefont {C.~O.}\
  \bibnamefont {Lousto}}, \ and\ \bibinfo {author} {\bibfnamefont
  {R.}~\bibnamefont {O'Shaughnessy}},\ }\Doi {10.1038/s41550-021-01568-w}
  {\bibfield  {journal} {\bibinfo  {journal} {Nature Astron.},\ }\textbf
  {\bibinfo {volume} {6}},\ \bibinfo {pages} {344} (\bibinfo {year} {2022})},\
  \Eprint {http://arxiv.org/abs/2009.05461} {arXiv:2009.05461 [astro-ph.HE]}
  \BibitemShut {NoStop}%
\bibitem [{\citenamefont {Romero-Shaw}\ \emph {et~al.}(2021)\citenamefont
  {Romero-Shaw}, \citenamefont {Lasky},\ and\ \citenamefont
  {Thrane}}]{Romero-Shaw:2021ual}%
  \BibitemOpen
  \bibfield  {author} {\bibinfo {author} {\bibfnamefont {I.~M.}\ \bibnamefont
  {Romero-Shaw}}, \bibinfo {author} {\bibfnamefont {P.~D.}\ \bibnamefont
  {Lasky}}, \ and\ \bibinfo {author} {\bibfnamefont {E.}~\bibnamefont
  {Thrane}},\ }\Doi {10.3847/2041-8213/ac3138} {\bibfield  {journal} {\bibinfo
  {journal} {Astrophys. J. Lett.},\ }\textbf {\bibinfo {volume} {921}},\
  \bibinfo {pages} {L31} (\bibinfo {year} {2021})},\ \Eprint
  {http://arxiv.org/abs/2108.01284} {arXiv:2108.01284 [astro-ph.HE]}
  \BibitemShut {NoStop}%
\bibitem [{\citenamefont {Romero-Shaw}\ \emph {et~al.}(2022)\citenamefont
  {Romero-Shaw}, \citenamefont {Lasky},\ and\ \citenamefont
  {Thrane}}]{Romero-Shaw:2022xko}%
  \BibitemOpen
  \bibfield  {author} {\bibinfo {author} {\bibfnamefont {I.~M.}\ \bibnamefont
  {Romero-Shaw}}, \bibinfo {author} {\bibfnamefont {P.~D.}\ \bibnamefont
  {Lasky}}, \ and\ \bibinfo {author} {\bibfnamefont {E.}~\bibnamefont
  {Thrane}},\ }\Doi {10.3847/1538-4357/ac9798} {\bibfield  {journal} {\bibinfo
  {journal} {Astrophys. J.},\ }\textbf {\bibinfo {volume} {940}},\ \bibinfo
  {pages} {171} (\bibinfo {year} {2022})},\ \Eprint
  {http://arxiv.org/abs/2206.14695} {arXiv:2206.14695 [astro-ph.HE]}
  \BibitemShut {NoStop}%
\bibitem [{\citenamefont {Romero-Shaw}\ \emph {et~al.}(2023)\citenamefont
  {Romero-Shaw}, \citenamefont {Gerosa},\ and\ \citenamefont
  {Loutrel}}]{Romero-Shaw:2022fbf}%
  \BibitemOpen
  \bibfield  {author} {\bibinfo {author} {\bibfnamefont {I.~M.}\ \bibnamefont
  {Romero-Shaw}}, \bibinfo {author} {\bibfnamefont {D.}~\bibnamefont {Gerosa}},
  \ and\ \bibinfo {author} {\bibfnamefont {N.}~\bibnamefont {Loutrel}},\ }\Doi
  {10.1093/mnras/stad031} {\bibfield  {journal} {\bibinfo  {journal} {Mon. Not.
  R. Astron. Soc.},\ }\textbf {\bibinfo {volume} {519}},\ \bibinfo {pages}
  {5352} (\bibinfo {year} {2023})},\ \Eprint {http://arxiv.org/abs/2211.07528}
  {arXiv:2211.07528 [astro-ph.HE]} \BibitemShut {NoStop}%
\bibitem [{\citenamefont {O'Shea}\ and\ \citenamefont
  {Kumar}(2021)}]{OShea:2021ugg}%
  \BibitemOpen
  \bibfield  {author} {\bibinfo {author} {\bibfnamefont {E.}~\bibnamefont
  {O'Shea}}\ and\ \bibinfo {author} {\bibfnamefont {P.}~\bibnamefont {Kumar}},\
  }\href@noop {} {} (\bibinfo {year} {2021}),\ \Eprint
  {http://arxiv.org/abs/2107.07981} {arXiv:2107.07981 [astro-ph.HE]}
  \BibitemShut {NoStop}%
\bibitem [{\citenamefont {Favata}\ \emph {et~al.}(2022)\citenamefont {Favata},
  \citenamefont {Kim}, \citenamefont {Arun}, \citenamefont {Kim},\ and\
  \citenamefont {Lee}}]{Favata:2021vhw}%
  \BibitemOpen
  \bibfield  {author} {\bibinfo {author} {\bibfnamefont {M.}~\bibnamefont
  {Favata}}, \bibinfo {author} {\bibfnamefont {C.}~\bibnamefont {Kim}},
  \bibinfo {author} {\bibfnamefont {K.~G.}\ \bibnamefont {Arun}}, \bibinfo
  {author} {\bibfnamefont {J.}~\bibnamefont {Kim}}, \ and\ \bibinfo {author}
  {\bibfnamefont {H.~W.}\ \bibnamefont {Lee}},\ }\Doi
  {10.1103/PhysRevD.105.023003} {\bibfield  {journal} {\bibinfo  {journal}
  {Phys. Rev. D},\ }\textbf {\bibinfo {volume} {105}},\ \bibinfo {pages}
  {023003} (\bibinfo {year} {2022})},\ \Eprint
  {http://arxiv.org/abs/2108.05861} {arXiv:2108.05861 [gr-qc]} \BibitemShut
  {NoStop}%
\bibitem [{\citenamefont {Saini}\ \emph {et~al.}(2022)\citenamefont {Saini},
  \citenamefont {Favata},\ and\ \citenamefont {Arun}}]{Saini:2022igm}%
  \BibitemOpen
  \bibfield  {author} {\bibinfo {author} {\bibfnamefont {P.}~\bibnamefont
  {Saini}}, \bibinfo {author} {\bibfnamefont {M.}~\bibnamefont {Favata}}, \
  and\ \bibinfo {author} {\bibfnamefont {K.~G.}\ \bibnamefont {Arun}},\ }\Doi
  {10.1103/PhysRevD.106.084031} {\bibfield  {journal} {\bibinfo  {journal}
  {Phys. Rev. D},\ }\textbf {\bibinfo {volume} {106}},\ \bibinfo {pages}
  {084031} (\bibinfo {year} {2022})},\ \Eprint
  {http://arxiv.org/abs/2203.04634} {arXiv:2203.04634 [gr-qc]} \BibitemShut
  {NoStop}%
\bibitem [{\citenamefont {Bhat}\ \emph {et~al.}(2023)\citenamefont {Bhat},
  \citenamefont {Saini}, \citenamefont {Favata},\ and\ \citenamefont
  {Arun}}]{Bhat:2022amc}%
  \BibitemOpen
  \bibfield  {author} {\bibinfo {author} {\bibfnamefont {S.~A.}\ \bibnamefont
  {Bhat}}, \bibinfo {author} {\bibfnamefont {P.}~\bibnamefont {Saini}},
  \bibinfo {author} {\bibfnamefont {M.}~\bibnamefont {Favata}}, \ and\ \bibinfo
  {author} {\bibfnamefont {K.~G.}\ \bibnamefont {Arun}},\ }\Doi
  {10.1103/PhysRevD.107.024009} {\bibfield  {journal} {\bibinfo  {journal}
  {Phys. Rev. D},\ }\textbf {\bibinfo {volume} {107}},\ \bibinfo {pages}
  {024009} (\bibinfo {year} {2023})},\ \Eprint
  {http://arxiv.org/abs/2207.13761} {arXiv:2207.13761 [gr-qc]} \BibitemShut
  {NoStop}%
\bibitem [{\citenamefont {Agathos}\ \emph {et~al.}(2014)\citenamefont
  {Agathos}, \citenamefont {Del~Pozzo}, \citenamefont {Li}, \citenamefont {Van
  Den~Broeck}, \citenamefont {Veitch},\ and\ \citenamefont
  {Vitale}}]{TIGER2014}%
  \BibitemOpen
  \bibfield  {author} {\bibinfo {author} {\bibfnamefont {M.}~\bibnamefont
  {Agathos}}, \bibinfo {author} {\bibfnamefont {W.}~\bibnamefont {Del~Pozzo}},
  \bibinfo {author} {\bibfnamefont {T.~G.~F.}\ \bibnamefont {Li}}, \bibinfo
  {author} {\bibfnamefont {C.}~\bibnamefont {Van Den~Broeck}}, \bibinfo
  {author} {\bibfnamefont {J.}~\bibnamefont {Veitch}}, \ and\ \bibinfo {author}
  {\bibfnamefont {S.}~\bibnamefont {Vitale}},\ }\Doi
  {10.1103/PhysRevD.89.082001} {\bibfield  {journal} {\bibinfo  {journal}
  {Phys. Rev. D},\ }\textbf {\bibinfo {volume} {89}},\ \bibinfo {pages}
  {082001} (\bibinfo {year} {2014})},\ \Eprint {http://arxiv.org/abs/1311.0420}
  {arXiv:1311.0420 [gr-qc]} \BibitemShut {NoStop}%
\bibitem [{\citenamefont {Meidam}\ \emph {et~al.}(2018)\citenamefont {Meidam}
  \emph {et~al.}}]{Meidam:2017dgf}%
  \BibitemOpen
  \bibfield  {author} {\bibinfo {author} {\bibfnamefont {J.}~\bibnamefont
  {Meidam}} \emph {et~al.},\ }\Doi {10.1103/PhysRevD.97.044033} {\bibfield
  {journal} {\bibinfo  {journal} {Phys. Rev. D},\ }\textbf {\bibinfo {volume}
  {97}},\ \bibinfo {pages} {044033} (\bibinfo {year} {2018})},\ \Eprint
  {http://arxiv.org/abs/1712.08772} {arXiv:1712.08772 [gr-qc]} \BibitemShut
  {NoStop}%
\bibitem [{\citenamefont {Mehta}\ \emph {et~al.}(2023)\citenamefont {Mehta},
  \citenamefont {Buonanno}, \citenamefont {Cotesta}, \citenamefont {Ghosh},
  \citenamefont {Sennett},\ and\ \citenamefont {Steinhoff}}]{Mehta:2022pcn}%
  \BibitemOpen
  \bibfield  {author} {\bibinfo {author} {\bibfnamefont {A.~K.}\ \bibnamefont
  {Mehta}}, \bibinfo {author} {\bibfnamefont {A.}~\bibnamefont {Buonanno}},
  \bibinfo {author} {\bibfnamefont {R.}~\bibnamefont {Cotesta}}, \bibinfo
  {author} {\bibfnamefont {A.}~\bibnamefont {Ghosh}}, \bibinfo {author}
  {\bibfnamefont {N.}~\bibnamefont {Sennett}}, \ and\ \bibinfo {author}
  {\bibfnamefont {J.}~\bibnamefont {Steinhoff}},\ }\Doi
  {10.1103/PhysRevD.107.044020} {\bibfield  {journal} {\bibinfo  {journal}
  {Phys. Rev. D},\ }\textbf {\bibinfo {volume} {107}},\ \bibinfo {pages}
  {044020} (\bibinfo {year} {2023})},\ \Eprint
  {http://arxiv.org/abs/2203.13937} {arXiv:2203.13937 [gr-qc]} \BibitemShut
  {NoStop}%
\bibitem [{\citenamefont {Mirshekari}\ \emph {et~al.}(2012)\citenamefont
  {Mirshekari}, \citenamefont {Yunes},\ and\ \citenamefont
  {Will}}]{Mirshekari:2011yq}%
  \BibitemOpen
  \bibfield  {author} {\bibinfo {author} {\bibfnamefont {S.}~\bibnamefont
  {Mirshekari}}, \bibinfo {author} {\bibfnamefont {N.}~\bibnamefont {Yunes}}, \
  and\ \bibinfo {author} {\bibfnamefont {C.~M.}\ \bibnamefont {Will}},\ }\Doi
  {10.1103/PhysRevD.85.024041} {\bibfield  {journal} {\bibinfo  {journal}
  {Phys. Rev. D},\ }\textbf {\bibinfo {volume} {85}},\ \bibinfo {pages}
  {024041} (\bibinfo {year} {2012})},\ \Eprint {http://arxiv.org/abs/1110.2720}
  {arXiv:1110.2720 [gr-qc]} \BibitemShut {NoStop}%
\bibitem [{\citenamefont {{Ghosh}}\ \emph {et~al.}(2016)\citenamefont
  {{Ghosh}}, \citenamefont {{Ghosh}}, \citenamefont {{Johnson-McDaniel}},
  \citenamefont {{Mishra}}, \citenamefont {{Ajith}}, \citenamefont {{Del
  Pozzo}}, \citenamefont {{Nichols}}, \citenamefont {{Chen}}, \citenamefont
  {{Nielsen}}, \citenamefont {{Berry}},\ and\ \citenamefont
  {{London}}}]{Ghosh:2016qgn}%
  \BibitemOpen
  \bibfield  {author} {\bibinfo {author} {\bibfnamefont {A.}~\bibnamefont
  {{Ghosh}}}, \bibinfo {author} {\bibfnamefont {A.}~\bibnamefont {{Ghosh}}},
  \bibinfo {author} {\bibfnamefont {N.~K.}\ \bibnamefont {{Johnson-McDaniel}}},
  \bibinfo {author} {\bibfnamefont {C.~K.}\ \bibnamefont {{Mishra}}}, \bibinfo
  {author} {\bibfnamefont {P.}~\bibnamefont {{Ajith}}}, \bibinfo {author}
  {\bibfnamefont {W.}~\bibnamefont {{Del Pozzo}}}, \bibinfo {author}
  {\bibfnamefont {D.~A.}\ \bibnamefont {{Nichols}}}, \bibinfo {author}
  {\bibfnamefont {Y.}~\bibnamefont {{Chen}}}, \bibinfo {author} {\bibfnamefont
  {A.~B.}\ \bibnamefont {{Nielsen}}}, \bibinfo {author} {\bibfnamefont
  {C.~P.~L.}\ \bibnamefont {{Berry}}}, \ and\ \bibinfo {author} {\bibfnamefont
  {L.}~\bibnamefont {{London}}},\ }\Doi {10.1103/PhysRevD.94.021101} {\bibfield
   {journal} {\bibinfo  {journal} {Phys. Rev. D},\ }\textbf {\bibinfo {volume}
  {94}},\ \bibinfo {pages} {021101(R)} (\bibinfo {year} {2016})},\ \Eprint
  {http://arxiv.org/abs/1602.02453} {arXiv:1602.02453 [gr-qc]} \BibitemShut
  {NoStop}%
\bibitem [{\citenamefont {Ghosh}\ \emph {et~al.}(2018)\citenamefont {Ghosh},
  \citenamefont {Johnson-McDaniel}, \citenamefont {Ghosh}, \citenamefont
  {Mishra}, \citenamefont {Ajith}, \citenamefont {Del~Pozzo}, \citenamefont
  {Berry}, \citenamefont {Nielsen},\ and\ \citenamefont
  {London}}]{Ghosh:2017gfp}%
  \BibitemOpen
  \bibfield  {author} {\bibinfo {author} {\bibfnamefont {A.}~\bibnamefont
  {Ghosh}}, \bibinfo {author} {\bibfnamefont {N.~K.}\ \bibnamefont
  {Johnson-McDaniel}}, \bibinfo {author} {\bibfnamefont {A.}~\bibnamefont
  {Ghosh}}, \bibinfo {author} {\bibfnamefont {C.~K.}\ \bibnamefont {Mishra}},
  \bibinfo {author} {\bibfnamefont {P.}~\bibnamefont {Ajith}}, \bibinfo
  {author} {\bibfnamefont {W.}~\bibnamefont {Del~Pozzo}}, \bibinfo {author}
  {\bibfnamefont {C.~P.~L.}\ \bibnamefont {Berry}}, \bibinfo {author}
  {\bibfnamefont {A.~B.}\ \bibnamefont {Nielsen}}, \ and\ \bibinfo {author}
  {\bibfnamefont {L.}~\bibnamefont {London}},\ }\Doi {10.1088/1361-6382/aa972e}
  {\bibfield  {journal} {\bibinfo  {journal} {Classical Quantum Gravity},\
  }\textbf {\bibinfo {volume} {35}},\ \bibinfo {pages} {014002} (\bibinfo
  {year} {2018})},\ \Eprint {http://arxiv.org/abs/1704.06784} {arXiv:1704.06784
  [gr-qc]} \BibitemShut {NoStop}%
\bibitem [{\citenamefont {Abbott}\ \emph {et~al.}(2020)\citenamefont {Abbott}
  \emph {et~al.}}]{Aasi:2013wya}%
  \BibitemOpen
  \bibfield  {author} {\bibinfo {author} {\bibfnamefont {B.~P.}\ \bibnamefont
  {Abbott}} \emph {et~al.} (\bibinfo {collaboration} {KAGRA Collaboration, LIGO
  Scientific Collaboration, and Virgo Collaboration}),\ }\Doi
  {10.1007/s41114-020-00026-9} {\bibfield  {journal} {\bibinfo  {journal}
  {Living Rev. Relativity},\ }\textbf {\bibinfo {volume} {23}},\ \bibinfo
  {pages} {3} (\bibinfo {year} {2020})},\ \bibinfo {note} {noise curves
  available from \url{https://dcc.ligo.org/LIGO-T2000012/public}},\ \Eprint
  {http://arxiv.org/abs/1304.0670} {arXiv:1304.0670 [gr-qc]} \BibitemShut
  {NoStop}%
\bibitem [{\citenamefont {Boyle}\ \emph {et~al.}(2019)\citenamefont {Boyle}
  \emph {et~al.}}]{Boyle:2019kee}%
  \BibitemOpen
  \bibfield  {author} {\bibinfo {author} {\bibfnamefont {M.}~\bibnamefont
  {Boyle}} \emph {et~al.},\ }\Doi {10.1088/1361-6382/ab34e2} {\bibfield
  {journal} {\bibinfo  {journal} {Classical Quantum Gravity},\ }\textbf
  {\bibinfo {volume} {36}},\ \bibinfo {pages} {195006} (\bibinfo {year}
  {2019})},\ \Eprint {http://arxiv.org/abs/1904.04831} {arXiv:1904.04831
  [gr-qc]} \BibitemShut {NoStop}%
\bibitem [{\citenamefont {Hinder}\ \emph {et~al.}(2018)\citenamefont {Hinder},
  \citenamefont {Kidder},\ and\ \citenamefont {Pfeiffer}}]{Hinder:2017sxy}%
  \BibitemOpen
  \bibfield  {author} {\bibinfo {author} {\bibfnamefont {I.}~\bibnamefont
  {Hinder}}, \bibinfo {author} {\bibfnamefont {L.~E.}\ \bibnamefont {Kidder}},
  \ and\ \bibinfo {author} {\bibfnamefont {H.~P.}\ \bibnamefont {Pfeiffer}},\
  }\Doi {10.1103/PhysRevD.98.044015} {\bibfield  {journal} {\bibinfo  {journal}
  {Phys. Rev. D},\ }\textbf {\bibinfo {volume} {98}},\ \bibinfo {pages}
  {044015} (\bibinfo {year} {2018})},\ \Eprint
  {http://arxiv.org/abs/1709.02007} {arXiv:1709.02007 [gr-qc]} \BibitemShut
  {NoStop}%
\bibitem [{Muk()}]{Mukesh_ICTS}%
  \BibitemOpen
  \href@noop {} {}\bibinfo {note} {M.~K.~Singh (private
  communication)}\BibitemShut {NoStop}%
\bibitem [{\citenamefont {Hannam}\ \emph {et~al.}(2014)\citenamefont {Hannam},
  \citenamefont {Schmidt}, \citenamefont {Boh{\'e}}, \citenamefont {Haegel},
  \citenamefont {Husa}, \citenamefont {Ohme}, \citenamefont {Pratten},\ and\
  \citenamefont {P{\"u}rrer}}]{Hannam:2013oca}%
  \BibitemOpen
  \bibfield  {author} {\bibinfo {author} {\bibfnamefont {M.}~\bibnamefont
  {Hannam}}, \bibinfo {author} {\bibfnamefont {P.}~\bibnamefont {Schmidt}},
  \bibinfo {author} {\bibfnamefont {A.}~\bibnamefont {Boh{\'e}}}, \bibinfo
  {author} {\bibfnamefont {L.}~\bibnamefont {Haegel}}, \bibinfo {author}
  {\bibfnamefont {S.}~\bibnamefont {Husa}}, \bibinfo {author} {\bibfnamefont
  {F.}~\bibnamefont {Ohme}}, \bibinfo {author} {\bibfnamefont {G.}~\bibnamefont
  {Pratten}}, \ and\ \bibinfo {author} {\bibfnamefont {M.}~\bibnamefont
  {P{\"u}rrer}},\ }\Doi {10.1103/PhysRevLett.113.151101} {\bibfield  {journal}
  {\bibinfo  {journal} {Phys. Rev. Lett.},\ }\textbf {\bibinfo {volume}
  {113}},\ \bibinfo {pages} {151101} (\bibinfo {year} {2014})},\ \Eprint
  {http://arxiv.org/abs/1308.3271} {arXiv:1308.3271 [gr-qc]} \BibitemShut
  {NoStop}%
\bibitem [{\citenamefont {Khan}\ \emph {et~al.}(2016)\citenamefont {Khan},
  \citenamefont {Husa}, \citenamefont {Hannam}, \citenamefont {Ohme},
  \citenamefont {P{\"u}rrer}, \citenamefont {Jim{\'e}nez~Forteza},\ and\
  \citenamefont {Boh{\'e}}}]{Khan:2015jqa}%
  \BibitemOpen
  \bibfield  {author} {\bibinfo {author} {\bibfnamefont {S.}~\bibnamefont
  {Khan}}, \bibinfo {author} {\bibfnamefont {S.}~\bibnamefont {Husa}}, \bibinfo
  {author} {\bibfnamefont {M.}~\bibnamefont {Hannam}}, \bibinfo {author}
  {\bibfnamefont {F.}~\bibnamefont {Ohme}}, \bibinfo {author} {\bibfnamefont
  {M.}~\bibnamefont {P{\"u}rrer}}, \bibinfo {author} {\bibfnamefont
  {X.}~\bibnamefont {Jim{\'e}nez~Forteza}}, \ and\ \bibinfo {author}
  {\bibfnamefont {A.}~\bibnamefont {Boh{\'e}}},\ }\Doi
  {10.1103/PhysRevD.93.044007} {\bibfield  {journal} {\bibinfo  {journal}
  {Phys. Rev. D},\ }\textbf {\bibinfo {volume} {93}},\ \bibinfo {pages}
  {044007} (\bibinfo {year} {2016})},\ \Eprint
  {http://arxiv.org/abs/1508.07253} {arXiv:1508.07253 [gr-qc]} \BibitemShut
  {NoStop}%
\bibitem [{\citenamefont {Boh{\'e}}\ \emph {et~al.}(2016)\citenamefont
  {Boh{\'e}}, \citenamefont {Hannam}, \citenamefont {Husa}, \citenamefont
  {Ohme}, \citenamefont {P{\"u}rrer},\ and\ \citenamefont
  {Schmidt}}]{Bohe:PPv2}%
  \BibitemOpen
  \bibfield  {author} {\bibinfo {author} {\bibfnamefont {A.}~\bibnamefont
  {Boh{\'e}}}, \bibinfo {author} {\bibfnamefont {M.}~\bibnamefont {Hannam}},
  \bibinfo {author} {\bibfnamefont {S.}~\bibnamefont {Husa}}, \bibinfo {author}
  {\bibfnamefont {F.}~\bibnamefont {Ohme}}, \bibinfo {author} {\bibfnamefont
  {M.}~\bibnamefont {P{\"u}rrer}}, \ and\ \bibinfo {author} {\bibfnamefont
  {P.}~\bibnamefont {Schmidt}},\ }\href@noop {} {\emph {\bibinfo {title}
  {PhenomPv2 - Technical Notes for LAL Implementation}}},\ \bibinfo {type}
  {Tech. Rep.}\ \bibinfo {number} {{LIGO}-T1500602}\ (\bibinfo  {institution}
  {{LIGO} Project},\ \bibinfo {year} {2016})\ \bibinfo {note}
  {\url{https://dcc.ligo.org/LIGO-T1500602/public}}\BibitemShut {NoStop}%
\bibitem [{\citenamefont {Cotesta}\ \emph {et~al.}(2018)\citenamefont
  {Cotesta}, \citenamefont {Buonanno}, \citenamefont {Boh\'e}, \citenamefont
  {Taracchini}, \citenamefont {Hinder},\ and\ \citenamefont
  {Ossokine}}]{Cotesta:2018fcv}%
  \BibitemOpen
  \bibfield  {author} {\bibinfo {author} {\bibfnamefont {R.}~\bibnamefont
  {Cotesta}}, \bibinfo {author} {\bibfnamefont {A.}~\bibnamefont {Buonanno}},
  \bibinfo {author} {\bibfnamefont {A.}~\bibnamefont {Boh\'e}}, \bibinfo
  {author} {\bibfnamefont {A.}~\bibnamefont {Taracchini}}, \bibinfo {author}
  {\bibfnamefont {I.}~\bibnamefont {Hinder}}, \ and\ \bibinfo {author}
  {\bibfnamefont {S.}~\bibnamefont {Ossokine}},\ }\Doi
  {10.1103/PhysRevD.98.084028} {\bibfield  {journal} {\bibinfo  {journal}
  {Phys. Rev. D},\ }\textbf {\bibinfo {volume} {98}},\ \bibinfo {pages}
  {084028} (\bibinfo {year} {2018})},\ \Eprint
  {http://arxiv.org/abs/1803.10701} {arXiv:1803.10701 [gr-qc]} \BibitemShut
  {NoStop}%
\bibitem [{\citenamefont {Cotesta}\ \emph {et~al.}(2020)\citenamefont
  {Cotesta}, \citenamefont {Marsat},\ and\ \citenamefont
  {P\"urrer}}]{Cotesta:2020qhw}%
  \BibitemOpen
  \bibfield  {author} {\bibinfo {author} {\bibfnamefont {R.}~\bibnamefont
  {Cotesta}}, \bibinfo {author} {\bibfnamefont {S.}~\bibnamefont {Marsat}}, \
  and\ \bibinfo {author} {\bibfnamefont {M.}~\bibnamefont {P\"urrer}},\ }\Doi
  {10.1103/PhysRevD.101.124040} {\bibfield  {journal} {\bibinfo  {journal}
  {Phys. Rev. D},\ }\textbf {\bibinfo {volume} {101}},\ \bibinfo {pages}
  {124040} (\bibinfo {year} {2020})},\ \Eprint
  {http://arxiv.org/abs/2003.12079} {arXiv:2003.12079 [gr-qc]} \BibitemShut
  {NoStop}%
\bibitem [{\citenamefont {Pratten}\ \emph {et~al.}(2021)\citenamefont {Pratten}
  \emph {et~al.}}]{Pratten:2020ceb}%
  \BibitemOpen
  \bibfield  {author} {\bibinfo {author} {\bibfnamefont {G.}~\bibnamefont
  {Pratten}} \emph {et~al.},\ }\Doi {10.1103/PhysRevD.103.104056} {\bibfield
  {journal} {\bibinfo  {journal} {Phys. Rev. D},\ }\textbf {\bibinfo {volume}
  {103}},\ \bibinfo {pages} {104056} (\bibinfo {year} {2021})},\ \Eprint
  {http://arxiv.org/abs/2004.06503} {arXiv:2004.06503 [gr-qc]} \BibitemShut
  {NoStop}%
\bibitem [{\citenamefont {Johnson-McDaniel}\ \emph {et~al.}(2022)\citenamefont
  {Johnson-McDaniel}, \citenamefont {Ghosh}, \citenamefont {Ghonge},
  \citenamefont {Saleem}, \citenamefont {Krishnendu},\ and\ \citenamefont
  {Clark}}]{TGR_relation}%
  \BibitemOpen
  \bibfield  {author} {\bibinfo {author} {\bibfnamefont {N.~K.}\ \bibnamefont
  {Johnson-McDaniel}}, \bibinfo {author} {\bibfnamefont {A.}~\bibnamefont
  {Ghosh}}, \bibinfo {author} {\bibfnamefont {S.}~\bibnamefont {Ghonge}},
  \bibinfo {author} {\bibfnamefont {M.}~\bibnamefont {Saleem}}, \bibinfo
  {author} {\bibfnamefont {N.~V.}\ \bibnamefont {Krishnendu}}, \ and\ \bibinfo
  {author} {\bibfnamefont {J.~A.}\ \bibnamefont {Clark}},\ }\Doi
  {10.1103/PhysRevD.105.044020} {\bibfield  {journal} {\bibinfo  {journal}
  {Phys. Rev. D},\ }\textbf {\bibinfo {volume} {105}},\ \bibinfo {pages}
  {044020} (\bibinfo {year} {2022})},\ \Eprint
  {http://arxiv.org/abs/2109.06988} {arXiv:2109.06988 [gr-qc]} \BibitemShut
  {NoStop}%
\bibitem [{\citenamefont {Gupta}\ \emph {et~al.}(2020)\citenamefont {Gupta},
  \citenamefont {Datta}, \citenamefont {Kastha}, \citenamefont {Borhanian},
  \citenamefont {Arun},\ and\ \citenamefont {Sathyaprakash}}]{Gupta:2020lxa}%
  \BibitemOpen
  \bibfield  {author} {\bibinfo {author} {\bibfnamefont {A.}~\bibnamefont
  {Gupta}}, \bibinfo {author} {\bibfnamefont {S.}~\bibnamefont {Datta}},
  \bibinfo {author} {\bibfnamefont {S.}~\bibnamefont {Kastha}}, \bibinfo
  {author} {\bibfnamefont {S.}~\bibnamefont {Borhanian}}, \bibinfo {author}
  {\bibfnamefont {K.~G.}\ \bibnamefont {Arun}}, \ and\ \bibinfo {author}
  {\bibfnamefont {B.~S.}\ \bibnamefont {Sathyaprakash}},\ }\Doi
  {10.1103/PhysRevLett.125.201101} {\bibfield  {journal} {\bibinfo  {journal}
  {Phys. Rev. Lett.},\ }\textbf {\bibinfo {volume} {125}},\ \bibinfo {pages}
  {201101} (\bibinfo {year} {2020})},\ \Eprint
  {http://arxiv.org/abs/2005.09607} {arXiv:2005.09607 [gr-qc]} \BibitemShut
  {NoStop}%
\bibitem [{\citenamefont {Datta}\ \emph {et~al.}(2021)\citenamefont {Datta},
  \citenamefont {Gupta}, \citenamefont {Kastha}, \citenamefont {Arun},\ and\
  \citenamefont {Sathyaprakash}}]{Datta:2020vcj}%
  \BibitemOpen
  \bibfield  {author} {\bibinfo {author} {\bibfnamefont {S.}~\bibnamefont
  {Datta}}, \bibinfo {author} {\bibfnamefont {A.}~\bibnamefont {Gupta}},
  \bibinfo {author} {\bibfnamefont {S.}~\bibnamefont {Kastha}}, \bibinfo
  {author} {\bibfnamefont {K.~G.}\ \bibnamefont {Arun}}, \ and\ \bibinfo
  {author} {\bibfnamefont {B.~S.}\ \bibnamefont {Sathyaprakash}},\ }\Doi
  {10.1103/PhysRevD.103.024036} {\bibfield  {journal} {\bibinfo  {journal}
  {Phys. Rev. D},\ }\textbf {\bibinfo {volume} {103}},\ \bibinfo {pages}
  {024036} (\bibinfo {year} {2021})},\ \Eprint
  {http://arxiv.org/abs/2006.12137} {arXiv:2006.12137 [gr-qc]} \BibitemShut
  {NoStop}%
\bibitem [{\citenamefont {Evans}\ \emph {et~al.}(2021)\citenamefont {Evans}
  \emph {et~al.}}]{Evans:2021gyd}%
  \BibitemOpen
  \bibfield  {author} {\bibinfo {author} {\bibfnamefont {M.}~\bibnamefont
  {Evans}} \emph {et~al.},\ }\href@noop {} {} (\bibinfo {year} {2021}),\
  \Eprint {http://arxiv.org/abs/2109.09882} {arXiv:2109.09882 [astro-ph.IM]}
  \BibitemShut {NoStop}%
\bibitem [{\citenamefont {Amaro-Seoane}\ \emph {et~al.}(2017)\citenamefont
  {Amaro-Seoane} \emph {et~al.}}]{LISA}%
  \BibitemOpen
  \bibfield  {author} {\bibinfo {author} {\bibfnamefont {P.}~\bibnamefont
  {Amaro-Seoane}} \emph {et~al.},\ }\href@noop {} {} (\bibinfo {year} {2017}),\
  \Eprint {http://arxiv.org/abs/1702.00786} {arXiv:1702.00786 [astro-ph.IM]}
  \BibitemShut {NoStop}%
\bibitem [{\citenamefont {{Shoom}}\ \emph {et~al.}(2023)\citenamefont
  {{Shoom}}, \citenamefont {{Gupta}}, \citenamefont {{Krishnan}}, \citenamefont
  {{Nielsen}},\ and\ \citenamefont {{Capano}}}]{Shoom:2021mdj}%
  \BibitemOpen
  \bibfield  {author} {\bibinfo {author} {\bibfnamefont {A.~A.}\ \bibnamefont
  {{Shoom}}}, \bibinfo {author} {\bibfnamefont {P.~K.}\ \bibnamefont
  {{Gupta}}}, \bibinfo {author} {\bibfnamefont {B.}~\bibnamefont {{Krishnan}}},
  \bibinfo {author} {\bibfnamefont {A.~B.}\ \bibnamefont {{Nielsen}}}, \ and\
  \bibinfo {author} {\bibfnamefont {C.~D.}\ \bibnamefont {{Capano}}},\ }\Doi
  {10.1007/s10714-023-03100-z} {\bibfield  {journal} {\bibinfo  {journal} {Gen.
  Relativ. Gravit.},\ }\textbf {\bibinfo {volume} {55}},\ \bibinfo {eid} {55}
  (\bibinfo {year} {2023})},\ \Eprint {http://arxiv.org/abs/2105.02191}
  {arXiv:2105.02191 [gr-qc]} \BibitemShut {NoStop}%
\bibitem [{\citenamefont {Saleem}\ \emph {et~al.}(2022)\citenamefont {Saleem},
  \citenamefont {Datta}, \citenamefont {Arun},\ and\ \citenamefont
  {Sathyaprakash}}]{Saleem:2021nsb}%
  \BibitemOpen
  \bibfield  {author} {\bibinfo {author} {\bibfnamefont {M.}~\bibnamefont
  {Saleem}}, \bibinfo {author} {\bibfnamefont {S.}~\bibnamefont {Datta}},
  \bibinfo {author} {\bibfnamefont {K.~G.}\ \bibnamefont {Arun}}, \ and\
  \bibinfo {author} {\bibfnamefont {B.~S.}\ \bibnamefont {Sathyaprakash}},\
  }\Doi {10.1103/PhysRevD.105.084062} {\bibfield  {journal} {\bibinfo
  {journal} {Phys. Rev. D},\ }\textbf {\bibinfo {volume} {105}},\ \bibinfo
  {pages} {084062} (\bibinfo {year} {2022})},\ \Eprint
  {http://arxiv.org/abs/2110.10147} {arXiv:2110.10147 [gr-qc]} \BibitemShut
  {NoStop}%
\bibitem [{\citenamefont {Datta}\ \emph {et~al.}(2022)\citenamefont {Datta},
  \citenamefont {Saleem}, \citenamefont {Arun},\ and\ \citenamefont
  {Sathyaprakash}}]{Datta:2022izc}%
  \BibitemOpen
  \bibfield  {author} {\bibinfo {author} {\bibfnamefont {S.}~\bibnamefont
  {Datta}}, \bibinfo {author} {\bibfnamefont {M.}~\bibnamefont {Saleem}},
  \bibinfo {author} {\bibfnamefont {K.~G.}\ \bibnamefont {Arun}}, \ and\
  \bibinfo {author} {\bibfnamefont {B.~S.}\ \bibnamefont {Sathyaprakash}},\
  }\href@noop {} {} (\bibinfo {year} {2022}),\ \Eprint
  {http://arxiv.org/abs/2208.07757} {arXiv:2208.07757 [gr-qc]} \BibitemShut
  {NoStop}%
\bibitem [{\citenamefont {Datta}(2023)}]{Datta:2023muk}%
  \BibitemOpen
  \bibfield  {author} {\bibinfo {author} {\bibfnamefont {S.}~\bibnamefont
  {Datta}},\ }\href@noop {} {} (\bibinfo {year} {2023}),\ \Eprint
  {http://arxiv.org/abs/2303.04399} {arXiv:2303.04399 [gr-qc]} \BibitemShut
  {NoStop}%
\bibitem [{\citenamefont {Ade}\ \emph {et~al.}(2016)\citenamefont {Ade} \emph
  {et~al.}}]{Ade:2015xua}%
  \BibitemOpen
  \bibfield  {author} {\bibinfo {author} {\bibfnamefont {P.~A.~R.}\
  \bibnamefont {Ade}} \emph {et~al.} (\bibinfo {collaboration} {Planck
  Collaboration}),\ }\Doi {10.1051/0004-6361/201525830} {\bibfield  {journal}
  {\bibinfo  {journal} {Astron. Astrophys.},\ }\textbf {\bibinfo {volume}
  {594}},\ \bibinfo {pages} {A13} (\bibinfo {year} {2016})},\ \Eprint
  {http://arxiv.org/abs/1502.01589} {arXiv:1502.01589 [astro-ph.CO]}
  \BibitemShut {NoStop}%
\bibitem [{\citenamefont {Ezquiaga}\ \emph {et~al.}(2022)\citenamefont
  {Ezquiaga}, \citenamefont {Hu}, \citenamefont {Lagos}, \citenamefont {Lin},\
  and\ \citenamefont {Xu}}]{Ezquiaga:2022nak}%
  \BibitemOpen
  \bibfield  {author} {\bibinfo {author} {\bibfnamefont {J.~M.}\ \bibnamefont
  {Ezquiaga}}, \bibinfo {author} {\bibfnamefont {W.}~\bibnamefont {Hu}},
  \bibinfo {author} {\bibfnamefont {M.}~\bibnamefont {Lagos}}, \bibinfo
  {author} {\bibfnamefont {M.-X.}\ \bibnamefont {Lin}}, \ and\ \bibinfo
  {author} {\bibfnamefont {F.}~\bibnamefont {Xu}},\ }\Doi
  {10.1088/1475-7516/2022/08/016} {\bibfield  {journal} {\bibinfo  {journal}
  {J. Cosmol. Astropart. Phys.},\ }\textbf {\bibinfo {volume} {08}},\ \bibinfo
  {pages} {016} (\bibinfo {year} {2022})},\ \Eprint
  {http://arxiv.org/abs/2203.13252} {arXiv:2203.13252 [gr-qc]} \BibitemShut
  {NoStop}%
\bibitem [{LAL()}]{LALSuite}%
  \BibitemOpen
  \href@noop {} {}\bibinfo {note} {{LVK Algorithm Library Suite (LALSuite),
  \url{https://doi.org/10.7935/GT1W-FZ16}}}\BibitemShut {NoStop}%
\bibitem [{\citenamefont {Bardeen}\ \emph {et~al.}(1972)\citenamefont
  {Bardeen}, \citenamefont {Press},\ and\ \citenamefont
  {Teukolsky}}]{Bardeen:1972fi}%
  \BibitemOpen
  \bibfield  {author} {\bibinfo {author} {\bibfnamefont {J.~M.}\ \bibnamefont
  {Bardeen}}, \bibinfo {author} {\bibfnamefont {W.~H.}\ \bibnamefont {Press}},
  \ and\ \bibinfo {author} {\bibfnamefont {S.~A.}\ \bibnamefont {Teukolsky}},\
  }\Doi {10.1086/151796} {\bibfield  {journal} {\bibinfo  {journal} {Astrophys.
  J.},\ }\textbf {\bibinfo {volume} {178}},\ \bibinfo {pages} {347} (\bibinfo
  {year} {1972})}\BibitemShut {NoStop}%
\bibitem [{\citenamefont {Hofmann}\ \emph {et~al.}(2016)\citenamefont
  {Hofmann}, \citenamefont {Barausse},\ and\ \citenamefont
  {Rezzolla}}]{Hofmann:2016yih}%
  \BibitemOpen
  \bibfield  {author} {\bibinfo {author} {\bibfnamefont {F.}~\bibnamefont
  {Hofmann}}, \bibinfo {author} {\bibfnamefont {E.}~\bibnamefont {Barausse}}, \
  and\ \bibinfo {author} {\bibfnamefont {L.}~\bibnamefont {Rezzolla}},\ }\Doi
  {10.3847/2041-8205/825/2/L19} {\bibfield  {journal} {\bibinfo  {journal}
  {Astrophys. J. Lett.},\ }\textbf {\bibinfo {volume} {825}},\ \bibinfo {pages}
  {L19} (\bibinfo {year} {2016})},\ \Eprint {http://arxiv.org/abs/1605.01938}
  {arXiv:1605.01938 [gr-qc]} \BibitemShut {NoStop}%
\bibitem [{\citenamefont {Healy}\ and\ \citenamefont
  {Lousto}(2017)}]{Healy:2016lce}%
  \BibitemOpen
  \bibfield  {author} {\bibinfo {author} {\bibfnamefont {J.}~\bibnamefont
  {Healy}}\ and\ \bibinfo {author} {\bibfnamefont {C.~O.}\ \bibnamefont
  {Lousto}},\ }\Doi {10.1103/PhysRevD.95.024037} {\bibfield  {journal}
  {\bibinfo  {journal} {Phys. Rev. D},\ }\textbf {\bibinfo {volume} {95}},\
  \bibinfo {pages} {024037} (\bibinfo {year} {2017})},\ \Eprint
  {http://arxiv.org/abs/1610.09713} {arXiv:1610.09713 [gr-qc]} \BibitemShut
  {NoStop}%
\bibitem [{\citenamefont {Jim\'enez-Forteza}\ \emph {et~al.}(2017)\citenamefont
  {Jim\'enez-Forteza}, \citenamefont {Keitel}, \citenamefont {Husa},
  \citenamefont {Hannam}, \citenamefont {Khan},\ and\ \citenamefont
  {P\"urrer}}]{Jimenez-Forteza:2016oae}%
  \BibitemOpen
  \bibfield  {author} {\bibinfo {author} {\bibfnamefont {X.}~\bibnamefont
  {Jim\'enez-Forteza}}, \bibinfo {author} {\bibfnamefont {D.}~\bibnamefont
  {Keitel}}, \bibinfo {author} {\bibfnamefont {S.}~\bibnamefont {Husa}},
  \bibinfo {author} {\bibfnamefont {M.}~\bibnamefont {Hannam}}, \bibinfo
  {author} {\bibfnamefont {S.}~\bibnamefont {Khan}}, \ and\ \bibinfo {author}
  {\bibfnamefont {M.}~\bibnamefont {P\"urrer}},\ }\Doi
  {10.1103/PhysRevD.95.064024} {\bibfield  {journal} {\bibinfo  {journal}
  {Phys. Rev. D},\ }\textbf {\bibinfo {volume} {95}},\ \bibinfo {pages}
  {064024} (\bibinfo {year} {2017})},\ \Eprint
  {http://arxiv.org/abs/1611.00332} {arXiv:1611.00332 [gr-qc]} \BibitemShut
  {NoStop}%
\bibitem [{\citenamefont {{Johnson-McDaniel}}\ \emph
  {et~al.}(2016)\citenamefont {{Johnson-McDaniel}}, \citenamefont {Gupta},
  \citenamefont {Ajith}, \citenamefont {Keitel}, \citenamefont {Birnholtz},
  \citenamefont {Ohme},\ and\ \citenamefont {Husa}}]{spinfit-T1600168}%
  \BibitemOpen
  \bibfield  {author} {\bibinfo {author} {\bibfnamefont {N.~K.}\ \bibnamefont
  {{Johnson-McDaniel}}}, \bibinfo {author} {\bibfnamefont {A.}~\bibnamefont
  {Gupta}}, \bibinfo {author} {\bibfnamefont {P.}~\bibnamefont {Ajith}},
  \bibinfo {author} {\bibfnamefont {D.}~\bibnamefont {Keitel}}, \bibinfo
  {author} {\bibfnamefont {O.}~\bibnamefont {Birnholtz}}, \bibinfo {author}
  {\bibfnamefont {F.}~\bibnamefont {Ohme}}, \ and\ \bibinfo {author}
  {\bibfnamefont {S.}~\bibnamefont {Husa}},\ }\href
  {https://dcc.ligo.org/T1600168/public} {\emph {\bibinfo {title} {Determining
  the final spin of a binary black hole system including in-plane spins: Method
  and checks of accuracy}}},\ \bibinfo {type} {Tech. Rep.}\ \bibinfo {number}
  {{LIGO}-T1600168}\ (\bibinfo  {institution} {{LIGO} Project},\ \bibinfo
  {year} {2016})\ \bibinfo {note}
  {\url{https://dcc.ligo.org/LIGO-T1600168/public/main}}\BibitemShut {NoStop}%
\bibitem [{tim()}]{timeline_graphic}%
  \BibitemOpen
  \href@noop {} {}\bibinfo {note} {Gravitational wave detector observing
  timeline, \url{https://dcc.ligo.org/G2002127-v19/public}}\BibitemShut
  {NoStop}%
\bibitem [{\citenamefont {{Abbott}}\ \emph {et~al.}(2016)\citenamefont
  {{Abbott}} \emph {et~al.}}]{GW150914}%
  \BibitemOpen
  \bibfield  {author} {\bibinfo {author} {\bibfnamefont {B.~P.}\ \bibnamefont
  {{Abbott}}} \emph {et~al.} (\bibinfo {collaboration} {LIGO Scientific
  Collaboration and Virgo Collaboration}),\ }\Doi
  {10.1103/PhysRevLett.116.061102} {\bibfield  {journal} {\bibinfo  {journal}
  {Phys. Rev. Lett.},\ }\textbf {\bibinfo {volume} {116}},\ \bibinfo {eid}
  {061102} (\bibinfo {year} {2016})},\ \Eprint
  {http://arxiv.org/abs/1602.03837} {arXiv:1602.03837 [gr-qc]} \BibitemShut
  {NoStop}%
\bibitem [{\citenamefont {Skilling}(2004)}]{Skilling2004a}%
  \BibitemOpen
  \bibfield  {author} {\bibinfo {author} {\bibfnamefont {J.}~\bibnamefont
  {Skilling}},\ }\Doi {10.1063/1.1835238} {\bibfield  {journal} {\bibinfo
  {journal} {AIP Conf. Proc.},\ }\textbf {\bibinfo {volume} {735}},\ \bibinfo
  {pages} {395} (\bibinfo {year} {2004})}\BibitemShut {NoStop}%
\bibitem [{\citenamefont {Veitch}\ \emph {et~al.}(2015)\citenamefont {Veitch}
  \emph {et~al.}}]{Veitch:2014wba}%
  \BibitemOpen
  \bibfield  {author} {\bibinfo {author} {\bibfnamefont {J.}~\bibnamefont
  {Veitch}} \emph {et~al.},\ }\Doi {10.1103/PhysRevD.91.042003} {\bibfield
  {journal} {\bibinfo  {journal} {Phys. Rev. D},\ }\textbf {\bibinfo {volume}
  {91}},\ \bibinfo {pages} {042003} (\bibinfo {year} {2015})},\ \Eprint
  {http://arxiv.org/abs/1409.7215} {arXiv:1409.7215 [gr-qc]} \BibitemShut
  {NoStop}%
\bibitem [{\citenamefont {Blanchet}\ \emph {et~al.}(2004)\citenamefont
  {Blanchet}, \citenamefont {Damour}, \citenamefont {Esposito-Farese},\ and\
  \citenamefont {Iyer}}]{Blanchet:2004ek}%
  \BibitemOpen
  \bibfield  {author} {\bibinfo {author} {\bibfnamefont {L.}~\bibnamefont
  {Blanchet}}, \bibinfo {author} {\bibfnamefont {T.}~\bibnamefont {Damour}},
  \bibinfo {author} {\bibfnamefont {G.}~\bibnamefont {Esposito-Farese}}, \ and\
  \bibinfo {author} {\bibfnamefont {B.~R.}\ \bibnamefont {Iyer}},\ }\Doi
  {10.1103/PhysRevLett.93.091101} {\bibfield  {journal} {\bibinfo  {journal}
  {Phys. Rev. Lett.},\ }\textbf {\bibinfo {volume} {93}},\ \bibinfo {pages}
  {091101} (\bibinfo {year} {2004})},\ \Eprint
  {http://arxiv.org/abs/gr-qc/0406012} {arXiv:gr-qc/0406012} \BibitemShut
  {NoStop}%
\bibitem [{\citenamefont {Buonanno}\ \emph {et~al.}(2009)\citenamefont
  {Buonanno}, \citenamefont {Iyer}, \citenamefont {Ochsner}, \citenamefont
  {Pan},\ and\ \citenamefont {Sathyaprakash}}]{Buonanno:2009zt}%
  \BibitemOpen
  \bibfield  {author} {\bibinfo {author} {\bibfnamefont {A.}~\bibnamefont
  {Buonanno}}, \bibinfo {author} {\bibfnamefont {B.~R.}\ \bibnamefont {Iyer}},
  \bibinfo {author} {\bibfnamefont {E.}~\bibnamefont {Ochsner}}, \bibinfo
  {author} {\bibfnamefont {Y.}~\bibnamefont {Pan}}, \ and\ \bibinfo {author}
  {\bibfnamefont {B.~S.}\ \bibnamefont {Sathyaprakash}},\ }\Doi
  {10.1103/PhysRevD.80.084043} {\bibfield  {journal} {\bibinfo  {journal}
  {Phys. Rev. D},\ }\textbf {\bibinfo {volume} {80}},\ \bibinfo {pages}
  {084043} (\bibinfo {year} {2009})},\ \Eprint {http://arxiv.org/abs/0907.0700}
  {arXiv:0907.0700 [gr-qc]} \BibitemShut {NoStop}%
\bibitem [{\citenamefont {Moore}\ \emph {et~al.}(2016)\citenamefont {Moore},
  \citenamefont {Favata}, \citenamefont {Arun},\ and\ \citenamefont
  {Mishra}}]{Moore:2016qxz}%
  \BibitemOpen
  \bibfield  {author} {\bibinfo {author} {\bibfnamefont {B.}~\bibnamefont
  {Moore}}, \bibinfo {author} {\bibfnamefont {M.}~\bibnamefont {Favata}},
  \bibinfo {author} {\bibfnamefont {K.~G.}\ \bibnamefont {Arun}}, \ and\
  \bibinfo {author} {\bibfnamefont {C.~K.}\ \bibnamefont {Mishra}},\ }\Doi
  {10.1103/PhysRevD.93.124061} {\bibfield  {journal} {\bibinfo  {journal}
  {Phys. Rev. D},\ }\textbf {\bibinfo {volume} {93}},\ \bibinfo {pages}
  {124061} (\bibinfo {year} {2016})},\ \Eprint
  {http://arxiv.org/abs/1605.00304} {arXiv:1605.00304 [gr-qc]} \BibitemShut
  {NoStop}%
\bibitem [{\citenamefont {Cutler}\ and\ \citenamefont
  {Vallisneri}(2007)}]{Cutler:2007mi}%
  \BibitemOpen
  \bibfield  {author} {\bibinfo {author} {\bibfnamefont {C.}~\bibnamefont
  {Cutler}}\ and\ \bibinfo {author} {\bibfnamefont {M.}~\bibnamefont
  {Vallisneri}},\ }\Doi {10.1103/PhysRevD.76.104018} {\bibfield  {journal}
  {\bibinfo  {journal} {Phys. Rev. D},\ }\textbf {\bibinfo {volume} {76}},\
  \bibinfo {pages} {104018} (\bibinfo {year} {2007})},\ \Eprint
  {http://arxiv.org/abs/0707.2982} {arXiv:0707.2982 [gr-qc]} \BibitemShut
  {NoStop}%
\bibitem [{\citenamefont {Cutler}\ and\ \citenamefont
  {Flanagan}(1994)}]{Cutler:1994ys}%
  \BibitemOpen
  \bibfield  {author} {\bibinfo {author} {\bibfnamefont {C.}~\bibnamefont
  {Cutler}}\ and\ \bibinfo {author} {\bibfnamefont {{\'E}.~{\'E}.}\
  \bibnamefont {Flanagan}},\ }\Doi {10.1103/PhysRevD.49.2658} {\bibfield
  {journal} {\bibinfo  {journal} {Phys. Rev. D},\ }\textbf {\bibinfo {volume}
  {49}},\ \bibinfo {pages} {2658} (\bibinfo {year} {1994})},\ \Eprint
  {http://arxiv.org/abs/gr-qc/9402014} {arXiv:gr-qc/9402014} \BibitemShut
  {NoStop}%
\bibitem [{\citenamefont {Poisson}\ and\ \citenamefont
  {Will}(1995)}]{Poisson:1995ef}%
  \BibitemOpen
  \bibfield  {author} {\bibinfo {author} {\bibfnamefont {E.}~\bibnamefont
  {Poisson}}\ and\ \bibinfo {author} {\bibfnamefont {C.~M.}\ \bibnamefont
  {Will}},\ }\Doi {10.1103/PhysRevD.52.848} {\bibfield  {journal} {\bibinfo
  {journal} {Phys. Rev. D},\ }\textbf {\bibinfo {volume} {52}},\ \bibinfo
  {pages} {848} (\bibinfo {year} {1995})},\ \Eprint
  {http://arxiv.org/abs/gr-qc/9502040} {arXiv:gr-qc/9502040} \BibitemShut
  {NoStop}%
\bibitem [{\citenamefont {Arun}\ \emph {et~al.}(2009)\citenamefont {Arun},
  \citenamefont {Buonanno}, \citenamefont {Faye},\ and\ \citenamefont
  {Ochsner}}]{Arun:2008kb}%
  \BibitemOpen
  \bibfield  {author} {\bibinfo {author} {\bibfnamefont {K.~G.}\ \bibnamefont
  {Arun}}, \bibinfo {author} {\bibfnamefont {A.}~\bibnamefont {Buonanno}},
  \bibinfo {author} {\bibfnamefont {G.}~\bibnamefont {Faye}}, \ and\ \bibinfo
  {author} {\bibfnamefont {E.}~\bibnamefont {Ochsner}},\ }\Doi
  {10.1103/PhysRevD.79.104023} {\bibfield  {journal} {\bibinfo  {journal}
  {Phys. Rev. D},\ }\textbf {\bibinfo {volume} {79}},\ \bibinfo {pages}
  {104023} (\bibinfo {year} {2009})},\ \bibinfo {note}
  {\href{https://doi.org/10.1103/PhysRevD.84.049901}{{\bf{84}}, 049901(E)
  (2011)}},\ \Eprint {http://arxiv.org/abs/0810.5336} {arXiv:0810.5336 [gr-qc]}
  \BibitemShut {NoStop}%
\bibitem [{\citenamefont {Mishra}\ \emph {et~al.}(2016)\citenamefont {Mishra},
  \citenamefont {Kela}, \citenamefont {Arun},\ and\ \citenamefont
  {Faye}}]{Mishra:2016whh}%
  \BibitemOpen
  \bibfield  {author} {\bibinfo {author} {\bibfnamefont {C.~K.}\ \bibnamefont
  {Mishra}}, \bibinfo {author} {\bibfnamefont {A.}~\bibnamefont {Kela}},
  \bibinfo {author} {\bibfnamefont {K.~G.}\ \bibnamefont {Arun}}, \ and\
  \bibinfo {author} {\bibfnamefont {G.}~\bibnamefont {Faye}},\ }\Doi
  {10.1103/PhysRevD.93.084054} {\bibfield  {journal} {\bibinfo  {journal}
  {Phys. Rev. D},\ }\textbf {\bibinfo {volume} {93}},\ \bibinfo {pages}
  {084054} (\bibinfo {year} {2016})},\ \Eprint
  {http://arxiv.org/abs/1601.05588} {arXiv:1601.05588 [gr-qc]} \BibitemShut
  {NoStop}%
\bibitem [{\citenamefont {Racine}(2008)}]{Racine:2008qv}%
  \BibitemOpen
  \bibfield  {author} {\bibinfo {author} {\bibfnamefont {{\'E}.}~\bibnamefont
  {Racine}},\ }\Doi {10.1103/PhysRevD.78.044021} {\bibfield  {journal}
  {\bibinfo  {journal} {Phys. Rev. D},\ }\textbf {\bibinfo {volume} {78}},\
  \bibinfo {pages} {044021} (\bibinfo {year} {2008})},\ \Eprint
  {http://arxiv.org/abs/0803.1820} {arXiv:0803.1820 [gr-qc]} \BibitemShut
  {NoStop}%
\bibitem [{\citenamefont {{Santamar{\'\i}a}}\ \emph {et~al.}(2010)\citenamefont
  {{Santamar{\'\i}a}}, \citenamefont {{Ohme}}, \citenamefont {{Ajith}},
  \citenamefont {{Br{\"u}gmann}}, \citenamefont {{Dorband}}, \citenamefont
  {{Hannam}}, \citenamefont {{Husa}}, \citenamefont {{M{\"o}sta}},
  \citenamefont {{Pollney}}, \citenamefont {{Reisswig}}, \citenamefont
  {{Robinson}}, \citenamefont {{Seiler}},\ and\ \citenamefont
  {{Krishnan}}}]{Santamaria:2010yb}%
  \BibitemOpen
  \bibfield  {author} {\bibinfo {author} {\bibfnamefont {L.}~\bibnamefont
  {{Santamar{\'\i}a}}}, \bibinfo {author} {\bibfnamefont {F.}~\bibnamefont
  {{Ohme}}}, \bibinfo {author} {\bibfnamefont {P.}~\bibnamefont {{Ajith}}},
  \bibinfo {author} {\bibfnamefont {B.}~\bibnamefont {{Br{\"u}gmann}}},
  \bibinfo {author} {\bibfnamefont {N.}~\bibnamefont {{Dorband}}}, \bibinfo
  {author} {\bibfnamefont {M.}~\bibnamefont {{Hannam}}}, \bibinfo {author}
  {\bibfnamefont {S.}~\bibnamefont {{Husa}}}, \bibinfo {author} {\bibfnamefont
  {P.}~\bibnamefont {{M{\"o}sta}}}, \bibinfo {author} {\bibfnamefont
  {D.}~\bibnamefont {{Pollney}}}, \bibinfo {author} {\bibfnamefont
  {C.}~\bibnamefont {{Reisswig}}}, \bibinfo {author} {\bibfnamefont {E.~L.}\
  \bibnamefont {{Robinson}}}, \bibinfo {author} {\bibfnamefont
  {J.}~\bibnamefont {{Seiler}}}, \ and\ \bibinfo {author} {\bibfnamefont
  {B.}~\bibnamefont {{Krishnan}}},\ }\Doi {10.1103/PhysRevD.82.064016}
  {\bibfield  {journal} {\bibinfo  {journal} {Phys. Rev. D},\ }\textbf
  {\bibinfo {volume} {82}},\ \bibinfo {pages} {064016} (\bibinfo {year}
  {2010})},\ \Eprint {http://arxiv.org/abs/1005.3306} {arXiv:1005.3306 [gr-qc]}
  \BibitemShut {NoStop}%
\bibitem [{\citenamefont {Schmidt}\ \emph {et~al.}(2015)\citenamefont
  {Schmidt}, \citenamefont {Ohme},\ and\ \citenamefont
  {Hannam}}]{Schmidt:2014iyl}%
  \BibitemOpen
  \bibfield  {author} {\bibinfo {author} {\bibfnamefont {P.}~\bibnamefont
  {Schmidt}}, \bibinfo {author} {\bibfnamefont {F.}~\bibnamefont {Ohme}}, \
  and\ \bibinfo {author} {\bibfnamefont {M.}~\bibnamefont {Hannam}},\ }\Doi
  {10.1103/PhysRevD.91.024043} {\bibfield  {journal} {\bibinfo  {journal}
  {Phys. Rev. D},\ }\textbf {\bibinfo {volume} {91}},\ \bibinfo {pages}
  {024043} (\bibinfo {year} {2015})},\ \Eprint {http://arxiv.org/abs/1408.1810}
  {arXiv:1408.1810 [gr-qc]} \BibitemShut {NoStop}%
\bibitem [{\citenamefont {London}\ and\ \citenamefont
  {Fauchon-Jones}(2019)}]{London:2018nxs}%
  \BibitemOpen
  \bibfield  {author} {\bibinfo {author} {\bibfnamefont {L.}~\bibnamefont
  {London}}\ and\ \bibinfo {author} {\bibfnamefont {E.}~\bibnamefont
  {Fauchon-Jones}},\ }\Doi {10.1088/1361-6382/ab2f11} {\bibfield  {journal}
  {\bibinfo  {journal} {Classical Quantum Gravity},\ }\textbf {\bibinfo
  {volume} {36}},\ \bibinfo {pages} {235015} (\bibinfo {year} {2019})},\
  \Eprint {http://arxiv.org/abs/1810.03550} {arXiv:1810.03550 [gr-qc]}
  \BibitemShut {NoStop}%
\bibitem [{Sai()}]{Saini_PC}%
  \BibitemOpen
  \href@noop {} {}\bibinfo {note} {P.~Saini (private
  communication)}\BibitemShut {NoStop}%
\bibitem [{\citenamefont {Klein}(2021)}]{Klein:2021jtd}%
  \BibitemOpen
  \bibfield  {author} {\bibinfo {author} {\bibfnamefont {A.}~\bibnamefont
  {Klein}},\ }\href@noop {} {} (\bibinfo {year} {2021}),\ \Eprint
  {http://arxiv.org/abs/2106.10291} {arXiv:2106.10291 [gr-qc]} \BibitemShut
  {NoStop}%
\bibitem [{\citenamefont {Nagar}\ \emph {et~al.}(2021)\citenamefont {Nagar},
  \citenamefont {Bonino},\ and\ \citenamefont {Rettegno}}]{Nagar:2021gss}%
  \BibitemOpen
  \bibfield  {author} {\bibinfo {author} {\bibfnamefont {A.}~\bibnamefont
  {Nagar}}, \bibinfo {author} {\bibfnamefont {A.}~\bibnamefont {Bonino}}, \
  and\ \bibinfo {author} {\bibfnamefont {P.}~\bibnamefont {Rettegno}},\ }\Doi
  {10.1103/PhysRevD.103.104021} {\bibfield  {journal} {\bibinfo  {journal}
  {Phys. Rev. D},\ }\textbf {\bibinfo {volume} {103}},\ \bibinfo {pages}
  {104021} (\bibinfo {year} {2021})},\ \Eprint
  {http://arxiv.org/abs/2101.08624} {arXiv:2101.08624 [gr-qc]} \BibitemShut
  {NoStop}%
\bibitem [{\citenamefont {Ramos-Buades}\ \emph {et~al.}(2022)\citenamefont
  {Ramos-Buades}, \citenamefont {Buonanno}, \citenamefont {Khalil},\ and\
  \citenamefont {Ossokine}}]{Ramos-Buades:2021adz}%
  \BibitemOpen
  \bibfield  {author} {\bibinfo {author} {\bibfnamefont {A.}~\bibnamefont
  {Ramos-Buades}}, \bibinfo {author} {\bibfnamefont {A.}~\bibnamefont
  {Buonanno}}, \bibinfo {author} {\bibfnamefont {M.}~\bibnamefont {Khalil}}, \
  and\ \bibinfo {author} {\bibfnamefont {S.}~\bibnamefont {Ossokine}},\ }\Doi
  {10.1103/PhysRevD.105.044035} {\bibfield  {journal} {\bibinfo  {journal}
  {Phys. Rev. D},\ }\textbf {\bibinfo {volume} {105}},\ \bibinfo {pages}
  {044035} (\bibinfo {year} {2022})},\ \Eprint
  {http://arxiv.org/abs/2112.06952} {arXiv:2112.06952 [gr-qc]} \BibitemShut
  {NoStop}%
\bibitem [{\citenamefont {Islam}\ \emph {et~al.}(2021)\citenamefont {Islam},
  \citenamefont {Varma}, \citenamefont {Lodman}, \citenamefont {Field},
  \citenamefont {Khanna}, \citenamefont {Scheel}, \citenamefont {Pfeiffer},
  \citenamefont {Gerosa},\ and\ \citenamefont {Kidder}}]{Islam:2021mha}%
  \BibitemOpen
  \bibfield  {author} {\bibinfo {author} {\bibfnamefont {T.}~\bibnamefont
  {Islam}}, \bibinfo {author} {\bibfnamefont {V.}~\bibnamefont {Varma}},
  \bibinfo {author} {\bibfnamefont {J.}~\bibnamefont {Lodman}}, \bibinfo
  {author} {\bibfnamefont {S.~E.}\ \bibnamefont {Field}}, \bibinfo {author}
  {\bibfnamefont {G.}~\bibnamefont {Khanna}}, \bibinfo {author} {\bibfnamefont
  {M.~A.}\ \bibnamefont {Scheel}}, \bibinfo {author} {\bibfnamefont {H.~P.}\
  \bibnamefont {Pfeiffer}}, \bibinfo {author} {\bibfnamefont {D.}~\bibnamefont
  {Gerosa}}, \ and\ \bibinfo {author} {\bibfnamefont {L.~E.}\ \bibnamefont
  {Kidder}},\ }\Doi {10.1103/PhysRevD.103.064022} {\bibfield  {journal}
  {\bibinfo  {journal} {Phys. Rev. D},\ }\textbf {\bibinfo {volume} {103}},\
  \bibinfo {pages} {064022} (\bibinfo {year} {2021})},\ \Eprint
  {http://arxiv.org/abs/2101.11798} {arXiv:2101.11798 [gr-qc]} \BibitemShut
  {NoStop}%
\bibitem [{\citenamefont {Ezquiaga}\ \emph {et~al.}(2021)\citenamefont
  {Ezquiaga}, \citenamefont {Holz}, \citenamefont {Hu}, \citenamefont {Lagos},\
  and\ \citenamefont {Wald}}]{Ezquiaga:2020gdt}%
  \BibitemOpen
  \bibfield  {author} {\bibinfo {author} {\bibfnamefont {J.~M.}\ \bibnamefont
  {Ezquiaga}}, \bibinfo {author} {\bibfnamefont {D.~E.}\ \bibnamefont {Holz}},
  \bibinfo {author} {\bibfnamefont {W.}~\bibnamefont {Hu}}, \bibinfo {author}
  {\bibfnamefont {M.}~\bibnamefont {Lagos}}, \ and\ \bibinfo {author}
  {\bibfnamefont {R.~M.}\ \bibnamefont {Wald}},\ }\Doi
  {10.1103/PhysRevD.103.064047} {\bibfield  {journal} {\bibinfo  {journal}
  {Phys. Rev. D},\ }\textbf {\bibinfo {volume} {103}},\ \bibinfo {pages}
  {064047} (\bibinfo {year} {2021})},\ \Eprint
  {http://arxiv.org/abs/2008.12814} {arXiv:2008.12814 [gr-qc]} \BibitemShut
  {NoStop}%
\bibitem [{\citenamefont {Vijaykumar}\ \emph {et~al.}(2022)\citenamefont
  {Vijaykumar}, \citenamefont {Mehta},\ and\ \citenamefont
  {Ganguly}}]{Vijaykumar:2022dlp}%
  \BibitemOpen
  \bibfield  {author} {\bibinfo {author} {\bibfnamefont {A.}~\bibnamefont
  {Vijaykumar}}, \bibinfo {author} {\bibfnamefont {A.~K.}\ \bibnamefont
  {Mehta}}, \ and\ \bibinfo {author} {\bibfnamefont {A.}~\bibnamefont
  {Ganguly}},\ }\href@noop {} {} (\bibinfo {year} {2022}),\ \Eprint
  {http://arxiv.org/abs/2202.06334} {arXiv:2202.06334 [gr-qc]} \BibitemShut
  {NoStop}%
\bibitem [{\citenamefont {Cardoso}\ and\ \citenamefont
  {Maselli}(2020)}]{Cardoso:2019rou}%
  \BibitemOpen
  \bibfield  {author} {\bibinfo {author} {\bibfnamefont {V.}~\bibnamefont
  {Cardoso}}\ and\ \bibinfo {author} {\bibfnamefont {A.}~\bibnamefont
  {Maselli}},\ }\Doi {10.1051/0004-6361/202037654} {\bibfield  {journal}
  {\bibinfo  {journal} {Astron. Astrophys.},\ }\textbf {\bibinfo {volume}
  {644}},\ \bibinfo {pages} {A147} (\bibinfo {year} {2020})},\ \Eprint
  {http://arxiv.org/abs/1909.05870} {arXiv:1909.05870 [astro-ph.HE]}
  \BibitemShut {NoStop}%
\bibitem [{\citenamefont {Barausse}\ \emph {et~al.}(2014)\citenamefont
  {Barausse}, \citenamefont {Cardoso},\ and\ \citenamefont
  {Pani}}]{Barausse:2014tra}%
  \BibitemOpen
  \bibfield  {author} {\bibinfo {author} {\bibfnamefont {E.}~\bibnamefont
  {Barausse}}, \bibinfo {author} {\bibfnamefont {V.}~\bibnamefont {Cardoso}}, \
  and\ \bibinfo {author} {\bibfnamefont {P.}~\bibnamefont {Pani}},\ }\Doi
  {10.1103/PhysRevD.89.104059} {\bibfield  {journal} {\bibinfo  {journal}
  {Phys. Rev. D},\ }\textbf {\bibinfo {volume} {89}},\ \bibinfo {pages}
  {104059} (\bibinfo {year} {2014})},\ \Eprint {http://arxiv.org/abs/1404.7149}
  {arXiv:1404.7149 [gr-qc]} \BibitemShut {NoStop}%
\bibitem [{\citenamefont {Bonvin}\ \emph {et~al.}(2017)\citenamefont {Bonvin},
  \citenamefont {Caprini}, \citenamefont {Sturani},\ and\ \citenamefont
  {Tamanini}}]{Bonvin:2016qxr}%
  \BibitemOpen
  \bibfield  {author} {\bibinfo {author} {\bibfnamefont {C.}~\bibnamefont
  {Bonvin}}, \bibinfo {author} {\bibfnamefont {C.}~\bibnamefont {Caprini}},
  \bibinfo {author} {\bibfnamefont {R.}~\bibnamefont {Sturani}}, \ and\
  \bibinfo {author} {\bibfnamefont {N.}~\bibnamefont {Tamanini}},\ }\Doi
  {10.1103/PhysRevD.95.044029} {\bibfield  {journal} {\bibinfo  {journal}
  {Phys. Rev. D},\ }\textbf {\bibinfo {volume} {95}},\ \bibinfo {pages}
  {044029} (\bibinfo {year} {2017})},\ \Eprint
  {http://arxiv.org/abs/1609.08093} {arXiv:1609.08093 [astro-ph.CO]}
  \BibitemShut {NoStop}%
\bibitem [{\citenamefont {Vijaykumar}\ \emph {et~al.}(2023)\citenamefont
  {Vijaykumar}, \citenamefont {Tiwari}, \citenamefont {Kapadia}, \citenamefont
  {Arun},\ and\ \citenamefont {Ajith}}]{Vijaykumar:2023tjg}%
  \BibitemOpen
  \bibfield  {author} {\bibinfo {author} {\bibfnamefont {A.}~\bibnamefont
  {Vijaykumar}}, \bibinfo {author} {\bibfnamefont {A.}~\bibnamefont {Tiwari}},
  \bibinfo {author} {\bibfnamefont {S.~J.}\ \bibnamefont {Kapadia}}, \bibinfo
  {author} {\bibfnamefont {K.~G.}\ \bibnamefont {Arun}}, \ and\ \bibinfo
  {author} {\bibfnamefont {P.}~\bibnamefont {Ajith}},\ }\href@noop {} {}
  (\bibinfo {year} {2023}),\ \Eprint {http://arxiv.org/abs/2302.09651}
  {arXiv:2302.09651 [astro-ph.HE]} \BibitemShut {NoStop}%
\bibitem [{\citenamefont {Hunter}(2007)}]{Hunter:2007ouj}%
  \BibitemOpen
  \bibfield  {author} {\bibinfo {author} {\bibfnamefont {J.~D.}\ \bibnamefont
  {Hunter}},\ }\Doi {10.1109/MCSE.2007.55} {\bibfield  {journal} {\bibinfo
  {journal} {Comput. Sci. Eng.},\ }\textbf {\bibinfo {volume} {9}},\ \bibinfo
  {pages} {90} (\bibinfo {year} {2007})}\BibitemShut {NoStop}%
\bibitem [{\citenamefont {Harris}\ \emph {et~al.}(2020)\citenamefont {Harris}
  \emph {et~al.}}]{Harris:2020xlr}%
  \BibitemOpen
  \bibfield  {author} {\bibinfo {author} {\bibfnamefont {C.~R.}\ \bibnamefont
  {Harris}} \emph {et~al.},\ }\Doi {10.1038/s41586-020-2649-2} {\bibfield
  {journal} {\bibinfo  {journal} {Nature (London)},\ }\textbf {\bibinfo
  {volume} {585}},\ \bibinfo {pages} {357} (\bibinfo {year} {2020})},\ \Eprint
  {http://arxiv.org/abs/2006.10256} {arXiv:2006.10256 [cs.MS]} \BibitemShut
  {NoStop}%
\bibitem [{\citenamefont {Hoy}\ and\ \citenamefont
  {Raymond}(2021)}]{Hoy:2020vys}%
  \BibitemOpen
  \bibfield  {author} {\bibinfo {author} {\bibfnamefont {C.}~\bibnamefont
  {Hoy}}\ and\ \bibinfo {author} {\bibfnamefont {V.}~\bibnamefont {Raymond}},\
  }\Doi {10.1016/j.softx.2021.100765} {\bibfield  {journal} {\bibinfo
  {journal} {SoftwareX},\ }\textbf {\bibinfo {volume} {15}},\ \bibinfo {pages}
  {100765} (\bibinfo {year} {2021})},\ \Eprint
  {http://arxiv.org/abs/2006.06639} {arXiv:2006.06639 [astro-ph.IM]}
  \BibitemShut {NoStop}%
\bibitem [{\citenamefont {Virtanen}\ \emph {et~al.}(2020)\citenamefont
  {Virtanen} \emph {et~al.}}]{Virtanen:2019joe}%
  \BibitemOpen
  \bibfield  {author} {\bibinfo {author} {\bibfnamefont {P.}~\bibnamefont
  {Virtanen}} \emph {et~al.},\ }\Doi {10.1038/s41592-019-0686-2} {\bibfield
  {journal} {\bibinfo  {journal} {Nat. Methods},\ }\textbf {\bibinfo {volume}
  {17}},\ \bibinfo {pages} {261} (\bibinfo {year} {2020})},\ \Eprint
  {http://arxiv.org/abs/1907.10121} {arXiv:1907.10121 [cs.MS]} \BibitemShut
  {NoStop}%
\bibitem [{\citenamefont {Waskom}(2021)}]{Waskom:2021psk}%
  \BibitemOpen
  \bibfield  {author} {\bibinfo {author} {\bibfnamefont {M.}~\bibnamefont
  {Waskom}},\ }\Doi {10.21105/joss.03021} {\bibfield  {journal} {\bibinfo
  {journal} {J. Open Source Softw.},\ }\textbf {\bibinfo {volume} {6}}
  (\bibinfo {year} {2021})},\ \doi {10.21105/joss.03021}\BibitemShut {NoStop}%
\end{thebibliography}%

\end{document}